\documentclass[structabstract]{aa}
\usepackage{natbib,graphicx}
\usepackage{lscape}
\usepackage{txfonts}
\usepackage{xspace}
\newcommand{\msun}{\rm{M}_\mathrm{\sun}}
\newcommand{\mpc}{$\msun\rm{pc}^{-3}$}
\newcommand{\mpcc}{$\msun\rm{pc}^{-2}$}
\newcommand{\tri}{TRILEGAL\xspace}
\newcommand{\besan}{Besan\c{c}on\xspace}
\newcommand{\juric}{Juri\'{c}\xspace}
\newcommand{\jj}{Just-Jahrei{\ss}\xspace}
\bibpunct{(}{)}{;}{a}{}{,} 
\usepackage{ctable}

\begin{document}
\title{Detailed comparison of Milky Way models based on stellar population
synthesis and SDSS star counts at the north Galactic pole}

\titlerunning{Comparison of the Milky Way model and SDSS data at the NGP}

\author{S. Gao\inst{1,2}
          \and
          A. Just\inst{1}
          \and
          E. K. Grebel\inst{1}}

\offprints{S. Gao, \email{sgao@nao.cas.cn}}

\institute{Astronomisches Rechen-Institut, Zentrum f\"{u}r Astronomie der
Universit\"{a}t Heidelberg, 
M\"{o}nchhofstr. 12--14, 69120 Heidelberg, Germany 
  \and National Astronomical Observatories, Chinese Academy of Sciences, 
  Datun Road 20A, Beijing 100012, China} 

\date{Received  October 2011 / Accepted }

\abstract
{This article investigates Sloan Digital Sky Survey (SDSS) star counts based on the \jj Galactic
disc model by analysing the TRI-dimensional modeL of thE GALaxy (\tri) and \besan models of the Milky Way.}
{We test the ability of the \tri {and \besan models} to reproduce the
colour-magnitude diagrams of SDSS data at the north
Galactic pole (NGP). We show that a Hess diagram analysis of colour-magnitude
diagrams is much more powerful than luminosity functions in determining the Milky
Way structure.}
{We derive a best-fitting \tri model to simulate the NGP field in the $(g-r,g)$
colour-magnitude diagram of SDSS filters via Hess diagrams. For the \besan
model, we simulate the luminosity functions (LF) and Hess diagrams in all SDSS
filters. We use a $\chi^2$ analysis and determine the median of the  relative
deviations in the  Hess diagrams to quantify the quality of the fits by the \tri
models and the \besan model in comparison and compare this to that of the \jj model. The input isochrones
in the colour-absolute magnitude diagrams of the thick disc and halo are tested
via the observed fiducial isochrones of globular clusters.}
{We find that the default parameter set lacking a thick disc
component gives the best representation of the LF in \tri. The Hess
diagram reveals that a metal-poor thick disc is needed. In the Hess
diagram, the median relative deviation of the \tri model and the SDSS
data amounts to 25 percent, whereas for the \jj model the deviation is
only 5.6 percent.  The isochrone analysis shows that the
representation of the main sequences of (at least metal-poor) stellar
populations in the $ugriz$ filter system is reliable. In contrast, the
red giant branches fail to match the observed fiducial sequences of
globular clusters. The \besan model shows a similar median relative
deviation of 26 percent in $(g-r,g)$. In the $u$ band, the deviations
are larger. There are significant offsets between the isochrone set
used in the \besan model and the observed fiducial isochrones.}
{In contrast to Hess diagrams, luminosity functions are insensitive to the detailed structure of
the Milky Way components due to the extended spatial distribution
along the line of sight. The flexibility
of the input parameters in the \tri model is insufficient to
successfully reproduce star count distributions in colour-magnitude
diagrams. In the \besan model, an improvement of the isochrones in the
$ugriz$ filter system is needed in order to improve the predicted
star counts.
}


\maketitle
%

\section{Introduction}

Since the seminal work of \citet{1980BSa,1980BSb,1984BS}, the method
of simulating star counts based on analytic Milky Way models has
evolved considerably. The structural and evolutionary parameters of
the Milky Way components have been refined in multi-population
scenarios. However, a ``concordance" model including the detailed
shape and parameters of the density profiles, star formation history
(SFH), age-velocity dispersion relation (AVR), etc., has not yet been
reached.

On the other hand, there is a rapidly increasing set of observations
consisting of more accurate and much deeper photometry to constrain the
parameters of the models of each population. The observations include
wider and deeper sky coverage and additional filter systems providing
results that may require modifications of the existing detailed Milky
Way models.  The Sloan Digital Sky Survey
\citep[SDSS,][]{SDSS,Stoughton2002} covers more than one third of the
celestial sphere providing photometric data in five bands of
$2.6\times10^8$ stars and spectra of more than 500,000 stars
\citep[for the most recent eighth data release (DR8), see][]{dr8}. The
SDSS has become one of the most widely used databases for Galactic
and extragalactic astronomy in recent years. The availability of
SDSS data enables us to make significant progress in constraining Galactic
model parameters.

Analytic models are a powerful tool to constrain either evolutionary
scenarios or assumptions about Galactic structure by
comparing model predictions with a large variety of
observational constraints such as star counts and kinematics. Models
can be used to create probability distributions in the parameter space
of observable or mock catalogues via Monte Carlo simulations.
Realistic parameters and details for further improvement can be
obtained by comparing model predictions with suitable observations,
such as multi-directional photometric star counts. Within these
parameters and functions, the density profiles, SFHs, and AVRs are
important inputs.

To simulate and predict star counts in every direction of the
sky in the bands of various photometric systems, several automated
tools are available on the web, e.g., the ``\besan'' model \footnote{http://model.obs-besancon.fr/}
by \citet{Robin} and the TRI-dimensional modeL of thE GALaxy (``\tri'') model \footnote{http://stev.oapd.inaf.it/cgi-bin/trilegal} 
(\citealt{1998Girardi,2002Girardi,2005Girardi}), which are able to deal
with isochrone construction, synthetic photometry, and simulations of
the stellar populations of the main Galactic components. \citet{Galaxia} presented {\em Galaxia}, a new tool which combines the
advantages of the \besan and the \tri model. The code generates very efficiently
 synthetic surveys and allows one to choose between a
much larger variety of input models. The first version of {\em
Galaxia} is expected to be presented online soon
\footnote{http://galaxia.sourceforge.net/}.

The ``\besan model'' \citep{Robin} is presently the most elaborate and
refined publicly available tool for predicting star counts. In this model,
the Milky Way is divided into four components (thin disc, thick disc,
spheroid, bulge), which are described by their spatial distribution,
SFHs, initial mass functions (IMFs), sets of evolutionary tracks,
kinematics, and metallicity characteristics, and include giants
and a white dwarf population. From Monte Carlo simulations, mock
catalogues including observable parameters as well as theoretical ones
are obtained.

The \tri model is a Monte Carlo simulation to predict the probability distributions of stars
across small sky areas and for special passbands systems, is still
work in progress. It is a population synthesis code to simulate
photometry of any field in the Milky Way \citep{2005Girardi}. It was
developed to model population synthesis star counts of the
Milky Way. It allows its users to create a pseudo stellar catalogue
including positions, multi-colour photometry, and other physical
parameters according to user-defined structural parameters.

On the basis of SDSS photometry from Data Release 6 \citep[DR6, ][]{dr6},
\citet{Juric} derived a three-dimensional (3D) stellar distribution of the
Milky Way by applying universal photometric parallaxes. They then
fitted two discs, each with exponential profiles in the radial and
vertical directions. They found vertical scale heights of 300 pc and
900 pc for the thin and thick disc, respectively. The local density
normalisation of the discs and basic parameters of the stellar halo
were also determined. Subtracting the resulting, smooth Milky Way
model, overdensities (and underdensities) were revealed that could be
investigated in more detail.

We have developed a new local Milky Way model based on local stellar
kinematics and star counts in the solar neighbourhood \citep[hereafter
``\jj model"]{jj}. Compared to the \besan and \tri models, the main
advantages of our approach are that vertical density profiles are
fully consistent with the SFH and AVR in the gravitational field and
that we use a well-calibrated empirical main-sequence (MS). In
the current state of the \jj model, only MS stars are used instead of
including the full Hertzsprung-Russell diagram for each stellar
sub-population. The comparison of the basic \jj disc model with SDSS
star counts in the NGP field with $b>80{\degr}$ enabled us to select a
best-fitting SFH and AVR \citep{jgv}. Predicted star-number densities
fit the data with a median relative deviation of $5.6\%$ across the
Hess diagram. We use the \jj model to judge the ability of the
\tri and the \besan model to reproduce SDSS data at the NGP.

In this article, we present a test of the ability of the \tri code to
reproduce the different Milky Way models mentioned above, since the
\tri model is presently the only code that allows an interactive
analysis of the impact of model parameter variations on star count
predictions. Additionally, we try to determine a best-fit model by a
systematic input-parameter optimisation in the \tri model. The \besan
model has no free parameters and we perform a multi-colour analysis
over all five filters in the $ugriz$ system. We restrict the analysis
to the SDSS star-count data of the NGP field, since these are very
high quality data and they give a direct measure of the controversial
vertical structures of thin and thick discs.

The output catalogues of the \tri and \besan models with photometric
parameters can be used to analyse the contributions of different
populations to the observed star counts and to quantify systematic
deviations in order to identify the weaknesses that need to be resolved.

Sect.\ 2 introduces the data used in this work. In Sect.\ 3, we
describe the details of the \jj, \tri, \besan, and the \juric models
used in our analysis. In Sect.\ 4, we present the results of the \tri
model concerning the isochrones, luminosity functions, and Hess
diagrams. In Sect.\ 5, we present the comparisons between the \besan
model and the data using similar methods. In Sect.\ 6, we summarise the
results of our study.

\section{Data}

The SDSS has become one of the most widely used databases for
Galactic and extragalactic astronomy in recent years. The
effective exposure time for imaging is approximately 54 seconds with a
limiting magnitude of $r \sim 22.6$ mag. The $g$ and $r$ band point-source imaging of the SDSS ranges from $14$ mag to $22.5$ mag. We selected
stars from the Catalog Archive Server (CAS) of the seventh data
release of the SDSS (DR7, \citealt{dr7}). The SDSS table
\emph{\textbf{Star}}, which we utilise, contains the photometric
parameters (no redshifts nor spectroscopic parameters) of all primary
point-like objects classified as stars from
\emph{\textbf{PhotoPrimary}}. We only consider the NGP field (Galactic
latitude $b>80\degr$) in this publication. A series of lower latitude
fields will be analysed in the future.

We use de-reddened magnitudes from the SDSS point-spread function
photometry based on the values from the extinction map of
\citet{Schlegel}. In the following, all magnitudes and colours are
de-reddened. We select all stars in DR7 within a colour range of
$(g-r)=[-0.2, 1.2]$ and magnitude limits of $g=[14, 20.5]$ with
magnitude errors below 0.2 mag in $g$ and in $r$. In these magnitude and
colour ranges, the error limits remove $0.88\%$ of the sample in the NGP
field. A total number of $N=276\,180$ stars remain in the final sample.

For our analysis, we produce Hess diagrams of the stellar number-density distribution in colour-magnitude bins normalised to 1 deg$^2$
sky area in the colour-magnitude diagram (CMD) for a given direction.
One obvious deficiency are the unknown stellar distances, hence the
vertical axis of the CMDs necessarily shows apparent magnitudes
instead of absolute magnitudes.  The combined distribution of
(apparent) stellar luminosity and colour shows the spatial structure
and the population properties in the chosen direction, which can then
be compared with model-generated Hess diagrams.

\begin{figure}
    \includegraphics[width=0.5\textwidth]{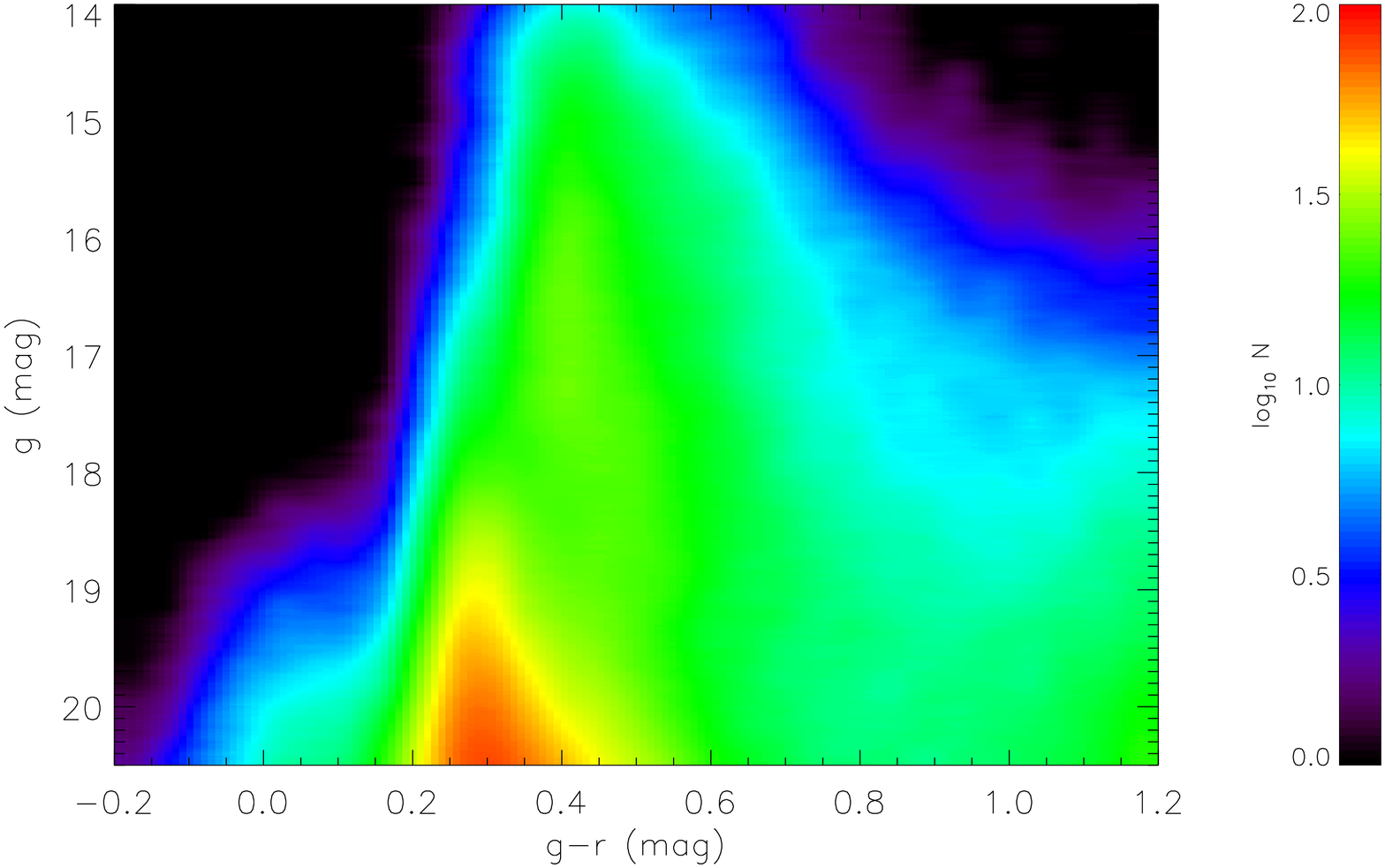}\\
    \includegraphics[width=0.5\textwidth,height=0.33\textwidth]{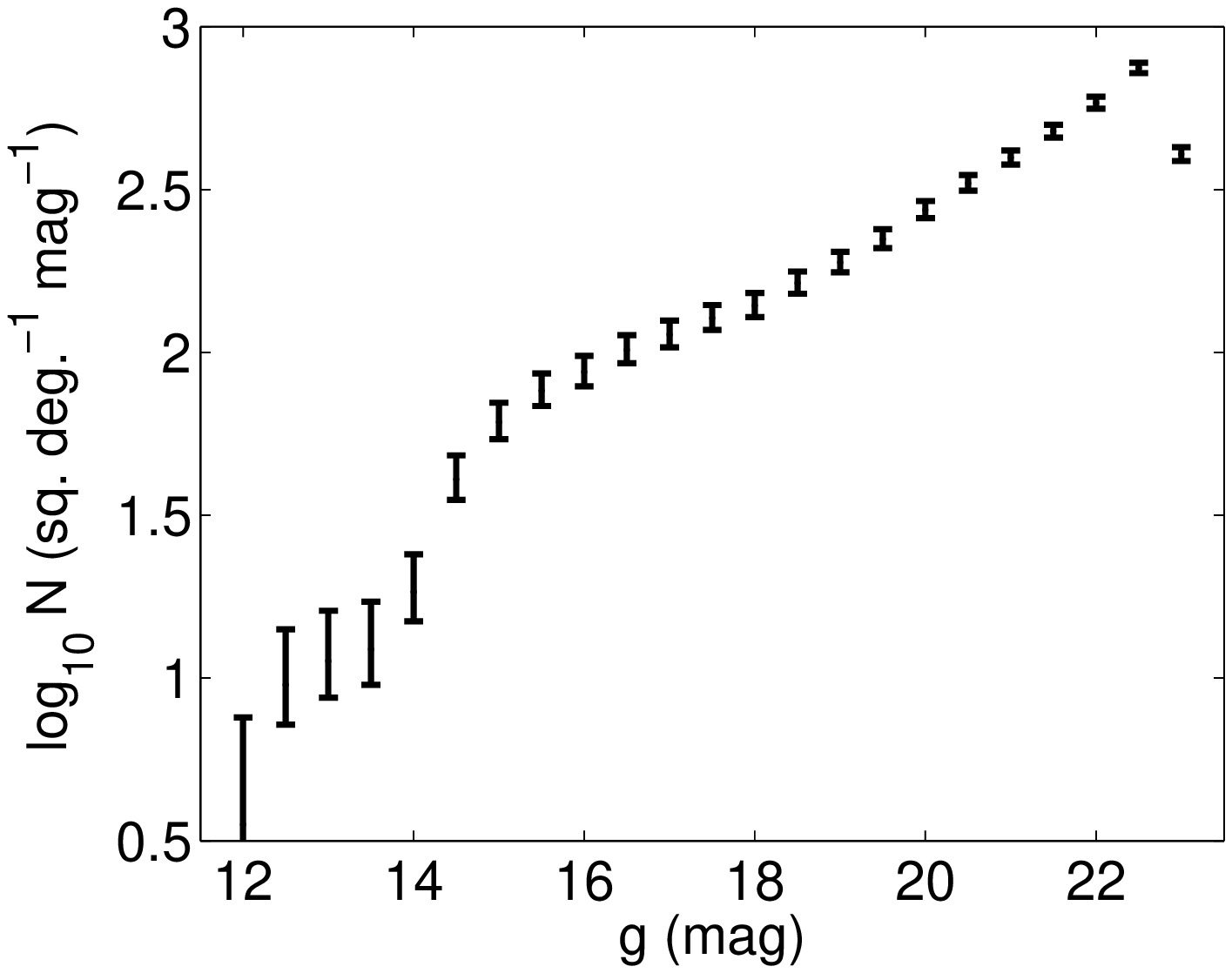}\\
    \caption{Top: Hess diagram of the NGP stellar sample taken from SDSS DR7 and
covering 313.36 deg$^2$. The x-axis is the de-reddened colour index $(g-r)$ and
the y-axis is the de-reddened apparent magnitude $g$. Number densities per
square degree, 1 mag in $g$ and 0.1 mag in $(g-r)$, are colour-coded on a
logarithmic scale. The stellar number density has been smoothed. Bottom:
De-reddened $g$ magnitude distribution of stars in the same colour range but
with an extended magnitude range. The bins of this luminosity function are the
logarithmic star numbers per 0.5 mag. The assumed Poissonian error bars are also
indicated. (A colour version is available on-line.)}
    \label{fig:NGP_SDSS_only_hess}
\end{figure}

In the top panel of Fig.\ \ref{fig:NGP_SDSS_only_hess}, the Hess diagram
in $(g-r,g)$ is shown. Number densities per square degree, 1 mag in
$g$ and 0.1 mag in $(g-r)$, are colour-coded on a logarithmic scale. The
bin size is $(0.05,0.5)$ in $(g-r,g)$ with additional smoothing in
steps of 0.01\,mag. From the upper right (purple/blue regions) to the lower
left (red triangular region) the thin disc, thick disc, and halo dominate,
respectively (see \citealt{jgv}). In the top left of the figure, no
stars are visible. In the bottom left, white dwarfs, blue horizonal
branch stars, and blue stragglers of the thick disc and halo may appear at
$(g-r)<0.2$, but the majority of objects are mis-identified point-like
extragalactic sources.

The luminosity function is obtained by projecting the Hess diagram
along the $(g-r)$-axis. In the lower panel of Fig.
\ref{fig:NGP_SDSS_only_hess}, the SDSS data are shown for the complete
range in the $g$ band in order to illustrate the total numbers of stars and the
cutoffs at bright and faint magnitudes. Poissonian error bars were
adopted. The incompleteness at $g>22$ mag is obvious and a significant
confusion with mis-identified extragalactic sources is likely to occur
here \citep{SDSS}. To be on the safe side, we choose for our analysis
$g=20.5$ mag as the faint limit. One significant jump appears at
$g=14$ mag, which is caused by the incompleteness due to the large magnitude errors.
The bright limit at 14 mag is set by the saturation of the stars in
the SDSS observations.

Apart from the DR7 catalogue data, we select the fiducial isochrones of
five globular clusters determined by \cite{An08} to compare the
photometric properties of the thick disc and halo populations in the
\tri and \besan models. Using the extinction parameters and distance
moduli of these globular clusters \citep{Harris1996}, the de-reddened
absolute magnitudes and colour indices are determined. The parameters
of the globular clusters are listed in Table \ref{tab:gcs}. These
clusters are selected to cover the metallicity range of the thick disc and
halo adopted in the \tri and the \besan models. 

\begin{table}\centering
\caption{Properties of five fiducial globular clusters observed in
the SDSS filter system.} \label{tab:gcs}
\begin{tabular}{lccc}
  \hline
  ID & $(m-M)_0$ & $\mathrm{E}(B-V)$ & $\mathrm{[Fe/H]}$  \\
  \hline
 M71      & 13.02 & 0.25 & -0.73  \\
 M5       & 14.37 & 0.03 & -1.27 \\
 M13      & 14.42 & 0.02 & -1.54   \\
 NGC 4147 & 16.42 & 0.02 & -1.83   \\
 M92      & 14.58 & 0.02 & -2.28   \\
  \hline
\end{tabular}
\tablefoot{Col.\ 1 lists the names of the globular clusters. Cols.\ 2
and 3 contain distance modulus and reddening ${\mathrm E(B-V)}$ from
\citet{An08}.  Col.\ 4 lists the metallicity of each cluster.
\citet{Harris1996} metallicity values are adopted here to ensure that we have a
consistent set. These clusters are used to enable us to compare with the \tri
and \besan models.}
\end{table}

%
%
\section{Models}\label{sec-models}

We describe the basic properties of the four models compared with the SDSS data.

%
%
\subsection{\jj model}\label{sub-jj}

\citet{jj} presented a local disc model where the thin disc is
composed of a continuous series of isothermal sub-populations
characterised by their SFH and AVR. A distinct isothermal thick disc
component can be added. The vertical density profiles
$\rho_\mathrm{s,j}(z)$ of all sub-populations $j$ are determined
self-consistently in the gravitational potential $\Phi(z)$ of the thin
disc, thick disc, gas component, and dark matter halo. 
The density profiles of the MS stars are calculated by adding the
profiles of the sub-populations, accounting for their respective MS
lifetimes. In a similar way, the thicknesses and the velocity
distribution functions of MS stars are derived as a function of
lifetime. 

For the thin disc of the \jj model, a factor of $1/g_{\rm eff}=1.53$
with $g_{\rm eff}=0.654$ needs to be applied in order to convert the
present-day stellar density to the density of ever formed stars. For the
thick disc and halo, the conversion factor is 1.8.

In \citet{jj}, we showed that the vertical profile of an isothermal
thick disc can be well-fitted by a
sech$^{\alpha_\mathrm{t}}(z/\alpha_\mathrm{t}h_\mathrm{t})$ function.
The power-law index $\alpha_\mathrm{t}$ and the scale height
$h_\mathrm{t}$ are connected via the velocity dispersion. 

The SFH and the AVR are the main input functions that are to be
determined as a pair of smooth continuous functions. For each given
SFH, the AVR is optimised to fit the velocity distribution functions of
MS stars in $V$-band magnitude bins in the solar neighbourhood. A metal
enrichment law (i.e., the age-metallicity relation (AMR)) is included that reproduces the local G-dwarf metallicity distribution.
Additionally, a consistent IMF is determined, that reproduces the
local luminosity function. In \citet{jj}, four different solutions with
similar minimum $\chi^2$ but very different SFHs were discussed.

In \citet{jgv}, we combined the models of \citet{jj} with the
empirically determined photometric properties of the local MS in the
$ugriz$ filter system.  By adopting a best-fitting procedure to determine the local
normalisations of thin disc, thick disc, and stellar halo as a function
of colour, we compared the star count predictions of the different
models for the NGP field. In this procedure, the thick disc parameters
$\alpha_t$ and $h_{t}$ were optimized and we added a simple stellar
halo with a flattened power-law-density distribution.

To measure the quality of the model and fit the data, we here
define a $\chi^2$ value for the Hess diagram. Each Hess diagram
contains $m\times n$ boxes of magnitudes and colours. The relative
difference (data $-$ model)/model of each Hess diagram is a matrix
$D_{ij}$ with the indices $i$ for colour and $j$ for magnitude.

The total $\chi^2$ is used to measure of the modelling quality,
which is our standard way of evaluating the goodness of a model. 
We apply the same definition of $\chi^2$ as \citet{jgv}
to measure the goodness of the \tri and \besan models.
We also define the median deviation $\Delta_{\rm{med}} =
{\rm{median}}(|D_{ij}|)$ of the relative differences $D_{ij}$ for each
Hess diagram to quantify the typical difference between model and data. 

In \citet{jgv} it turned out that the \citet{jj} ``model A'' is
clearly superior to the other models. We therefore select ``model A''
as the fiducial \jj model. Its basic density profiles and parameters
are listed in the last column of Table \ref{tab:tri:def} for
reference.

\subsection{\tri}

The main goal of the \tri model is the simulation of Milky Way star counts in several photometric systems using simple profiles. For this purpose, \citet{2002Girardi} developed a detailed method to convert theoretical stellar spectral libraries to photometrically measurable luminosities. These authors can produce these transformations routinely for any new photometric system, such as the five-colour filters of the SDSS \citep{2004Girardi}. In addition, \tri has been used to predict stellar photometry in different systems. 

Being able to simulate both very deep and very shallow surveys is one of the most difficult goals of \tri. For shallower surveys such as Hipparcos data, MS, red giant branch, and red clump stars with partly very low number statistics need to be generated. On the other hand, for more sensitive surveys an extended lower-luminosity MS, which 

The measurements of stellar age ($\tau$) and metallicity ($Z$) provide parameters such as bolometric magnitude $M_{bol}$, effective temperature $T_{\rm{eff}}$, surface gravity $g$,
core mass, and surface chemical composition.

The IMF is a crucial ingredient in Galactic modelling, because it determines the relative number of stars with different photometric properties and lifetimes. For comparison and convenience, the default shape of the IMF is a log-normal function in \tri, following \citet{2001Chabrier}. 

The SFH describes the age distribution of the stars. For old components, the shape of the SFH is quite poorly constrained, because the observed star counts are quite insensitive to it. For the thin disc, the SFH has not yet been well-determined (e.g., \citet{aumer,jj}). The interface of \tri allows only the choice between a constant star formation rate (SFR) up to an age of 11 Gyr and a two-step SFH with a 1.5 times enhancement in the SFR between the ages of 1 and 4 Gyr.

The AMR is also an important ingredient of the Galactic model, because it can help modify
the stellar CMDs. For the thin disc, the abundances of
\citet{Rocha-Pinto} are adopted in \tri.

There are five Galactic components in \tri: thin disc, thick disc, halo, bulge, and an extinction layer in the disc.

The mass density of the thin disc decreases exponentially with Galactocentric radius $R$ projected onto the plane of the disc. The Galactic thin disc is radially exponential and locally not isothermal, because the scale height $h_{\rm{d}}$ is a function of
stellar age and the thin disc is not isolated. In \tri, one can choose
between an exponential and a sech$^2$ profile for the vertical densities
of both the thin and the thick disc. The typical height of the
disc sub-populations depends on stellar age. The stellar density in the
solar neighbourhood $\rho_d(\sun)$ is the mass density of all stars
ever born there.

The density of the thick disc can be described as either a double exponential or vertically with a sech$^2$ form similar to that of the thin disc. However $h_{\rm{t}}$ is not a function of stellar age because the thick disc is predominantly an old population. The SFH is constant over an age range of 11--12 Gyr with a metallicity [Fe/H] $=-0.67 \pm 0.1$ dex and an $\alpha$-enhancement of [$\alpha$/Fe] $\sim 0.3$.

The mass density $\rho_{\rm{h}}$ of the halo follows a de Vaucouleurs $r^{1/4}$ law \citep{Vaucouleurs} and allows for a spheroidal flattening \citep{Gilmore}. The density profile of the stellar halo can be described by the parameters effective radius, oblateness, and local density \citep{Young}. The SFH is constant over an age range of 12---13\,Gyr and there are two choices for the metallicity distribution. We fix the metallicity to be [Fe/H] $=-1.6 \pm 1$\,dex with a corresponding $\alpha$-enhancement of [$\alpha$/Fe] $=0.3$.

The bulge has not yet been calibrated well in \tri because of a lack of observational constraints.  Since the bulge does not contribute to high Galactic-latitude data, we do not discuss its parameters any further.

We do not use the dust extinction model of \tri but instead we compare star counts with de-reddened observational data as described in Sect.\ 2.

Each simulated photometric catalogue in \tri has a given direction that is a conical area with a maximum sky coverage of 10 deg$^2$. The centre of the area can be specified in either Galactic coordinates $(l, b)$ or equatorial coordinates $(\alpha, \delta)$. The resolution in magnitudes, $\triangle{m}$, in \tri's calculations can be modified by the user.  The default value is $\triangle{m}=0.1$ mag and can be reduced to a minimum of $0.05$ mag.  Any data point with a resolution less than $\triangle{m}$ will not be presented. The output catalogues of \tri are based on a random number generator, which allows one to investigate the effect of noise.

Since the maximum sky area is much smaller than the observed field (313.36 deg$^2$ for the NGP field), we generate catalogues for 37 different sight lines (Table \ref{tab:tri_fields}) to cover the NGP field. We add up all the star counts and then normalise to star counts per deg$^2$.


\subsection{\besan model}

\citet{Robin} developed the population synthesis approach to simulate the structure and evolution of the Milky Way. The \besan model is based on different well-calibrated data sets and the structural parameters are optimised to reproduce a set of selected fields on the sky. The construction of each component is based on a set of evolutionary tracks and assumptions about stellar number-density distributions, which are both constrained by either dynamical considerations or empirical data and guided by a scenario of a continuous formation and evolution of stellar populations.

The \besan model deals with the thin disc is composed of seven isothermal sub-populations with different age ranges, kinematics parameters ($W$ velocity dispersion $\sigma_W$), scale heights, and local densities. The thick disc is considered as a single population with an age of 11 Gyr and a local density value ($1.34\times 10^{-3}\mathrm{M}_\odot \mathrm{pc}^{-3}$). The halo is modelled by a single population of older age (14 Gyr) and lower local density ($2.2\times 10^{-5}\mathrm{M}_\odot \mathrm{pc}^{-3}$ including white dwarfs).  This prescription is similar to our \jj model \citep{jj}. The main differences are that 1) in the \jj model, we adopt a continuous series of isothermal sub-populations in the thin disc; 2) the density profile and local density of each component are determined by solving iteratively Poisson and Jeans equations that are consistently based on the local kinematic data and the SDSS photometric data; and 3) we use a mean empirically calibrated MS instead of a full Hertzsprung-Russell 
diagram (HRD) based on stellar evolution.

Since we limit our targets to the NGP field with a Galactic latitude $b>80 \degr$, stars belonging to the Galactic bulge can be ignored. The stellar populations in the NGP field can be divided into three distinct components: the thin disc, the thick disc, and stellar halo. For details of the properties of the Galactic components, we refer to \citet{Robin}. With the NGP field, we can test the vertical structure of the Galactic components and the calibration of the underlying isochrones in the SDSS filter system.

The \besan model outputs $u^*g'r'i'z'$ photometry. To obtain SDSS magnitudes in $g$, $r$, $i$, and $z$ and the colours $u-g$, $g-r$, $r-i$ and $i-z$ from the \besan model, we transform the photometric system from the $u^*$, $g'$, $r'$, $i'$ and $z'$ filter set using a two-step method. In the first step, we transform from the special $u^*$ filter to $u'$ \citep{Clem08}. In the second step, we transform from the $u'$, $g'$, $r'$, $i'$, and $z'$ filters to $u$, $g$, $r$, $i$, and $z$ following the transformations in \citet{Tucker2006} and \citet{Ju08}.


\subsection{\juric model}

In \cite{Juric}, a map of the 3D number density distribution in the Milky Way derived by photometric parallaxes that were based on star count data of SDSS DR6 is presented. The stellar distances in the sample were determined by the colour-absolute magnitude relation of MS stars. Most stars are at high latitudes and cover a distance range from 100 pc to 20 kpc and 6500 deg$^2$ on the sky in the sample.

The density distribution is fitted using two double-exponential discs and the sample implies the existence of an oblate halo. The estimated errors in the model parameters are smaller than $\sim 20\%$ and the errors in the scale heights of discs are smaller than $\sim 10\%$. By adjusting the fraction of binaries and the bias correction, the scale heights of the discs are found to be $h_{\rm{d}}=300$ pc and $h_{\rm{t}}=900$ pc, and the local calibration is $\rho_{\rm{t}}(\sun)/\rho_{\rm{d}}(\sun)=0.12$.  The best-fit of the spheroid leads to an oblateness of 0.64, a power-law radial profile of $\rho_{\rm{h}}\propto r^{-2.8}$, and a local calibration of $\rho_{\rm{h}}(\sun)/\rho_{\rm{d}}(\sun)=0.005$. After subtraction of the smooth background distribution, there are still overdensity areas, in particular areas that are explained as tidal streams including the ``Virgo overdensity'' \citep{duffau2006} and the Monoceros stream \citep{yanny2003}.

The solar offset $Z_{\sun}$ is $(25\pm 5)$ pc or $(24\pm 5)$ pc at the bright and faint ends, respectively, which is consistent with the value of $24.2$ pc adopted by \tri.


\section{Simulations and results of \tri}
\begin{table}
\caption{The central Galactic coordinates of 37 selected fields simulated using \tri in order to be compared with SDSS observations.}
\label{tab:tri_fields}
\begin{tabular}{ll}
  \hline
  $b$ (deg) & $l$ (deg)  \\
  \hline
  90&   0\\
  87&   0, 60, 120, 180, 240, 300\\
  84&   0, 30, 60, 90, 120, 150, 180, 210, 240, 270, 300, 330\\
  81&   0, 20, 40, 60, 80, 100, 120, 140, 160, 180, 200, 220, \\
    &   240, 260, 280, 300, 320, 340\\
  \hline
\end{tabular}
\end{table}

The model \tri can be accessed via a web-based interface. Users can simulate the stellar populations in a special volume defined by the luminosity range and direction on the sky using different parameter sets. Because the parameter sets and profile shapes are relatively simple and the web-based programme can return the results in real time, we can compare the output interactively with the data.

In general, it is difficult to find the ``best'' input parameters for the components by using the web interface interactively, since the transitions of thin to thick disc and thick disc to halo in the CMD are not known a priori. We cannot search independently for the best-fitting parameters in each component.  We therefore first test both the parameters of the default model and those that reproduce best the other three Milky Way models described in Sect.\ \ref{sec-models}.

We fix some parameters and input functions in the \tri model in order to facilitate the comparison with observational data. We fix the IMF to a Chabrier log-normal function and use the default binary distribution with a binary fraction of 0.3 and a mass-ratio range of 0.7--1. The solar position is at the default value of $(R_{\sun}, z_{\sun})=(8700, 24.2)$~pc and we fix the SFH for the thin disc to be constant. For the other input parameters in the \tri model, we selected five sets of parameters as given in Table \ref{tab:tri:def}. The ``set 1'' parameters in this table are from the default input of \tri in the web interface, ``set 2'' is best adjusted to the \jj model \citep{jj}, ``set 3'' to the \besan model \citep{Robin}, ``set 4'' to the \juric model \citep{Juric}, and ``set 5'' is optimised (as described in Sect.\ \ref{sec-set5}).

Because in the \tri model, we have to choose between either an  exponential or a sech$^2$ shape of the disc profiles
\begin{eqnarray}\label{eq:exp:thick}
 \rho_{\rm{t}}(z) &=& \rho_{\rm{t}}(\sun)\exp( -\frac{z}{h_{\rm{t}}}), \\
 \rho_{\rm{t}}(z) &=& \rho_{\rm{t}}(\sun)\mathrm{sech}^2(\frac{z}{2h_{\rm{t}}}),
\end{eqnarray}
we determine for all combinations the minimum $\chi^2$ values and then select the profiles with the lowest $\chi^2$ value.

At colours bluer than $g-r=0.2$ mag, the SDSS data are unreliable owing to contamination by extragalactic objects and all selected models are unable to generate sufficient stars in this region. Therefore, the mean $\chi^2$ value $\langle\chi^2\rangle$ of the reduced colour range $0.2<g-r<1.2$ is treated as the standard way of determining the quality of the model. For completeness, the values $\langle\chi^2\rangle^\star$ for the full colour range are also listed in Table \ref{tab:tri:def}.

\begin{landscape}
\begin{table}\centering
\caption{The sets of input parameters in \tri and the $\chi^2$ values of the Hess diagram comparisons.}
\label{tab:tri:def}
\begin{tabular}{llccccccc}
\hline
  Quantity             &Unit & Set 1   & Set 2  & Set 3  & Set 4 & Set 5 & \besan model  & \jj model \\
                       &     & Default & \jj input & \besan input & Juri\'c input  & Optimization &  &\\
\hline
 $\langle\chi^2\rangle^\star$&- &  275.76   & 289.07   &  255.39  & 243.64  & 
208.38 & 549.95& -  \\
 $\langle\chi^2\rangle$&-& 151.98   & 214.65   &  207.60  & 137.88  &  104.19 & 408.15& 4.31 \\
 $\Delta_{\rm{med}}$      &-&  0.3035   & 0.2993       &  0.3445    &  0.4639 & 0.2576 & 0.2724 &0.0563\\
 \hline
 $z_0$                 & pc  &  95.00  & 60.64  & 13.61  & 300.00& 90.00 &- &-\\
 $t_0$                 & Gyr &  4.4    & 0.7118 &0.02016 & -     &  4.4 &- &- \\
 $\alpha$              &  -  &  1.6666 & 0.7576 & 0.5524 & 0     &  1.3 &- &- \\
 $h_{\rm{d,max}}$       & pc  &  767    & 506    &  443   & 300   &  459 &- & 670 \\
 $\rho_{\rm{d}}({\sun})$  &\mpc &         & 0.0297 &  0.030 & 0.029 &  -   & 0.0394 &0.037  \\
 $\Sigma_{\rm{d}}({\sun})$&\mpcc&  59.00  & 49.36  &  25.20 & 17.40 &  50.00 & - &29.4 \\
 form                 &  -  &sech$^2$ &sech$^2$& $\exp$ & $\exp$& sech$^2$ & Einasto &-\\
\hline
 $h_{\rm{t}}$       & pc  &  -      &  793   &  910   & 900   &  750 &- &800 \\  $\rho_{\rm{t}}({\sun})$&\mpc &  -      & 0.0044 & 0.0015 & 0.0034&  0.0014 & 0.00134 &0.0022\\
$\Sigma_{\rm{t}}({\sun})$&\mpcc &  -      & 7.08   &  2.82  & 6.12  &  4.2 & &5.3  \\
 form               &  -  &  -      & $\exp$ & $\exp$ & $\exp$&  sech$^2$ & cored $\exp$ &sech$^{1.16}$\\
\hline
  $r_{\rm{eff}}$/$\alpha_{\rm{h}}$  & pc/-  &   2800  & 11994   & 39436  & 18304 & 22000 &-2.44 &-3.0 \\
  $q_{\rm{h}}$          & -   &   0.65  & 0.65    &  0.71  & 0.59  & 0.61 & 0.76 & 0.7 \\
$\rho_{\rm{h}}({\sun})$ &\mpc&0.00015 &0.000158 &0.00009 &0.000145& 0.0001 & 0.00000932 & 0.00015\\
\hline
\end{tabular}
\tablefoot{Cols.\ 1 and 2 list the parameters of each Galactic component in \tri and their units. Cols.\ 3 to 7 list the adopted values of the five input sets for \tri. Set 1 is the default input of the web-based \tri interface. Set 2, 3, and 4 are from \tri fits with three models: \cite{jj}, \cite{Robin}, and \cite{Juric}. Set 5 (the last column) lists our best input set determined by the optimisation based on the results of Sets 1 to 4. We provide two mean $\chi^2$  values and one median absolute relative difference of the Hess diagram comparisons of the five input sets: $\langle\chi^2\rangle^\star$ is with all stars covering the full colour range in the Hess diagram, $\langle\chi^2\rangle$ is only stars with $0.2<g-r<1.2$, $\Delta_{\rm{med}}$ gives each median value of the absolute relative difference between the data and the model covering only for stars with $0.2<g-r<1.2$ in the Hess diagram. The scale-height parameters of the thin disc $h_{\rm{d}}$ and the local thin disc's density $\rho_{\rm{d}}({\sun}
)$ as well as the surface density $\Sigma_{\rm{d}}(\sun)$ are listed. $h_{\rm{t}}$ is the scale height of the thick disc. $\rho_{\rm{t}}(\sun)$ and $\Sigma_{\rm{t}}(\sun)$ are the locally calibrated volume and surface densities of the thick disc. The $r_{\rm{eff}}$ is the effective radius of the halo along the major axis with an oblateness $q_{\rm{h}}$. The $\alpha_{\rm{h}}$ is the index in the power law of the halo profile. One value of $r_{\rm{eff}}$ and $\alpha_{\rm{h}}$ is adopted based on the form of the halo profile. The $\Omega_{\rm{h}}(\sun)$ is the local volume density of the halo. Cols.\ 8 and 9 list the corresponding parameters of the \besan and the \jj models for reference.}
\end{table}
\end{landscape}

\begin{figure}
\includegraphics[width=0.5\textwidth,height=0.33\textwidth]{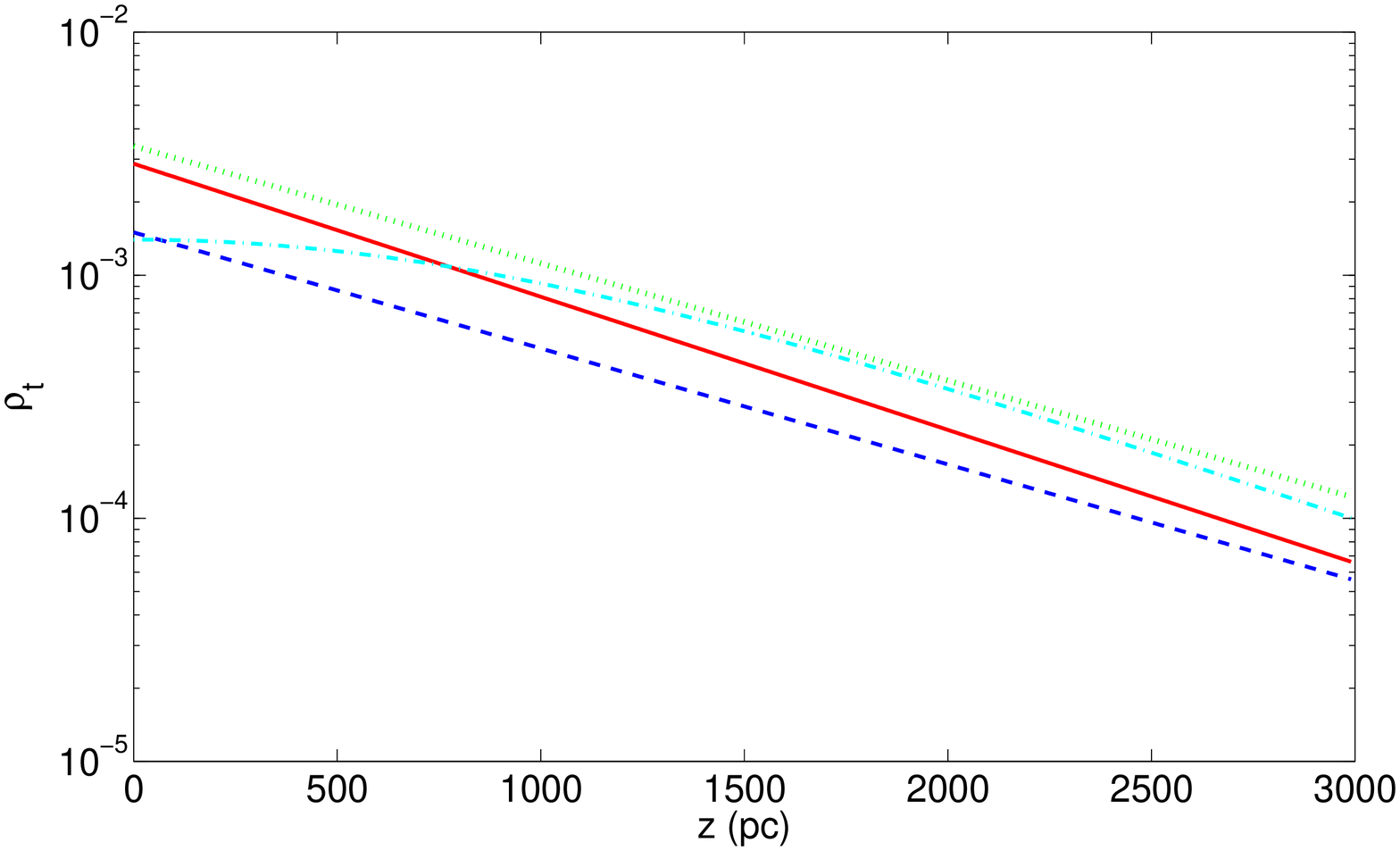}\\
\includegraphics[width=0.5\textwidth,height=0.33\textwidth]{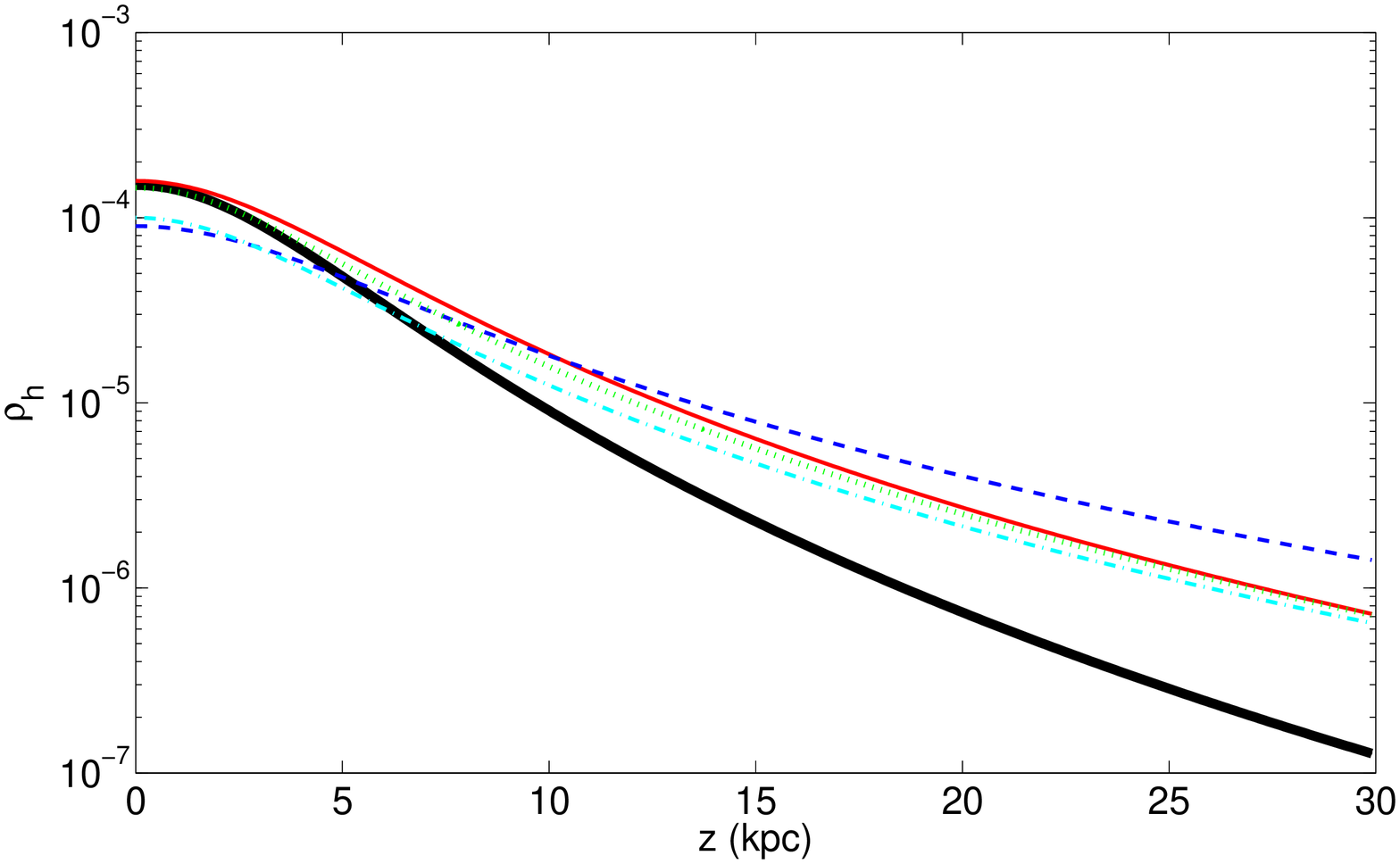}
    \caption{The vertical density profiles of thick disc (top panel) and halo (bottom panel) of parameter set 1 (black, thick solid line), set 2 (red, solid line), set 3 (blue, dashed line), set 4 (green, dotted line), and set 5 (cyan, dashed-dotted line). There is no thick disc in set 1. The $z$ ranges of the panels are different. The vertical axis $\rho_{\rm{t}}$ and $\rho_{\rm{h}}$ are the densities of the thick disc and the halo in units of \mpc . (A colour version is available on-line.)}
    \label{fig:thick:profile}
\end{figure}

Fig. \ref{fig:thick:profile} compares the vertical density profiles of the thick disc and halo for the different sets. There are significant differences between the models, even in the distance regime where the thick disc component dominates the star counts in the Hess diagram. We discuss this point later.

\begin{figure}\centering
    \includegraphics[width=0.5\textwidth]{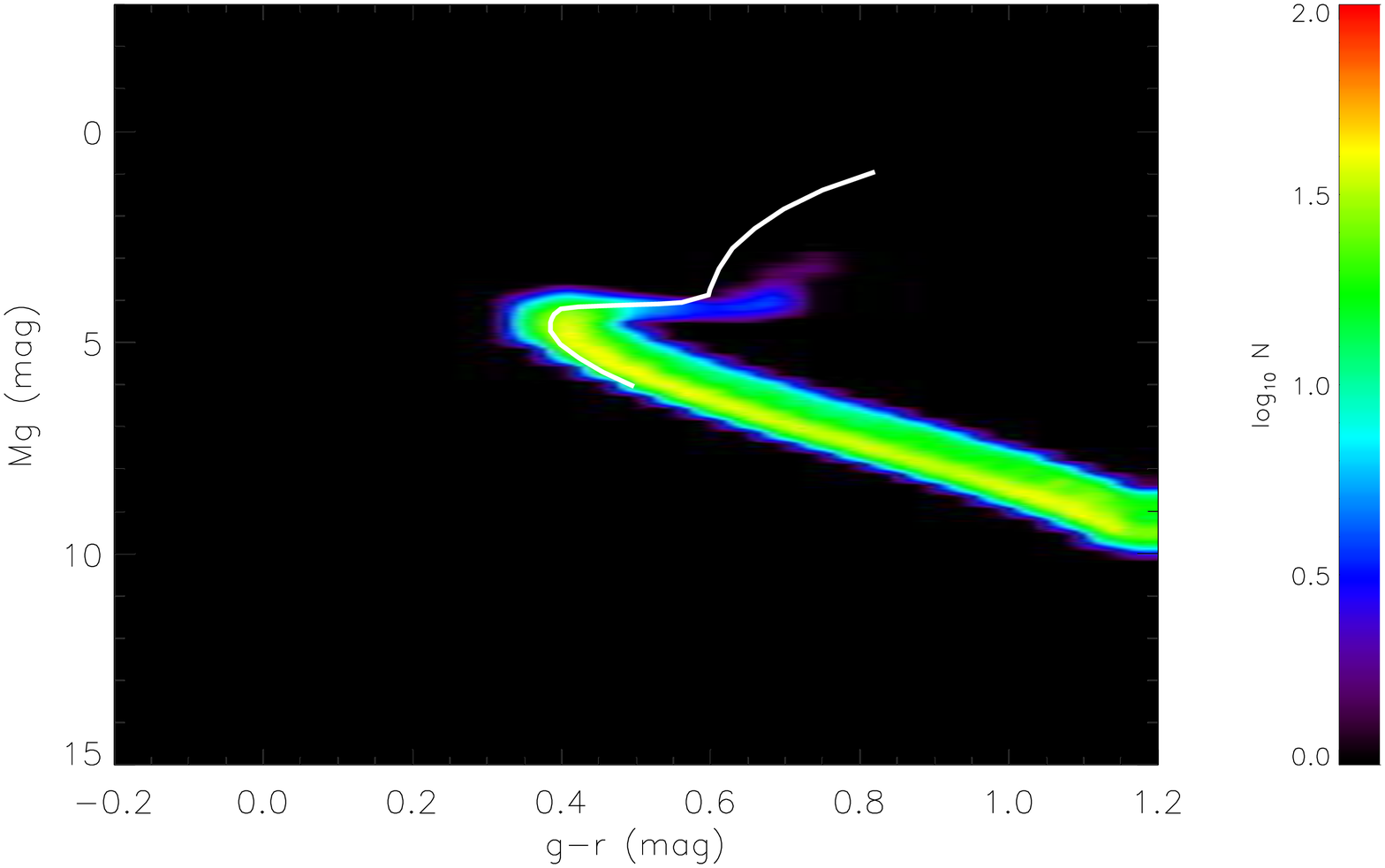}\\
    \includegraphics[width=0.5\textwidth]{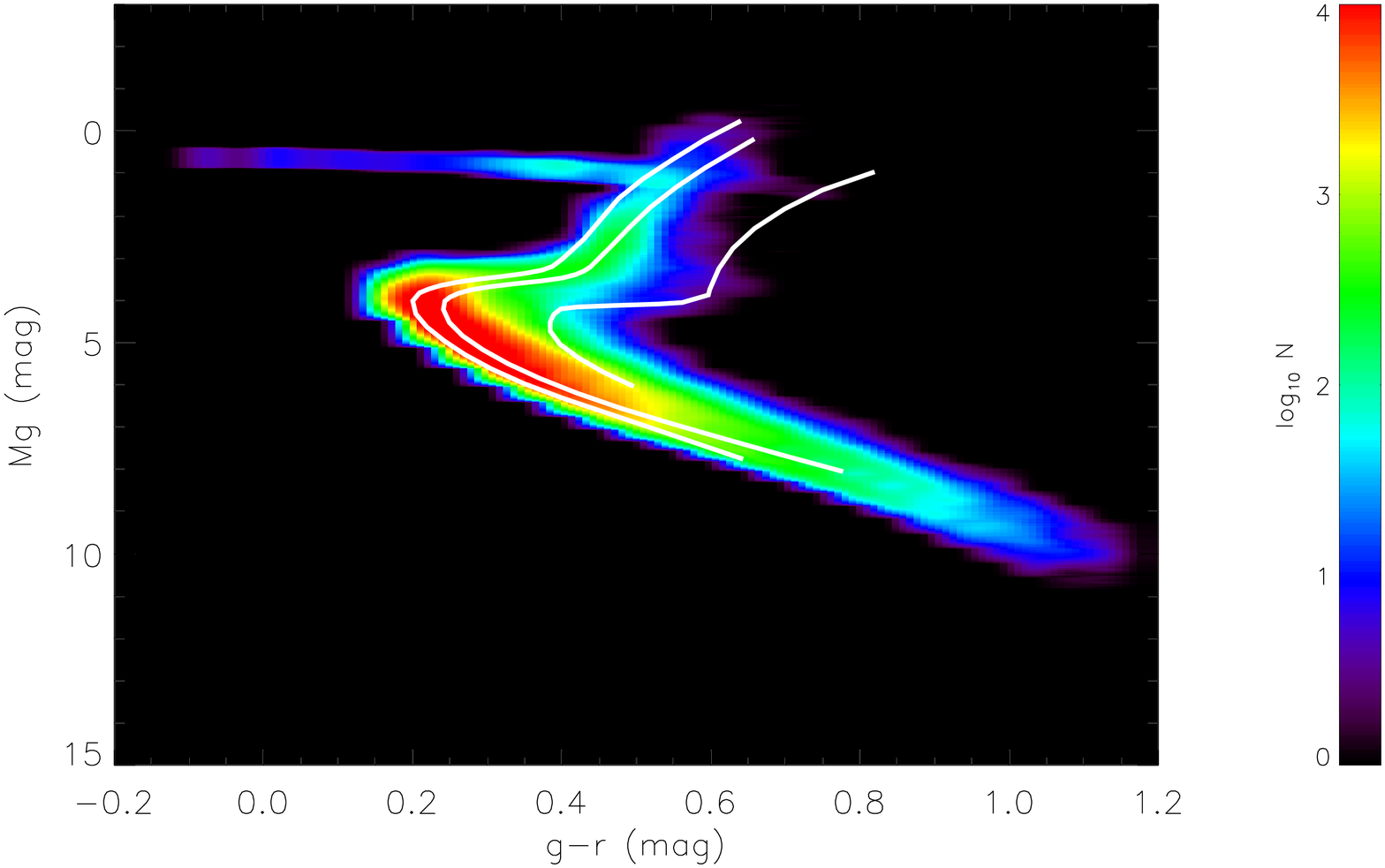}
    \caption{The fiducial isochrones of globular clusters from \citet{An08} are overplotted in the Hess diagrams of absolute magnitude. The upper panel shows the thick disc simulated by \tri using parameter set 2. The fiducial isochrone of the globular clusters M71 is overplotted as a white line. The lower panel shows the corresponding halo populations. From left to right, white lines represent the fiducial isochrones of the three globular clusters M71, M13, and M92 (Table \ref{tab:gcs}) covering the halo's metallicity range. (A colour version is available on-line.)}
    \label{fig:iso:tri}
\end{figure}

To separate the impact of the synthetic colours and luminosities of the underlying stellar populations from the structural parameters of the Milky Way model, we compare the colour-absolute magnitude diagrams provided by \tri with the observed fiducial isochrones of globular clusters with corresponding metallicities. This comparison is only useful for both thick disc and halo, which are essentially simple stellar populations with small ranges in age and metallicity. The colour-absolute magnitude diagrams of Set 2 are plotted in Fig.\ \ref{fig:iso:tri}. The cluster M71 ($\rm{[Fe/H]}=-0.73$) is used to provide fiducial isochrones of the thick disc (see \citealt{An08}). The three clusters M71, M13, and M92 are used to provide fiducial isochrones of the halo whose mean metallicity is given as $-1.6$ \citep{2005Girardi}. The consistency between the model and the fiducial isochrones for the location of the MS and the turn-off implies that the stellar population synthesis of \tri is satisfactory for predicting the 
dominant 
contribution to the star counts. However, there are some systematic inconsistencies concerning the location of the giant branches of the stellar isochrones.


\begin{figure*}\centering
    \includegraphics[width=0.45\textwidth,height=0.3\textwidth]{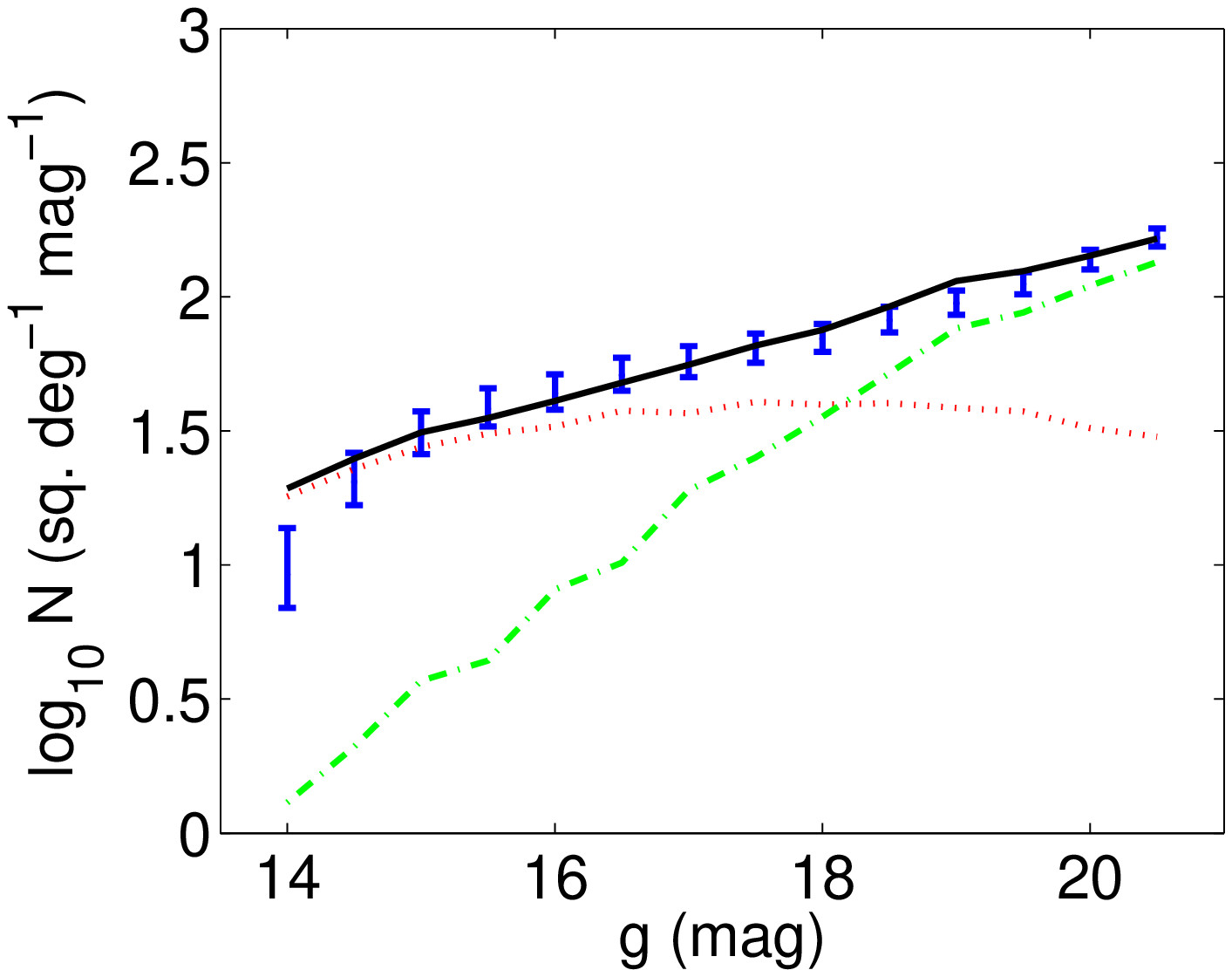}
    \includegraphics[width=0.45\textwidth,height=0.3\textwidth]{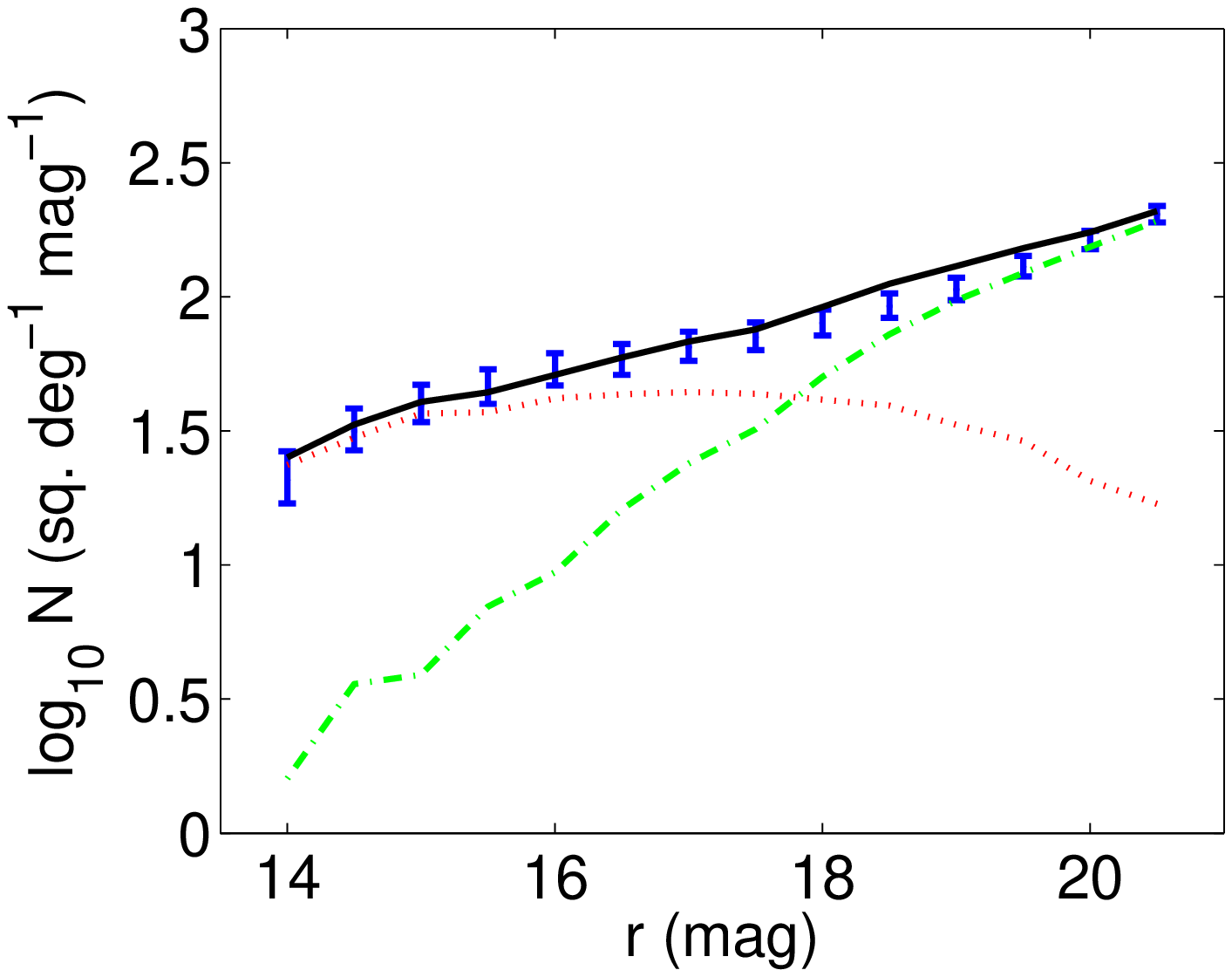}\\
    \includegraphics[width=0.45\textwidth,height=0.3\textwidth]{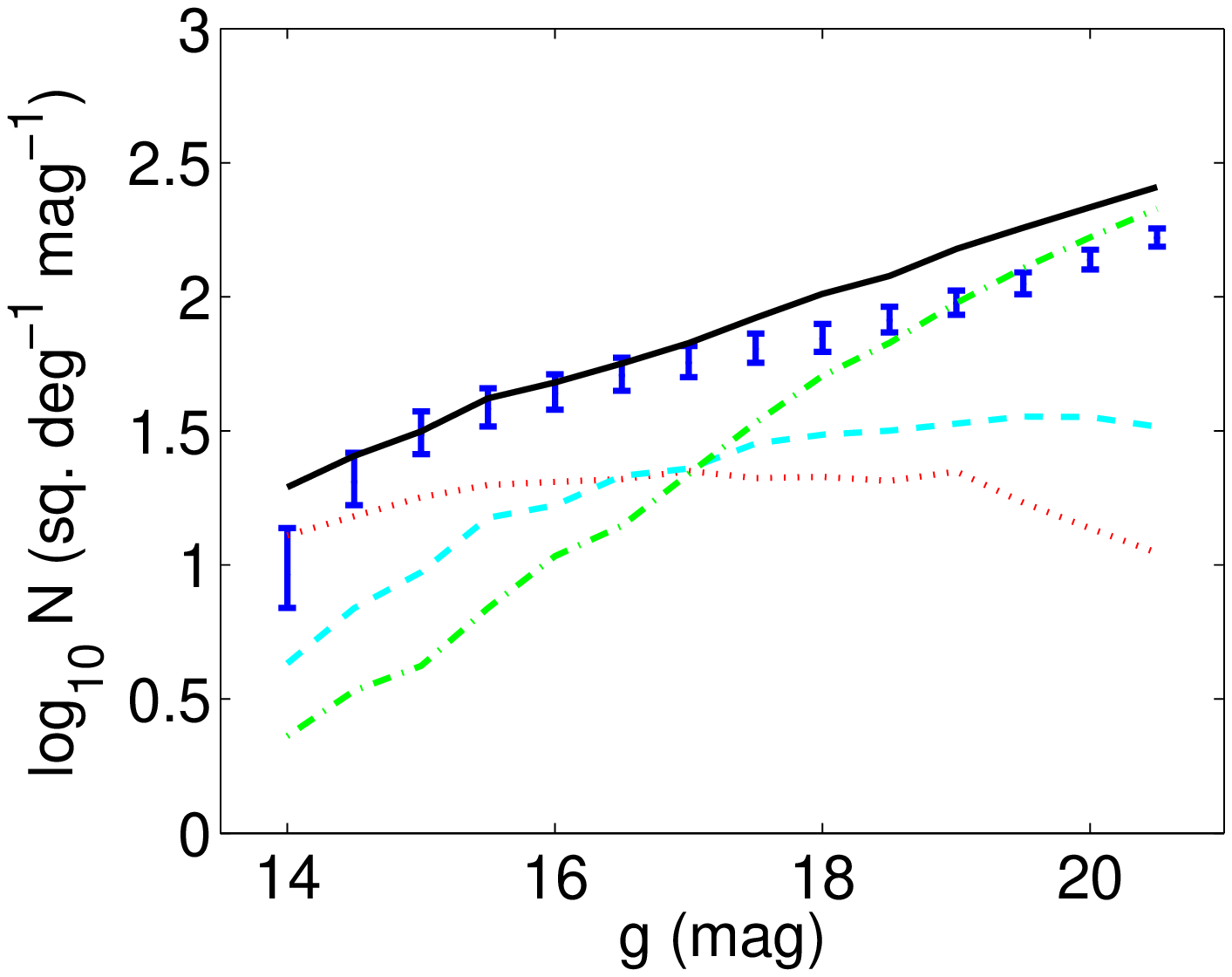}
    \includegraphics[width=0.45\textwidth,height=0.3\textwidth]{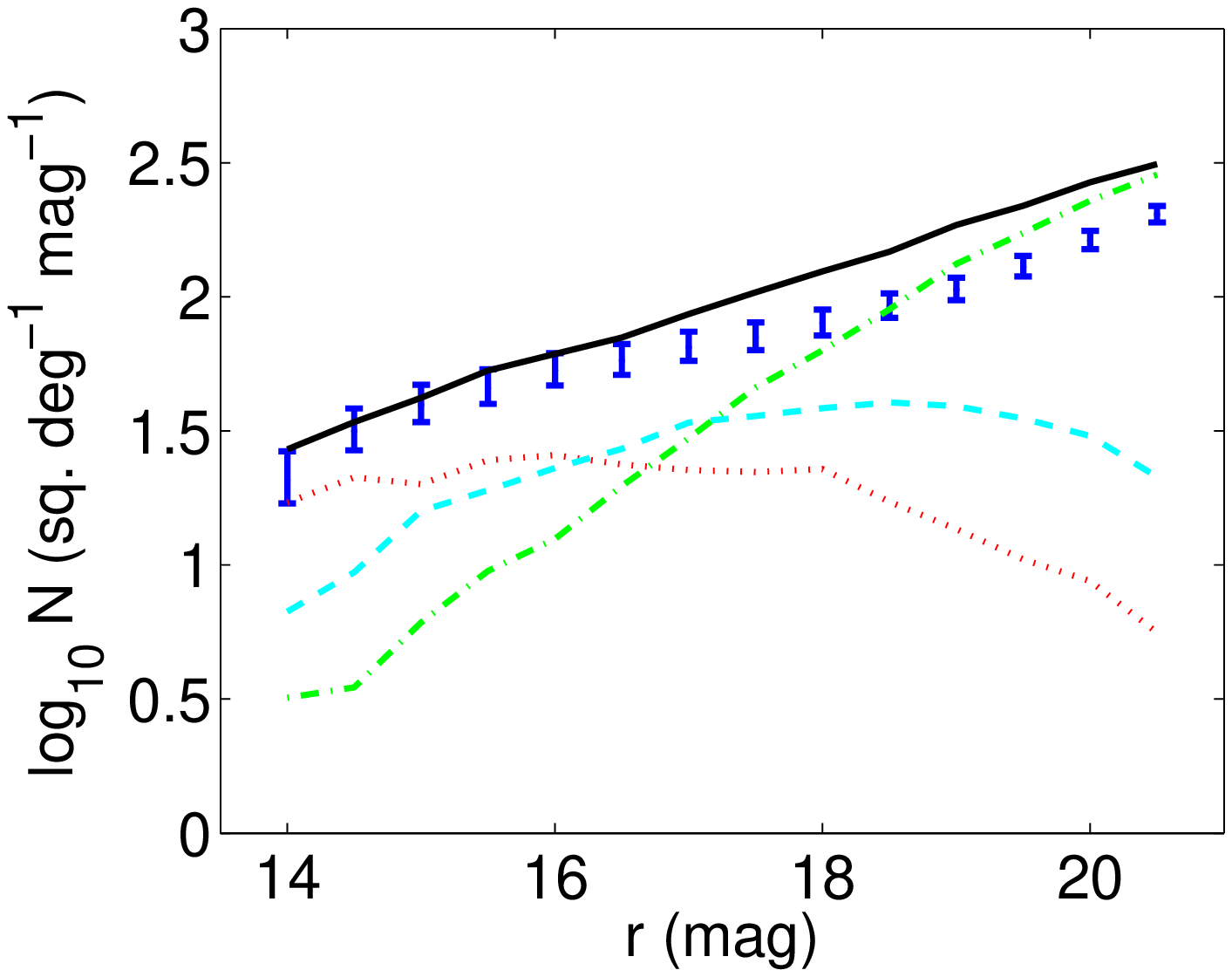}\\
    \includegraphics[width=0.45\textwidth,height=0.3\textwidth]{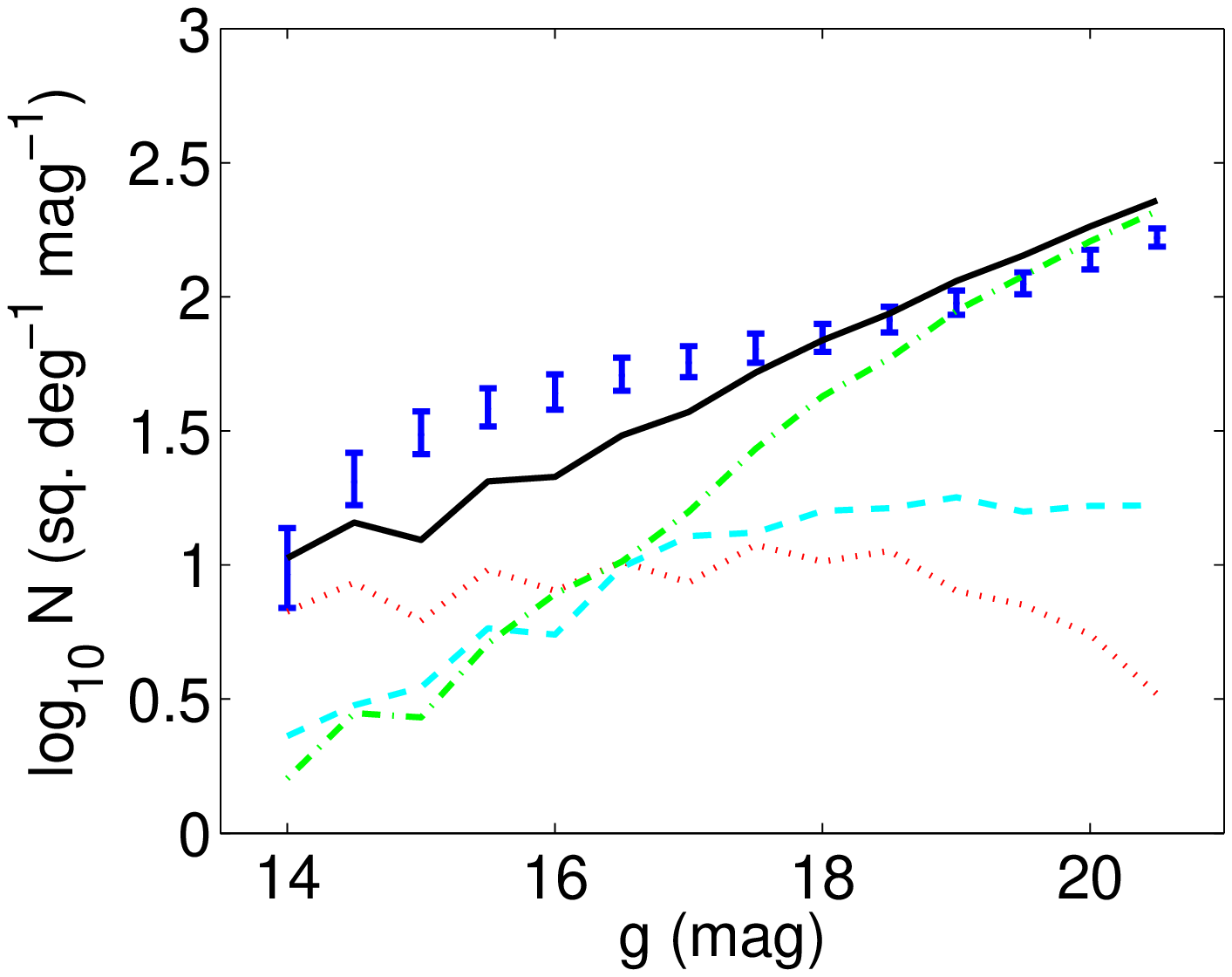}
    \includegraphics[width=0.45\textwidth,height=0.3\textwidth]{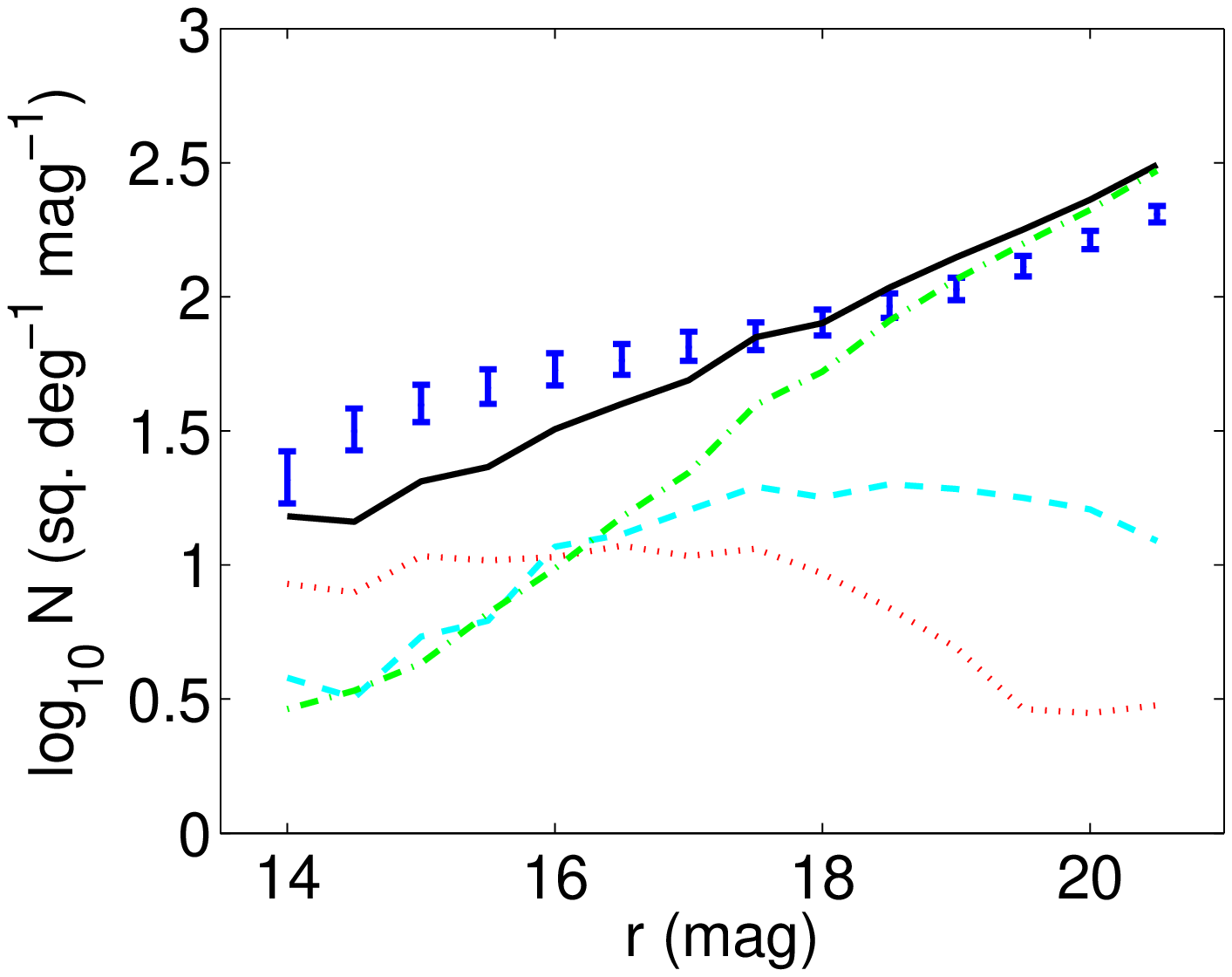}\\
    \includegraphics[width=0.45\textwidth,height=0.3\textwidth]{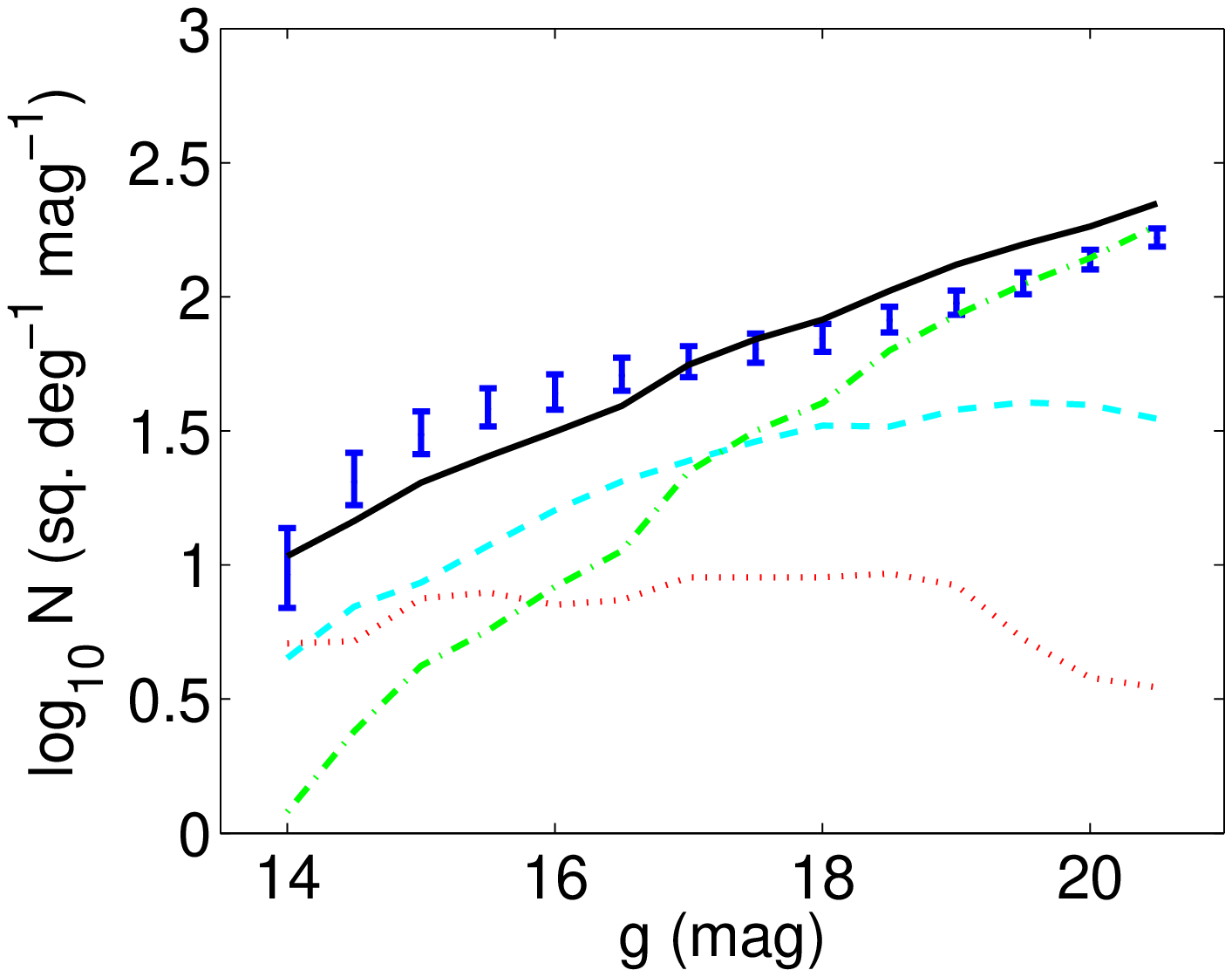}
    \includegraphics[width=0.45\textwidth,height=0.3\textwidth]{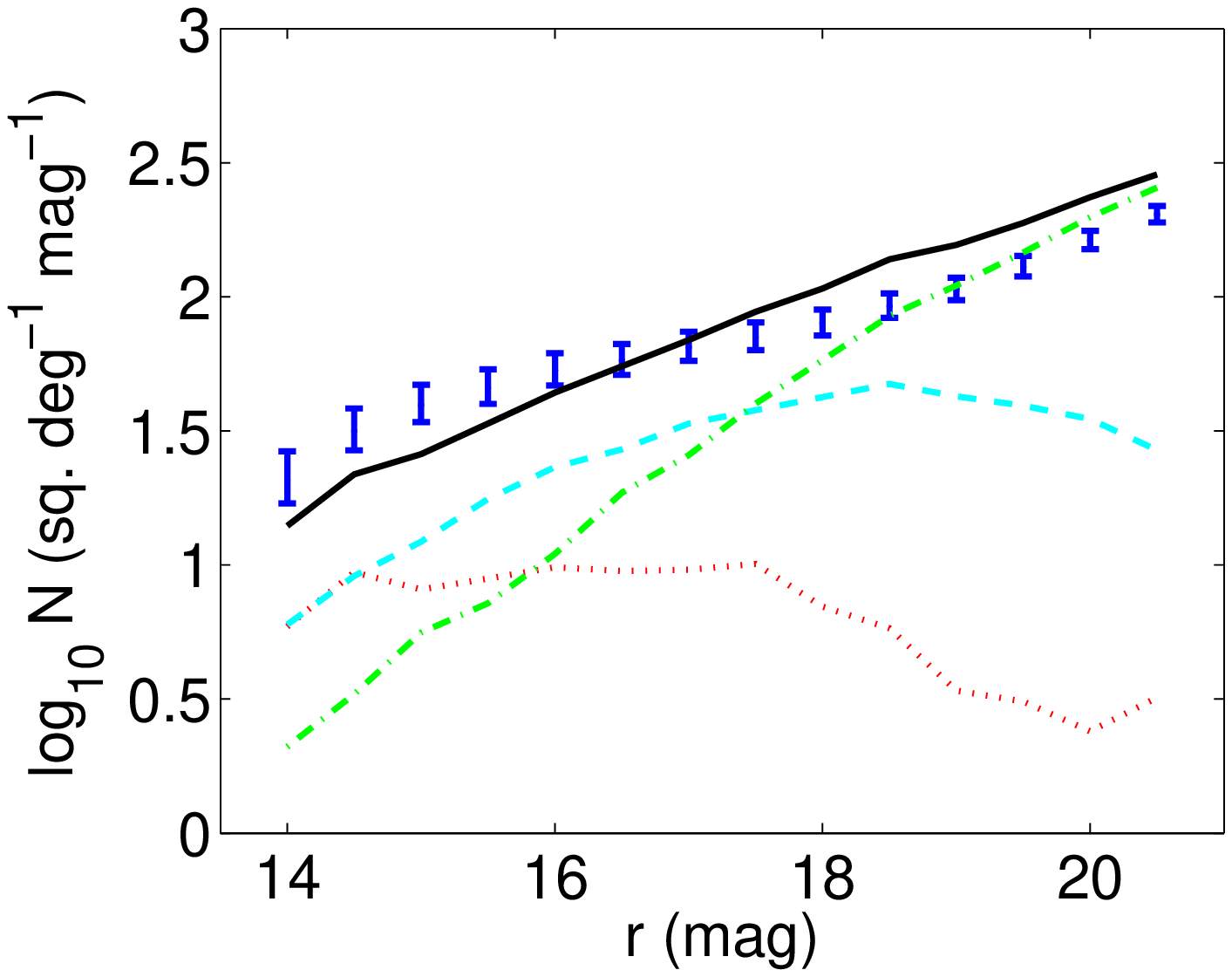}\\
    \caption{The magnitude distributions of the SDSS observations and the \tri simulations generated with parameter sets 1 to 4 (top to bottom). The x- and y-axes are similar to the bottom panel of Fig.\ \ref{fig:NGP_SDSS_only_hess}. The error bars (blue) are from SDSS DR7 data with the colour limit of $0.2\leq(g-r)<1.2$. The black solid line represents the model star count predictions of \tri. The dotted (red), dashed (cyan), and dot-dashed (green) lines stand for the thin disc, thick disc, and halo of the \tri model, respectively. (A colour version is available on-line.)}
    \label{fig:mag_distri_SDSS_set1234}
\end{figure*}

\begin{figure*}\centering
  \includegraphics[width=0.32\textwidth]{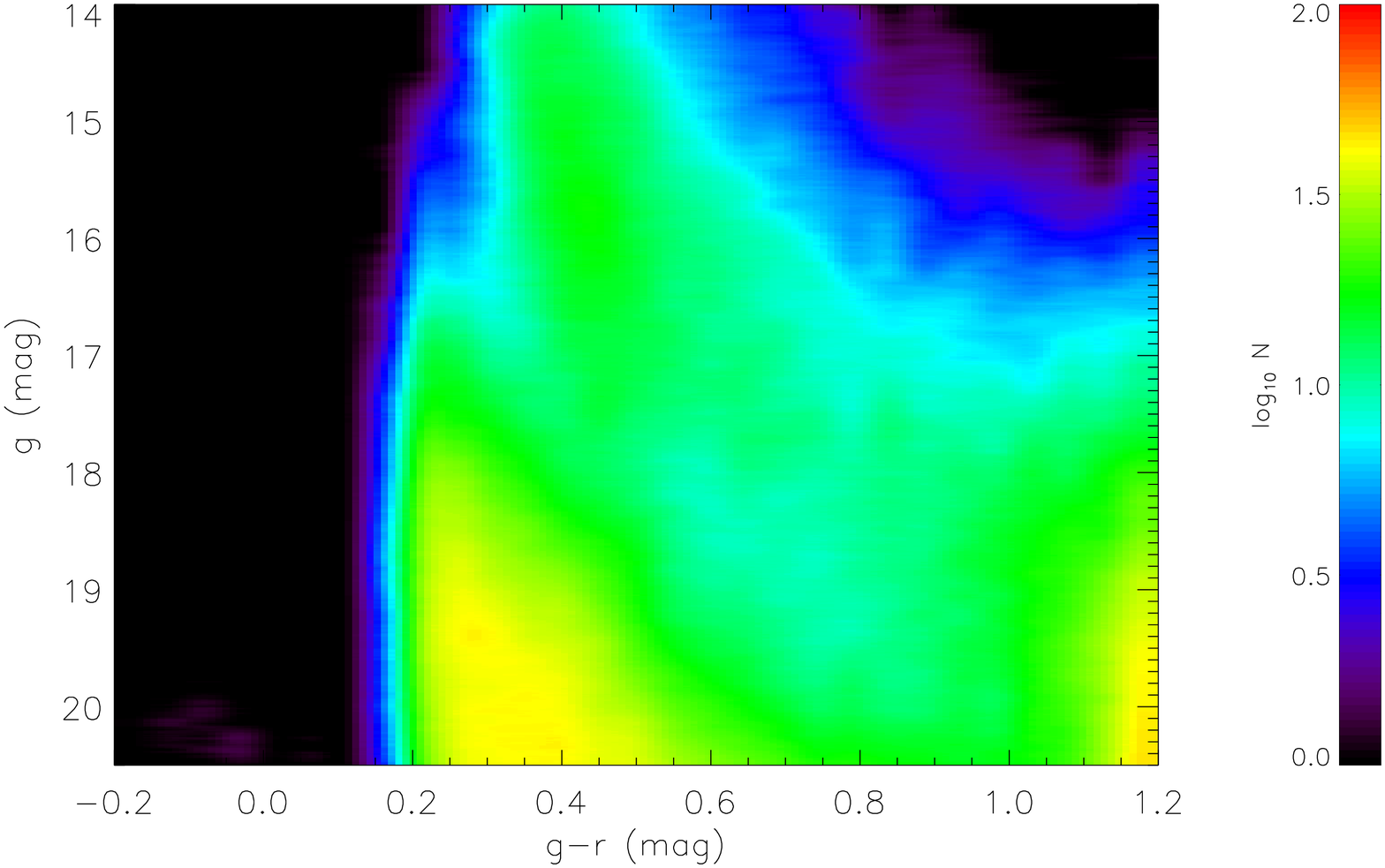}
  \includegraphics[width=0.32\textwidth]{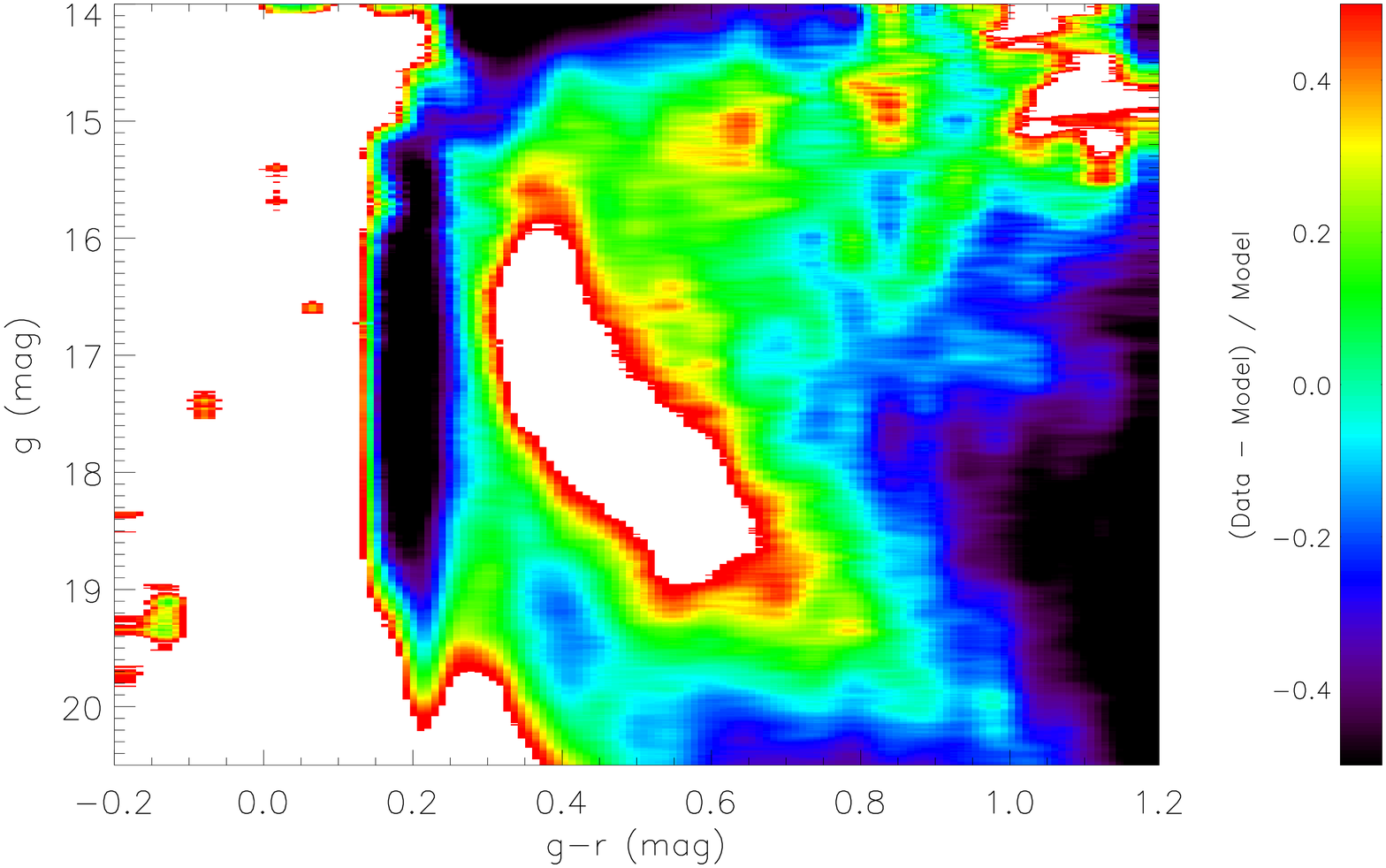}
  \includegraphics[width=0.32\textwidth]{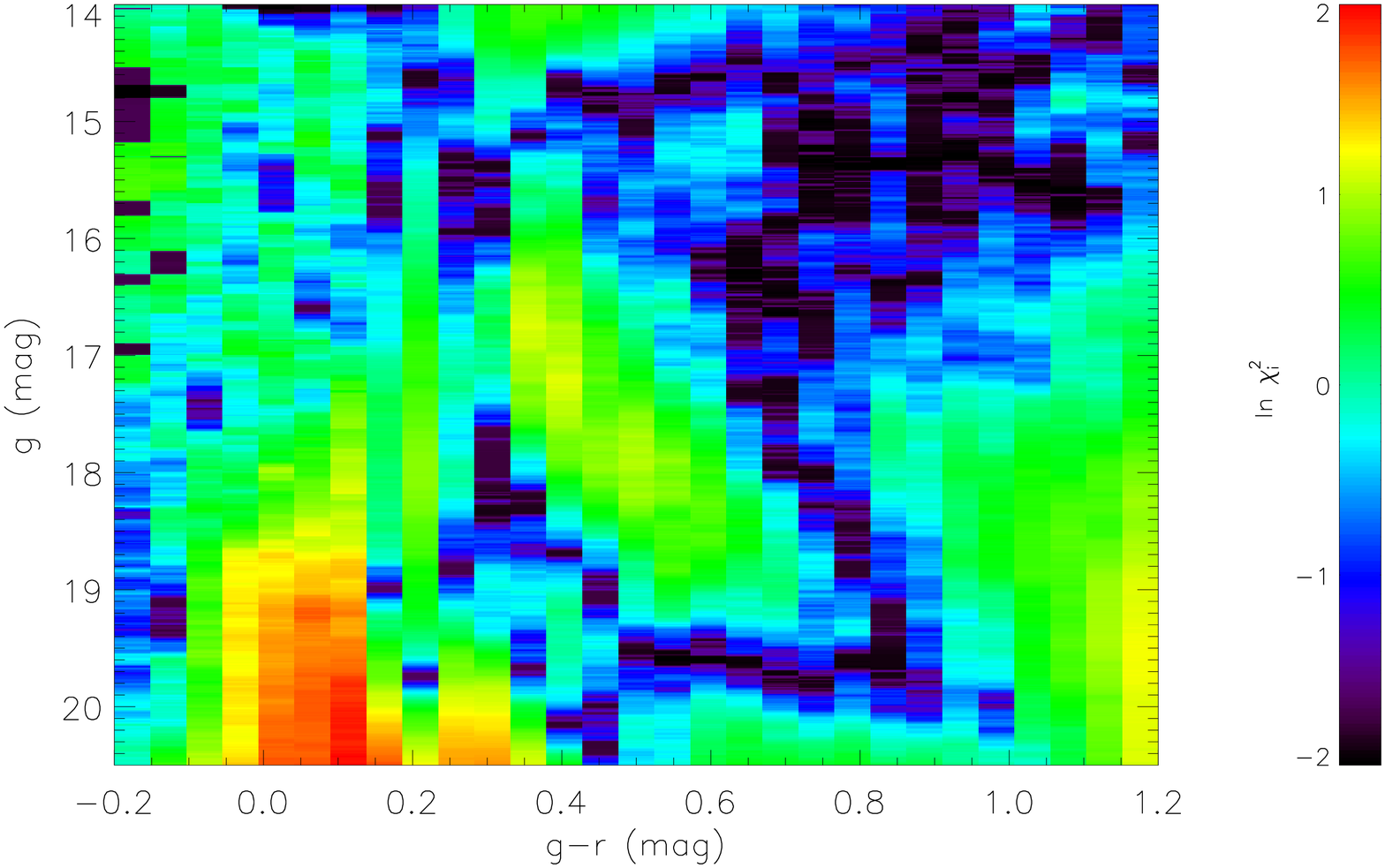}\\
  \includegraphics[width=0.32\textwidth]{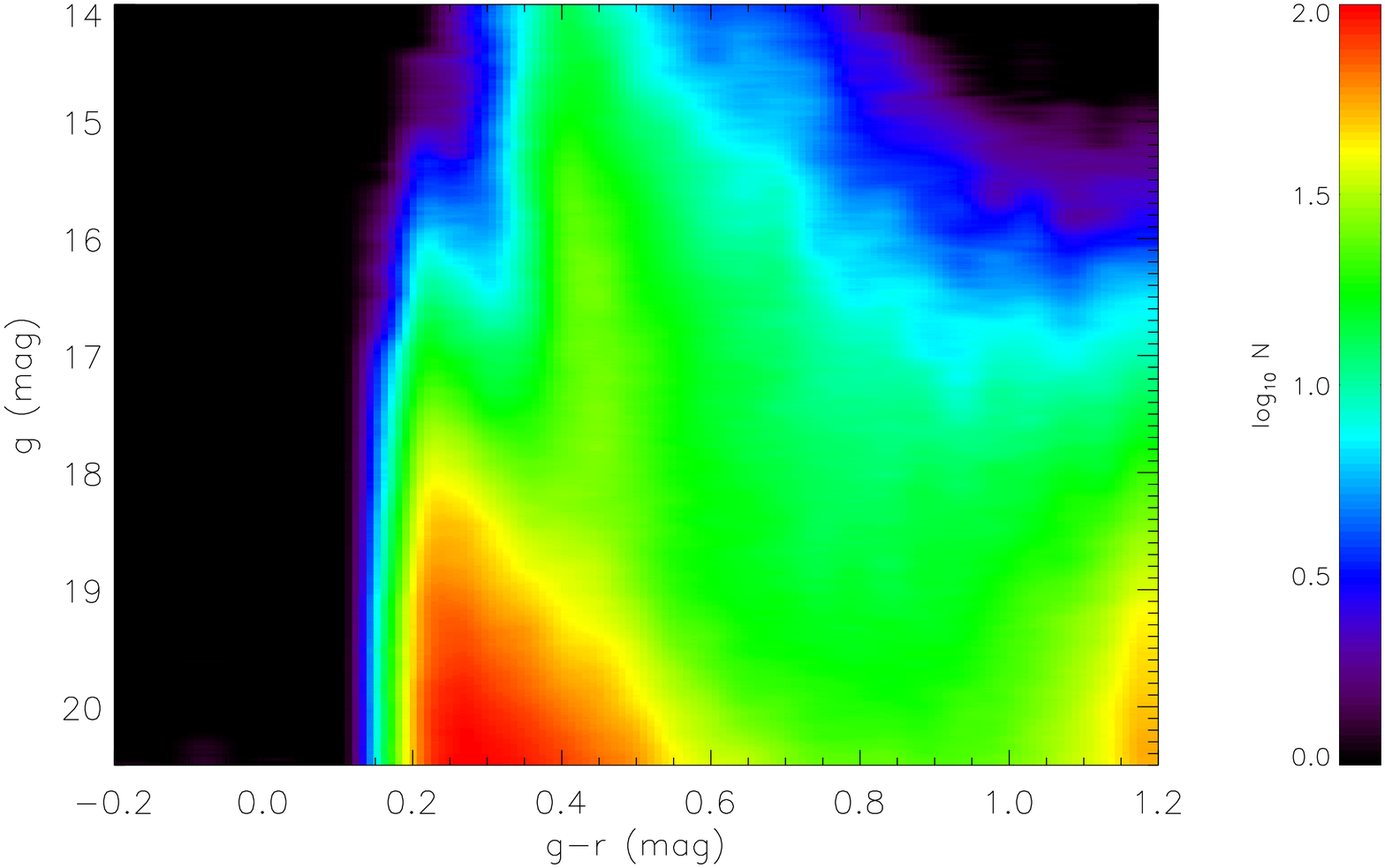}
  \includegraphics[width=0.32\textwidth]{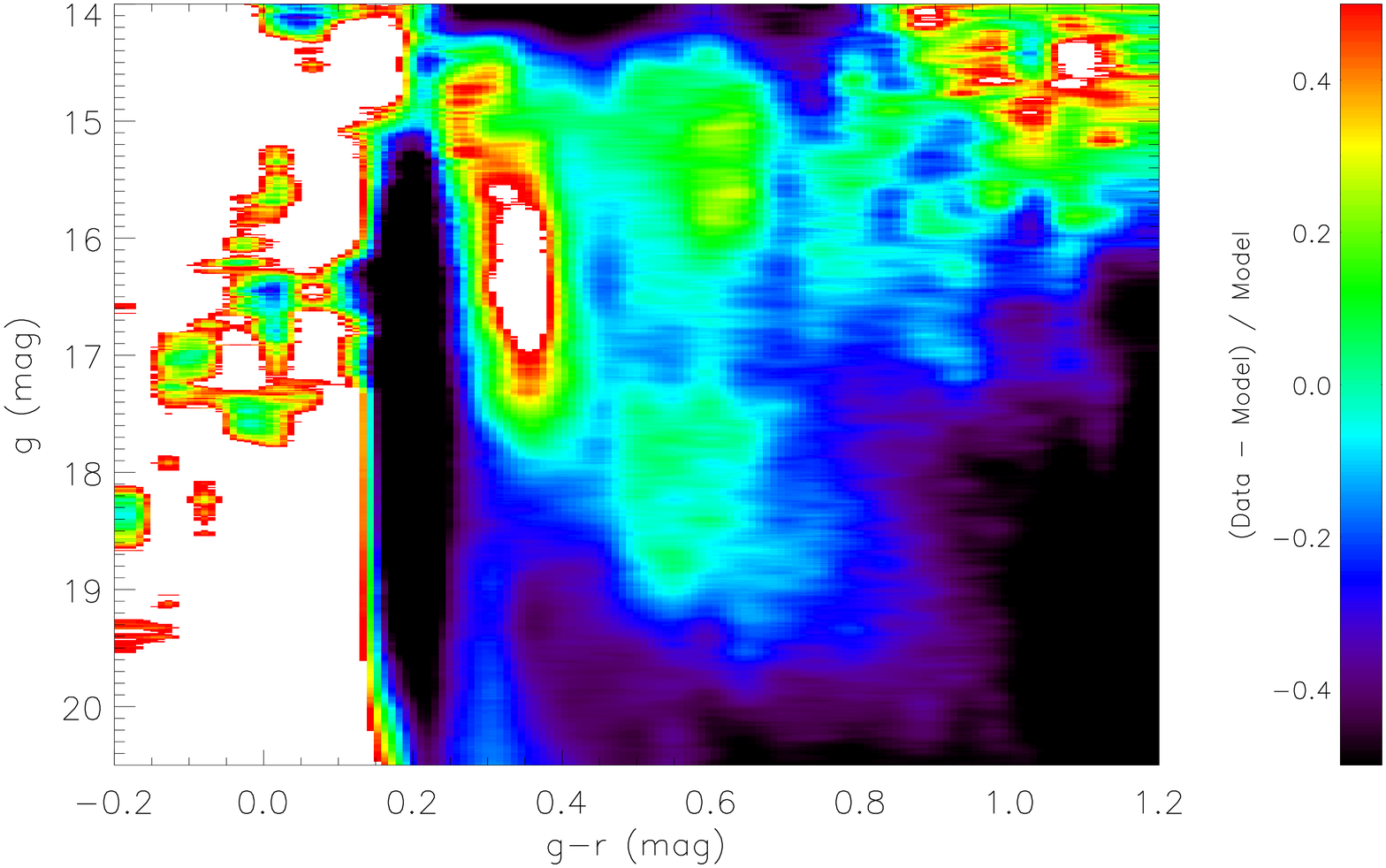}
  \includegraphics[width=0.32\textwidth]{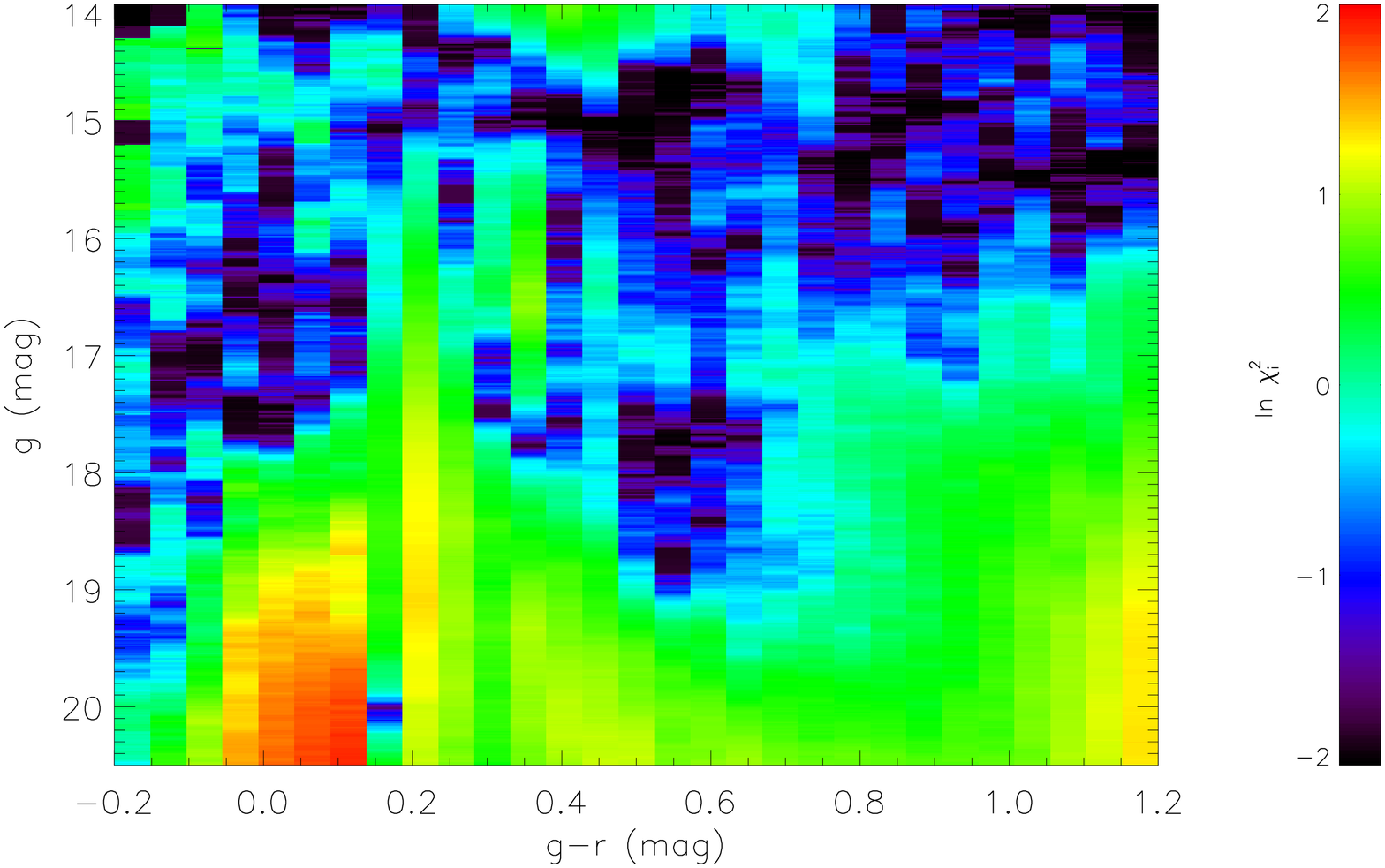}\\
  \includegraphics[width=0.32\textwidth]{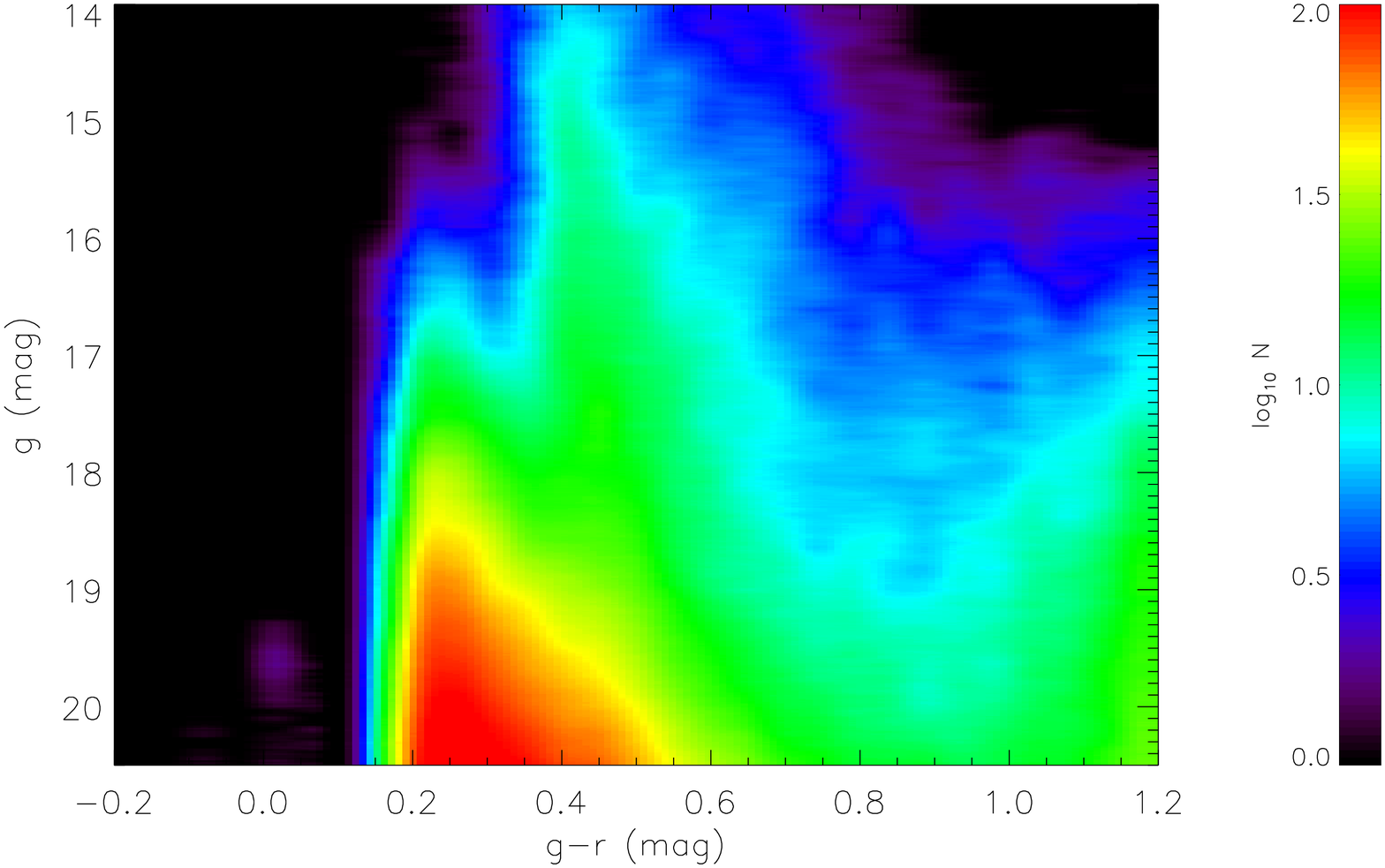}
  \includegraphics[width=0.32\textwidth]{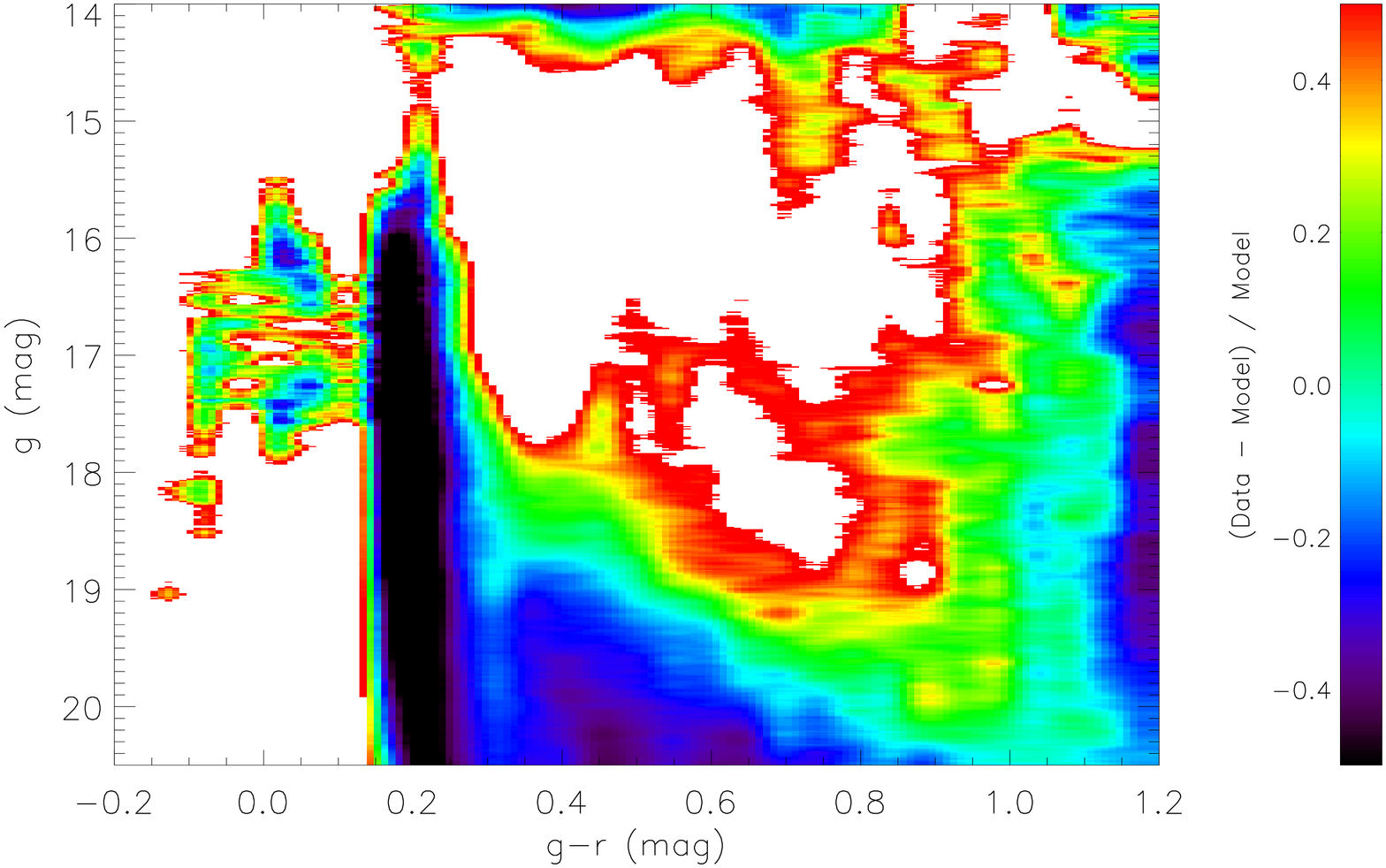}
  \includegraphics[width=0.32\textwidth]{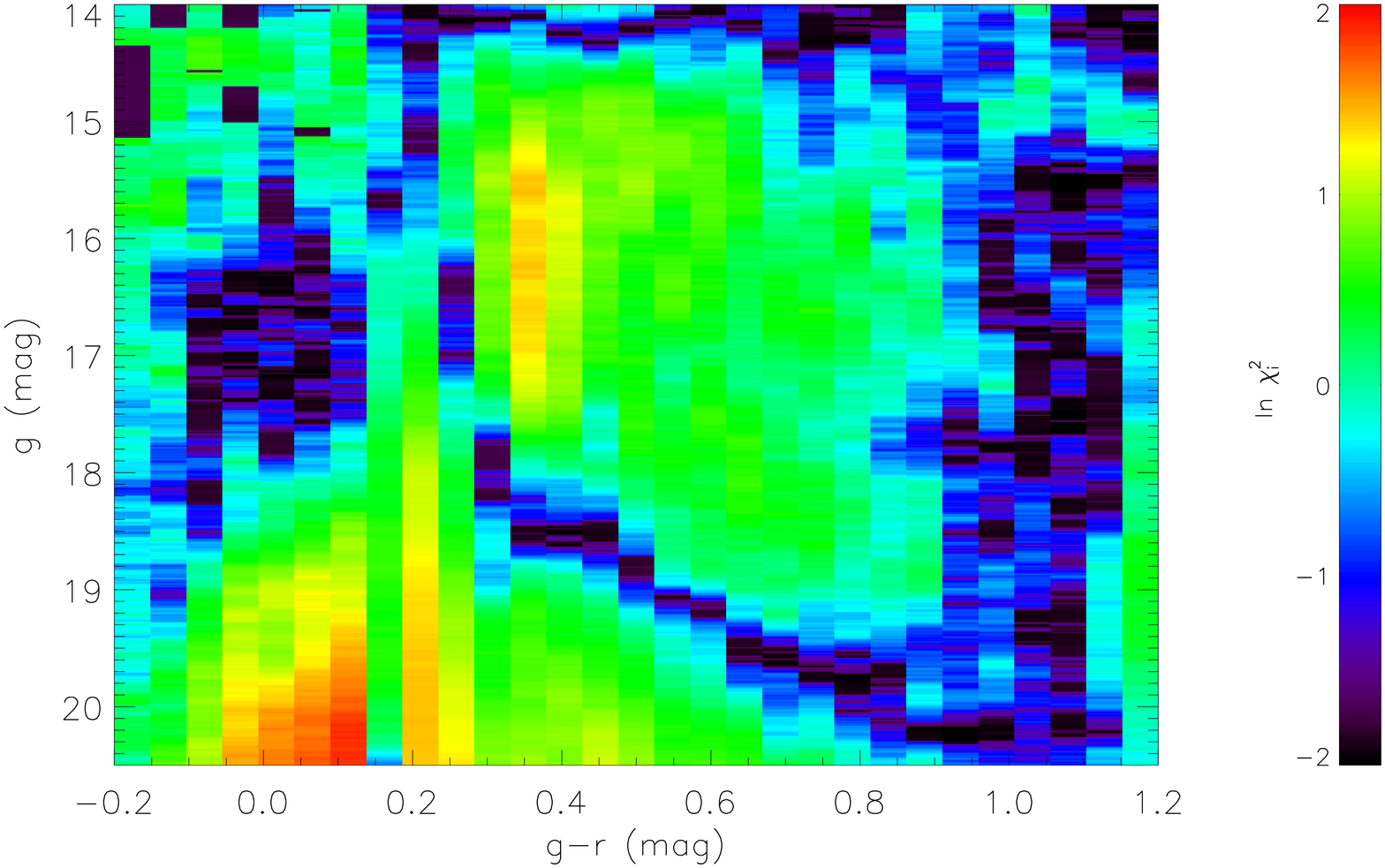}\\
  \includegraphics[width=0.32\textwidth]{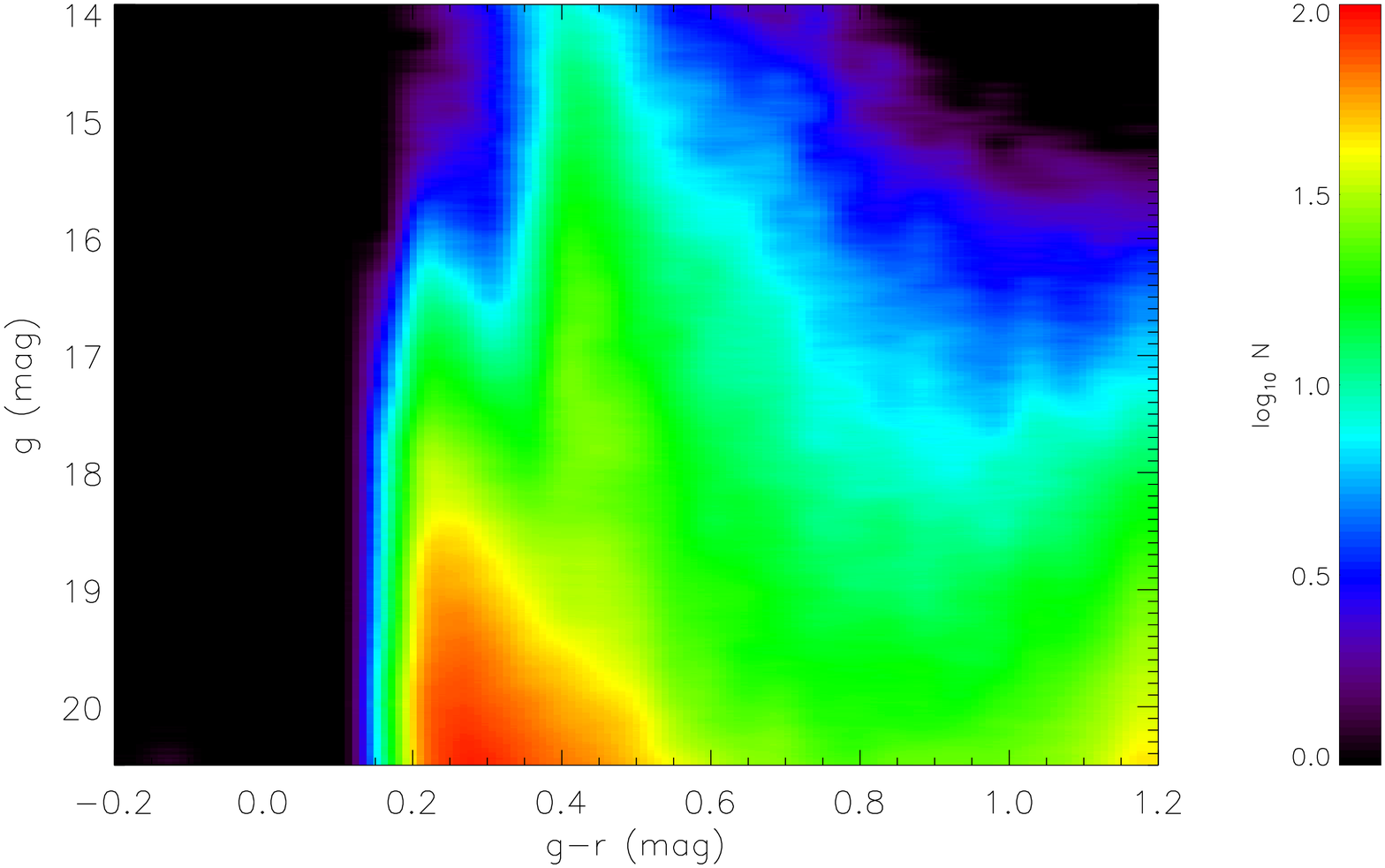}
  \includegraphics[width=0.32\textwidth]{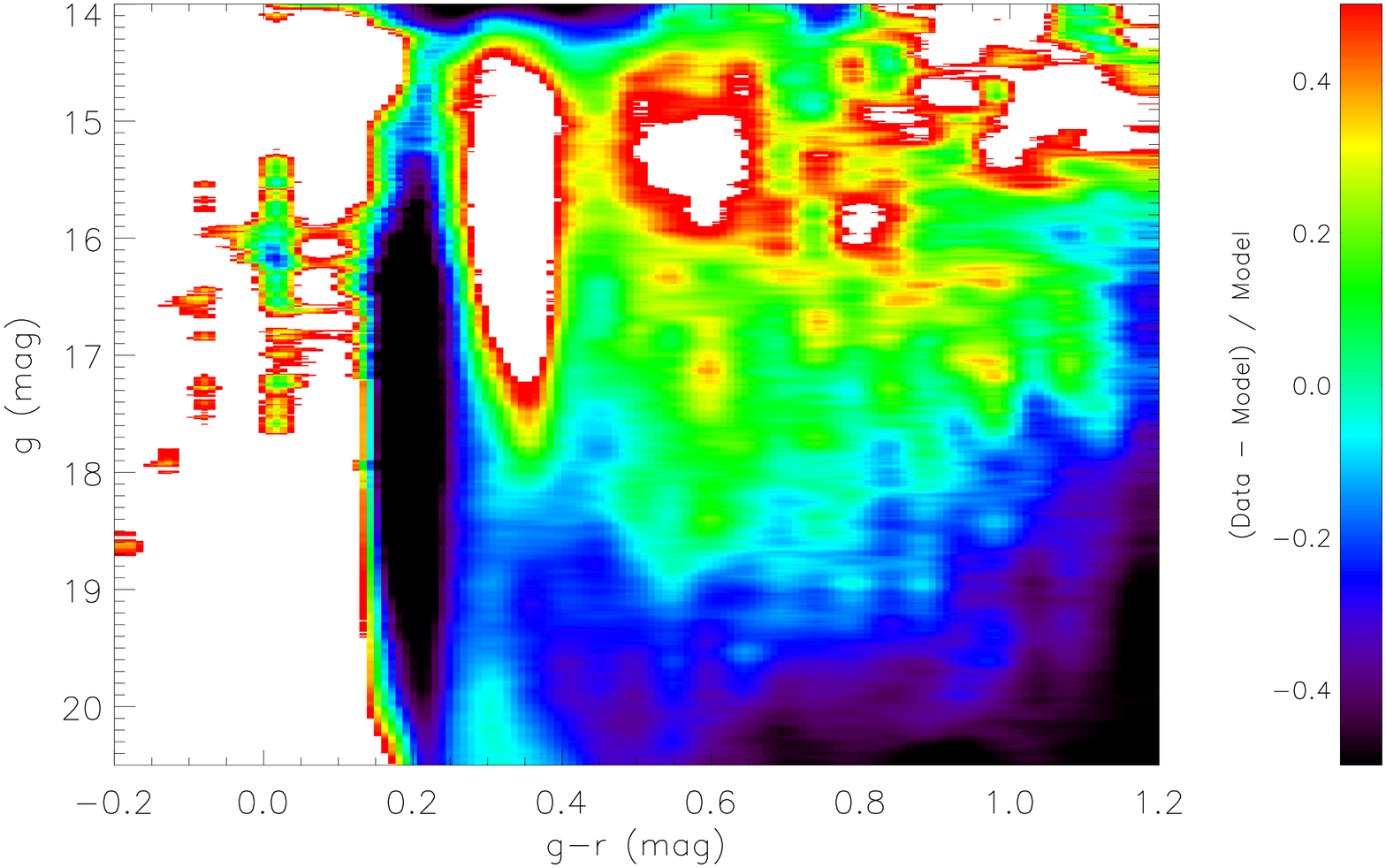}
  \includegraphics[width=0.32\textwidth]{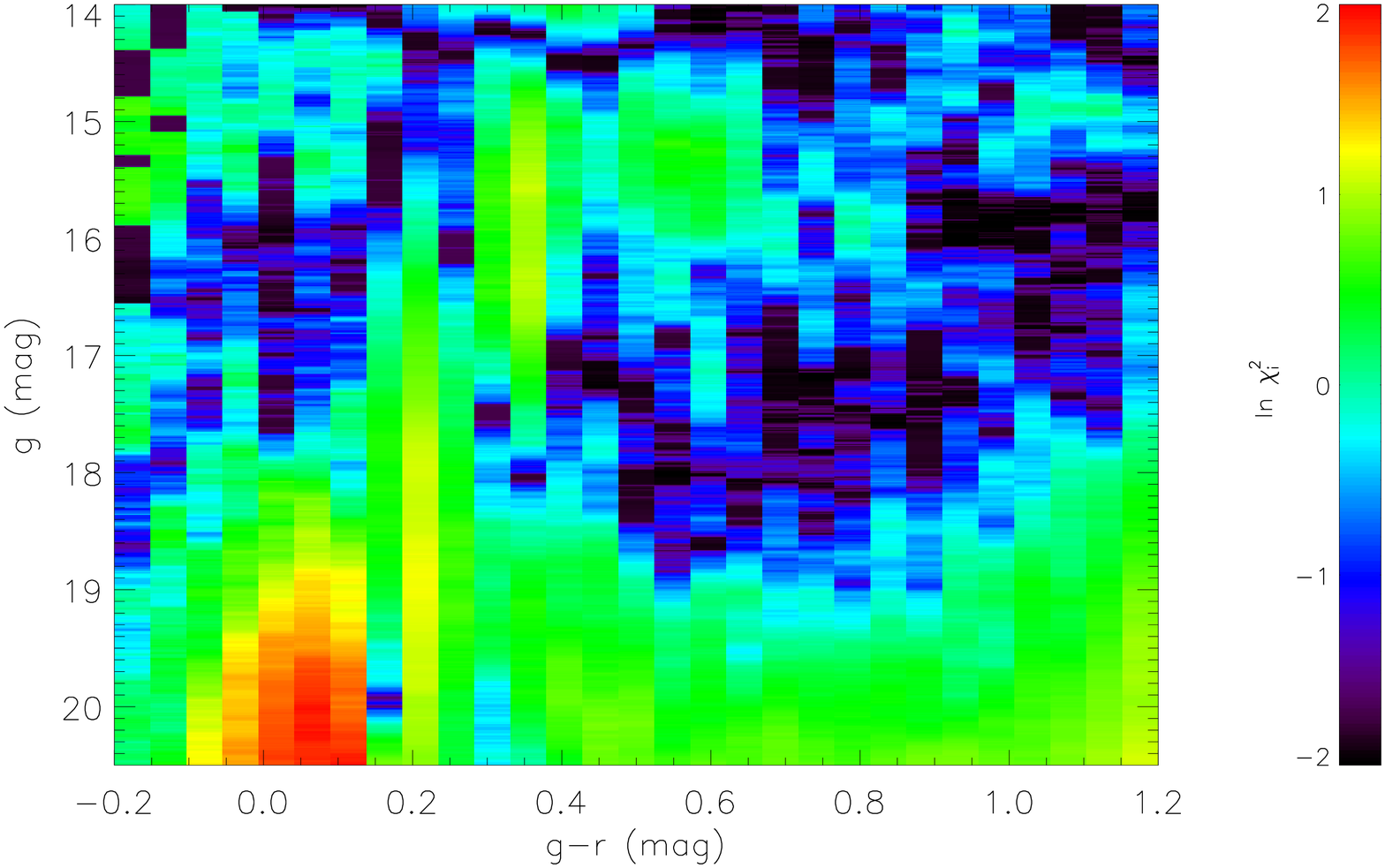}
    \caption{Hess diagrams (first column), relative differences (second column)  between the Hess diagrams of the models and the SDSS observations, and the $\chi^2$ diagrams (last column). From top to bottom, we show the comparisons based on the parameter sets 1 to 4, respectively. The x- and y-axes correspond to the colour index $-0.2\leqslant g-r\leqslant1.2$ and de-reddened apparent magnitude $14\leqslant g\leqslant 20.5$. The left column shows the stellar density using a colour coding from 0 to 100 stars per square degree per 1 mag and 0.1 colour mag on a logarithmic scale. The middle panels are the relative differences between data and models with a colour coding from $-50\%$ to $+50\%$. The right panels are the $\ln\chi^2$ distributions with a colour coding from $-2$ to $+2$. Note that the colour ranges in the middle and right panels cover much larger ranges than in Fig.\ 5 of \citet{jgv}. (A colour version is available on-line.)}
    \label{fig:NGP_SDSS_hess_set}
\end{figure*}

%
%
\subsection{Default input (set 1)}

The default input of \tri is adopted directly as parameter set 1. The parameters are listed in Col.\ 3 of Table \ref{tab:tri:def}. The default set 1 has no thick disc. For the vertical structure pointing to the NGP, set 1 simulates star counts using only thin disc and halo.

The top row of Fig.\ \ref{fig:mag_distri_SDSS_set1234} shows the luminosity functions in the $g$ and $r$ bands of the SDSS data and of the models in the colour range $0.2<g-r<1.2$. The default model (set 1) fits the data very well in the selected magnitude range 14 -- 20.5 mag.

In the CMD, the picture is different. We construct a Hess diagram that is similar to the
observed Hess diagram in Fig.\ \ref{fig:NGP_SDSS_only_hess}.  The top row of
Fig.\ \ref{fig:NGP_SDSS_hess_set} shows the analysis of the Hess diagram. The
left panel shows the simulated Hess diagram, which we compare directly to the
observed Hess diagram. The middle panel shows the median absolute relative
differences $D_\mathrm{ij}$ with a colour-coded range of $\pm 0.5$. The right
panel gives the unreduced $\chi^2$ distribution on a logarithmic scale. From the
relative differences, some clear features can be identified (middle panel of the
top row of Fig.\ \ref{fig:NGP_SDSS_hess_set}):
\begin{enumerate}
  \item The large white ``island'' in the middle of the diagram is a consequence of the lack of a thick disc in the model.
  \item The missing stars at the top-right corner of $g-r\sim 1.1$ and $g\sim 15$ mag are a hint of an underestimation of the local thin-disc density caused by the strongly flattened sech$^2$ profile close to the Galactic plane.
  \item The underestimated number counts at the faint blue end of $0.2<g-r<0.4$ and $g>19.5$ mag show that the halo density profile falls off too steeply.
  \item The horizontally adjacent black-and-white coloured areas in the colour range $0.15<g-r<0.45$ at $g\sim 17$\,mag show that the (thick disc) population is replaced in the model by a more metal-poor (halo) population with a bluer F-turnoff region. This is additional strong evidence that a thick disc component of intermediate metallicity is missing in the model. 
  \item At the faint end ($0.15<g-r<0.25$ and $15<g<19$), the stars missing in blue colours (halo) are compensated for by too many (or slightly too blue) M dwarfs of the thin disc in the luminosity function.
\end{enumerate}

The value of the mean reduced $\langle\chi^2\rangle= 151.98$  is significantly
larger than that of the \jj model ($=4.31$, Table \ref{tab:tri:def}). The median
relative deviation of the data from the model in the Hess diagram $\Delta_{\rm{med}}$ is
$0.3035$, which is 5.4 times larger than for the \jj model.

A comparison of Figs.\ \ref{fig:mag_distri_SDSS_set1234} and \ref{fig:NGP_SDSS_hess_set} indicates that the analysis of the luminosity function integrated over a large colour range alone may result in misleading conclusions about the underlying stellar populations.

\subsection{\jj input (set 2)}

For parameter set 2, we derive the best representation of the input function of the \jj model with the \tri parameters. With this ``model of a model'', we investigate the capability of \tri to reproduce a Milky Way model suggested by other authors.

\subsubsection{Input parameters}

For parameter set 2, which is based on the \jj model, we derived a best-fit \tri profile
of the thick disc to the sech$^{\alpha_\mathrm{t}}$ profile. Both the
exponential and the sech$^2$ profiles are fitted with a standard routine and the
unreduced $\chi^2$ values are used to select the best fit. The exponential form results in $\chi^2=4.0\times10^{-8}$, which is a better fit
than the sech$^2$ profile with $\chi^2=6.6\times 10^{-8}$. We therefore use the
exponential profile. The scale height of the thick disc is $793\pm6$ pc and the
local density is $0.0029\pm0.000015$ \mpc. The derived local surface density of
the thick disc is $\Sigma_{\rm{t}}(\sun)=4.63$ \mpcc.

In \tri, the scale height of the thin disc is a function of both age and the
parameters $z_0$, $t_0$, and $\alpha$ \citep{2005Girardi}. The parameters are
determined by fitting the numerical \jj model relationship of the thin disc. We
fit the thin disc parameters in exponential and sech$^2$ form to the \jj model,
respectively. The latter fit results in a smaller $\chi^2$ (namely $4.7\times10^{-6}$) than the former one ($8.3\times10^{-6}$), hence we choose the sech$^2$ form as
the profile of the thin disc. The derived local surface density of the thin disc
is $\Sigma_{\rm{d}}(\sun)=32.26$ \mpcc, which is $10\%$ higher than the surface
density of the thin disc of the \jj model.

The halo of the \jj model is fitted by the \tri form in the oblate $r^{1/4}$
spheroid. The best-fit parameters are local density
$\rho_{\rm{h}}(\sun)=0.001576$ \mpc, oblateness $q_{\rm{h}}=0.65$, and effective
radius $r_{\rm{eff}}=11994$ pc.

To take into account the estimated total local density of all stars ever
formed, the present-day densities of the \jj model were corrected
accordingly (see Sect. \ref{sub-jj}).

\subsubsection{Star counts}

The second row in Figs.\ \ref{fig:mag_distri_SDSS_set1234} and \ref{fig:NGP_SDSS_hess_set} shows the star count results of \tri based on the \jj parameters. The luminosity functions are of poorer quality than that for the default parameter set 1, because the halo is now over-represented by about 0.2 dex. In the CMD, the thick disc fills the minimum in the centre of the Hess diagram seen in the default model. The F-turnoff of the thick disc is still slightly too red (metal-rich). In the faint (distant) regime, there are significantly too many stars in all components.

The overall fit of the Hess diagram with $\langle\chi^2\rangle= 214.65$ is worse than for the default model owing to the poor fit in the faint magnitude range. The median absolute relative difference in Hess diagram $\Delta_{\rm{med}}$ is $0.2993$.

%
%
%
%
\subsection{\besan input (set 3)}
\subsubsection{Model parameters}

The thick disc of \citet{Robin} is described by a parabolic shape up to $|z|=400$ pc and an exponential profile beyond this height.

We fit this piecewise-defined function using a sech$^2$ and an exponential function, respectively. The exponential function with a smaller fitting error is adopted in Col.\ 5 of Table \ref{tab:tri:def}. The scale height of 910 pc and a local density of 0.0015 \mpc yield a local surface density $\Sigma(\sun)$=2.82 \mpcc.

There are seven sub-populations in the thin disc of the \besan model. Each sub-population has its own age range, metallicity, and spatial vertical distribution. We fit scale heights and local densities of each sub-population (see Table \ref{tab:set3:thin}). Exponential profiles yield smaller $\chi^2$ values.

\begin{table}\centering
\caption{The fitting results of the thin disc sub-populations in the \besan model.}
\label{tab:set3:thin}
\begin{tabular}{cccc}
  \hline
  Age range & $t$ & $h_i$ & $\rho_i(\sun)$ \\
  (Gyr)     & (Gyr)    & (pc)  & \mpc \\
  \hline
  0-0.15 & 0.075 & 32.5 & 0.0019 \\
  0.15-1 & 0.575 & 97.8 & 0.0066 \\
  1-2    & 1.5   & 120.0 & 0.0073 \\
  2-3    & 2.5   & 195.9 & 0.0043 \\
  3-5    & 4.0   & 273.5 & 0.0049 \\
  5-7    & 6.0   & 328.3 & 0.0030 \\
  7-10   & 8.5   & 370.9 & 0.0020 \\
  \hline
\end{tabular}
\tablefoot{Col.\ 1 lists the age range of each sub-population in the \besan model. Col.\ 2 is the simple arithmetic mean age in each bin. We fit the vertical profile of each sub-population via two parameters: scale height $h_i$ (Col.\ 3) and local density $\rho_i(\sun)$ (Col.\ 4). The relationship of $t$ to $h_i$ is determined by fitting a power-law function.}
\end{table}

In the next step, the scale heights as a function of age are fitted by the \tri law. The best-fit parameters $z_0$, $t_0$, and $\alpha$ are 13.61 pc, 0.02016 Gyr, and 0.5524, respectively. The total local density $\rho_{\rm{d}}(\sun)$ is the sum of $\rho_{i}(\sun)$ of each population, namely 0.030 \mpc. The derived local surface density of the thin disc is 25.20 \mpcc. The density was corrected by the factor $1/g_{\rm eff}=1.53$ to obtain the mass of ever formed stars.

The local density of the halo in the \besan model is determined by the SFH and the IMF. The adopted value is much lower than our local calibration established using SDSS data. We therefore fit the halo of the \besan model with the effective radius, oblateness, and local density of the \tri de Vaucouleur profile in two steps:
\begin{enumerate}
   \item We fit the vertical profile to find the effective radius and the oblateness;
   \item We use the local density as a new fitting parameter to reproduce the halo star counts of the SDSS data and then use this value to fit a model of the parameter set 3 (Table \ref{tab:tri:def}).
\end{enumerate}

In the first step, we find the best-fit parameters $q_{\rm{h}}=0.71$and $r_{\rm{eff}}=39.436$ kpc. In the second step, we determine the local density of the halo by fixing $q_{\rm{h}}$ and $r_{\rm{eff}}$ and fitting the SDSS data in a restricted area of the Hess diagram where the halo dominates (i.e., in the dashed triangle of Fig.\ \ref{fig:triangle}, see Section \ref{sec-set5}). The best-fit local density of the stellar halo is $\rho_{\rm{h}}(\sun)=0.00009$ \mpc, which is about ten times higher than the value used in the \besan model.

\subsubsection{Star counts}

The third row in Fig.\ \ref{fig:mag_distri_SDSS_set1234} and \ref{fig:NGP_SDSS_hess_set} shows the star count predictions of \tri based on the \besan parameters. The luminosity functions appear to disagree significantly with these predictions at bright magnitudes primarily owing to the presences of the thin and thick discs, but be a reasonably fit at the faint end. In the CMD, the thin and thick disc regime are strongly under-represented in the model. The halo with the original local density also appears to be under-represented.

The overall fit to the Hess diagram with $\langle\chi^2\rangle= 207.60$ is tighter than that of parameter set 2 but worse than that of the default set (set 1). The $\Delta_{\rm{med}}$ is $0.3445$, which is worse than for sets 1 and 2.

This result is not a judgement on the \besan model compared to the \jj or \tri model. It is a measure of the flexibility of the \tri model. A comparison of the \besan model with the NGP data is later discussed in more detail.

%
\subsection{\juric input (set 4)}
\subsubsection{Model parameters}

We do not consider the scale height as a function of the age of the subpopulation in the \juric case ($\alpha=0$). \cite{Juric} fit the basic parameters of their Galactic model via star count observations from the SDSS. Only the final parameters of their discs are considered here. \cite{Juric} describe the vertical profiles of the thin and thick disc by simple exponentials with $h_{\rm{d}}=300$ pc and $\rho_{\rm{d}}(\sun)=0.029$ \mpc for the thin disc and $h_{\rm{t}}=900$ pc, $\rho_{\rm{t}}(\sun)=0.0034$ \mpc for the thick disc. This corresponds to $12\%$ of the local density of the thin disc given by \cite{Juric}. The derived local surface density of the thick disc is $\Sigma_{\rm{t}}(\sun)=6.12$ \mpcc. The density has been corrected to the mass of ever formed stars with the factor $g_{\rm eff}$ for both the thin and thick discs.

The halo density profile is fitted by \tri with three parameters. The best-fit parameters are an oblateness of $q_{\rm{h}}=0.59$, a local density of $\rho_{\rm{h}}(\sun)=0.000145$, and an effective radius of $r_{\rm{eff}}=18304$ pc.

\subsubsection{Star counts}

The bottom rows in Fig.\ \ref{fig:mag_distri_SDSS_set1234} and \ref{fig:NGP_SDSS_hess_set} show the star count results of \tri based on the \juric parameters. The luminosity functions are slightly more clisely reproduced than for set 3. They show a significant deficiency at bright magnitudes dominated by the thin and thick disc and too many stars at the faint end. In the CMD, the fit of the  thick disc regime and the halo is reasonable in the model. At the red end, the bright part is under-represented and the faint end is over-represented.

The overall fit of the Hess diagram with $\langle\chi^2\rangle= 137.88$ is the best of the four parameter sets but has the worst $\Delta_{\rm{med}}$ ($0.4639$).

\subsection{Discussion}
An overall comparison of sets 1 to 4 shows that the default parameter set fits best the luminosity functions in the $g$ and $r$ band. On the one hand, this confirms the fitting procedure used in \tri to determine these parameters mainly by comparing luminosity functions in different fields. On the other hand, it is remarkable that the thick disc component, which becomes obvious in the Hess diagram analysis, is extremely poorly fit.

All other model adaptations fail to closely reproduce the luminosity functions. No model with a thin disc, thick disc, and halo can reproduce the shallow regime of the luminosity function in the range $17 \rm{mag}<g<19 \rm{mag}$. The analysis of star counts in the CMD shows that disagreements between the data and models exceed $\pm 50\%$ in some regions of the Hess diagram. In this sense, all models fail to provide a good fit of the CMD.

The default model shows that luminosity functions alone are inappropriate for determining the structure of the Milky Way owing to the large distance range in each component. Nevertheless, the total $\chi^2$ value for the Hess diagram fit is relatively small compared to the other sets.

In the next section, we try to optimise the \tri parameters to find the best model for the CMD at the NGP.

%
%
%

\subsection{Optimisation (set 5)}\label{sec-set5}

To find the best-fit \tri model, it is necessary to identify the best-fitting parameters. Because there are so many parameters that can be adjusted within \tri, we cannot determine them simultaneously. However since the thin disc, thick disc, and the halo do not completely overlap in the $(g-r,g)$ CMD, it is possible to optimise the parameters for the halo, thick disc, and thin disc sequentially.

Fig.\ \ref{fig:triangle} gives the schematic diagram for distinguishing areas where only the halo or both the halo and thick disc contribute significantly to the star counts, and where only the thick and thin discs contribute in this manner. We define the separating lines in terms of the relative densities of these three components based on the \jj model. In the resulting triangle (dashed line), the halo dominates over the other two components because the relative density of the discs to the halo is less than $1\%$. In the area below the solid line, the thin disc can be neglected, because the relative density of thin disc to thick disc is less than $1\%$. We fit first the halo parameters in the triangle. We then fix the halo parameters and fit the thick disc parameters below the full line by including the star counts of the halo. Finally, we investigate the thin disc parameters by taking into account all three components in the full CMD.

%
%
%
%
\subsubsection{Halo fitting}\label{sec:set5:halo}

\begin{figure}
    \includegraphics[width=0.5\textwidth,height=0.33\textwidth]{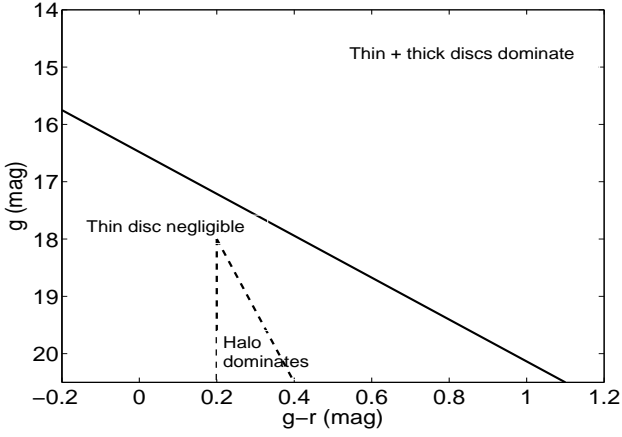}\\
    \caption{A schematic diagram of the CMD illustrating the distinction between thin and thick disc and halo stars. The dashed triangle represents the area where the halo dominates. The solid line is the upper boundary below which the thin disc contribution is negligible.}
    \label{fig:triangle}
\end{figure}

Locating an area with halo stars but very few disc stars in the CMD is the first step. Our choice is marked by the dashed triangle in Fig. \ref{fig:triangle}.  The discs stars in this selected region can be neglected, i.e.,
$(\rho_{\rm{thin}}+\rho_{\rm{thick}})/\rho_{\rm{halo}}<0.01$ as estimated by the \jj model. The boundary conditions to determine the halo parameters are $g-r\geq0.2$ and $g\geq 15.5+12.5(g-r)$. We choose an oblate $r^{1/4}$ spheroid and search for a best fit in the parameter ranges
\begin{description}
    \item[] $0.00005 ~\rm{M_{\sun}/pc^3} \leqslant \Omega_{\rm{h}}({\sun}) \leqslant 0.00035 \rm{M_{\sun}/pc^3}$,
    \item[] $2000 ~\rm{pc} \leqslant r_{\rm{eff}} \leqslant 30000 ~\rm{pc}$,
    \item[] $0.57 \leqslant q_{\rm{h}} \leqslant 0.70$.
\end{description}

\begin{figure} 
   \includegraphics[width=0.5\textwidth]{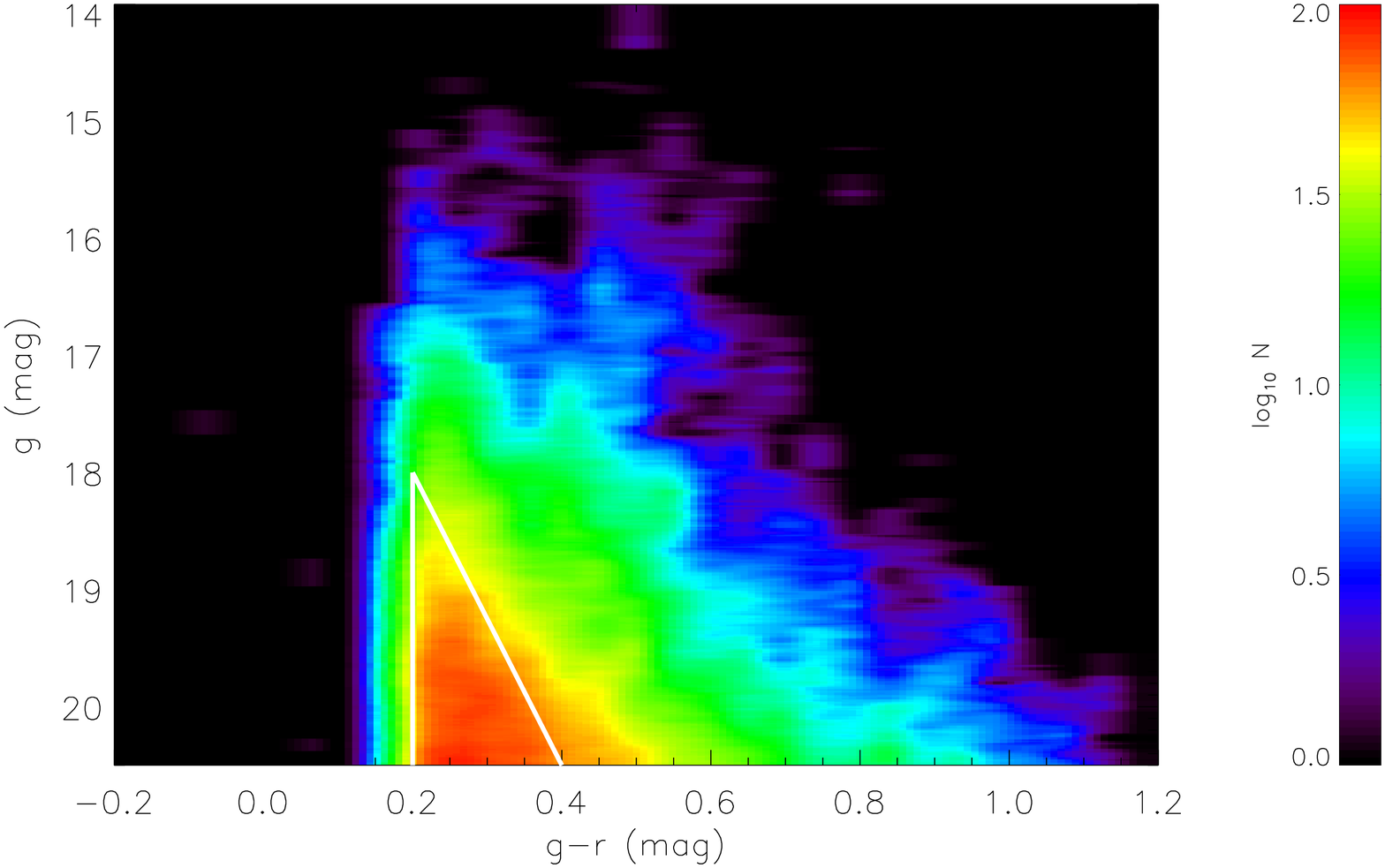}\\
   \includegraphics[width=0.5\textwidth]{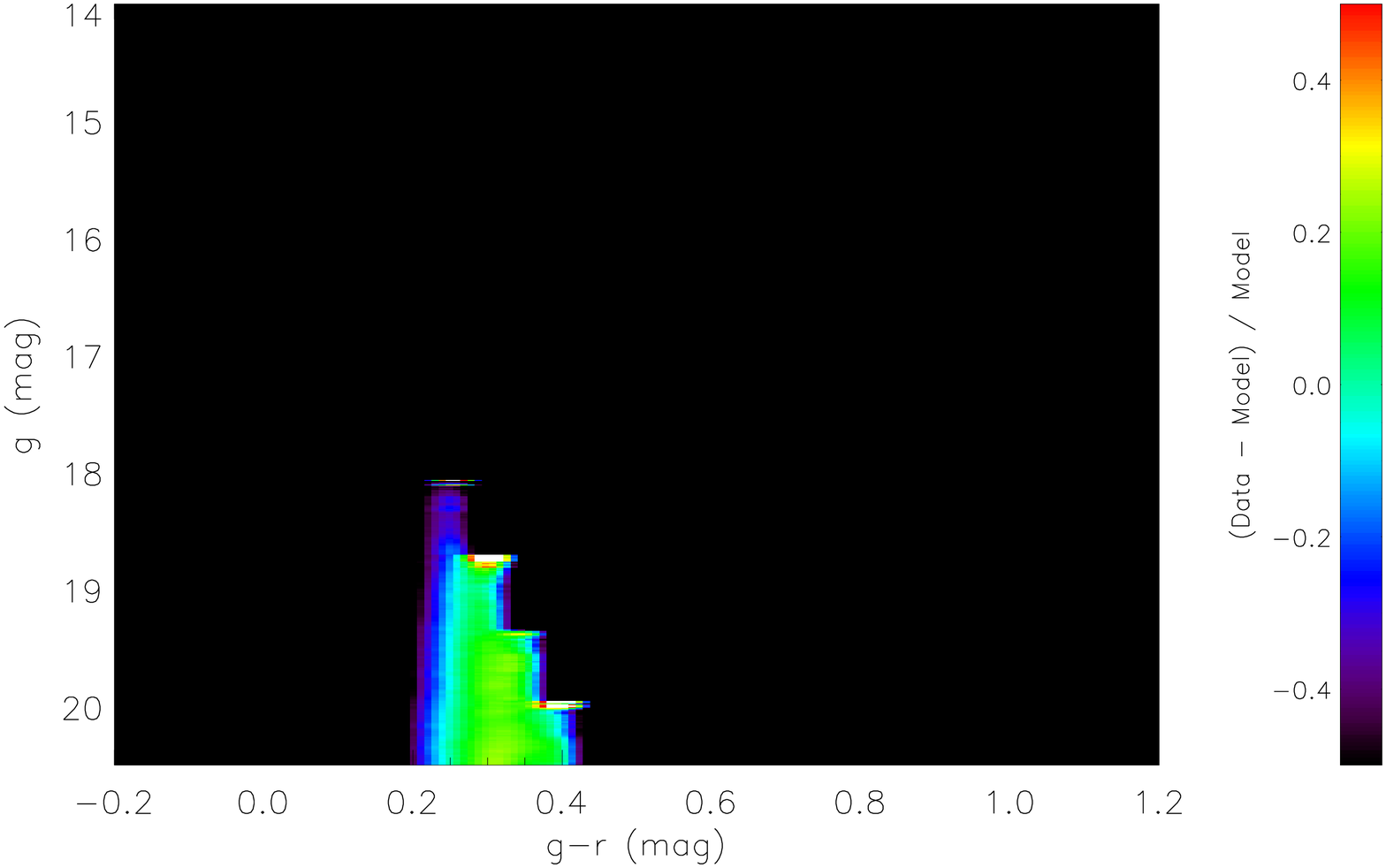}\\
   \includegraphics[width=0.5\textwidth]{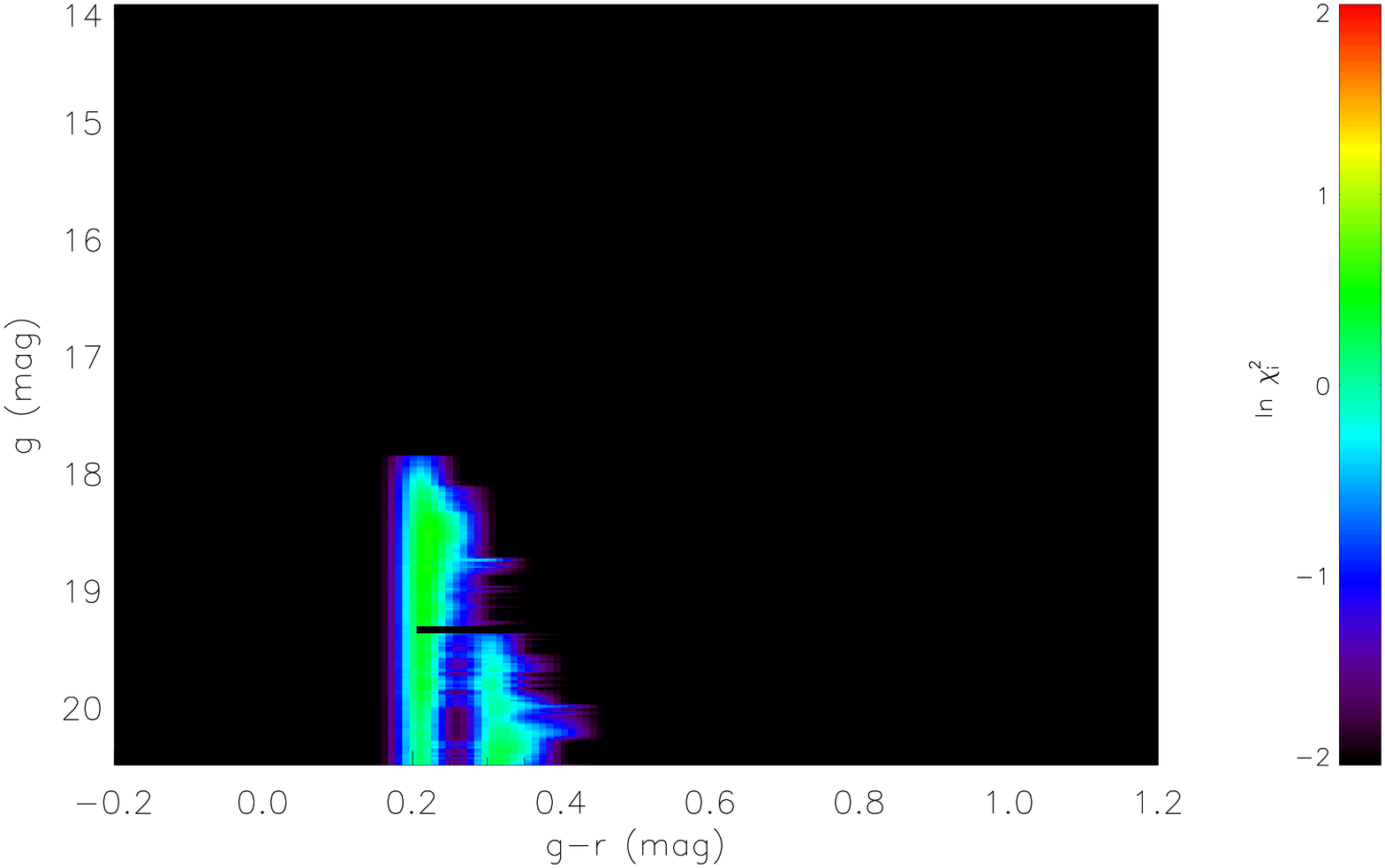}\\
   \caption{Hess diagram, relative differences, and $\chi^2$ distribution (top to bottom) of the best-fit halo in \tri for parameter set 5. The upper panel is the Hess diagram of the halo of set 5. The white thick lines enclose the triangle mentioned in Fig.\ \ref{fig:triangle}. Two lower panels are the relative difference and $\chi^2$ distribution restricted in the halo area. The coding is the same as in the rows of Fig.\ \ref{fig:NGP_SDSS_hess_set}. (A colour version is available on-line.)}
   \label{fig:set5:halo}
\end{figure}

The fitting result is shown in Fig.\ \ref{fig:set5:halo}. The best-fit halo parameters are listed in the last column (set 5) of Table \ref{tab:tri:def} with $\langle\chi^2\rangle=4.86$.

\subsubsection{Thick disc fitting}

We fix the parameters of the halo determined in Sect.\ \ref{sec:set5:halo} to fit the parameters of the thick disc in the area dominated by the thick disc and halo (below the solid line determined by $g\geq 16.5+3.64(g-r)$ in Fig.\ \ref{fig:triangle}). In this area, only relatively few thin-disc stars exist, i.e., $\rho_{\rm{thin}}/(\rho_{\rm{thick}}+\rho_{\rm{halo}})<0.01$.

\begin{figure} 
   \includegraphics[width=0.5\textwidth]{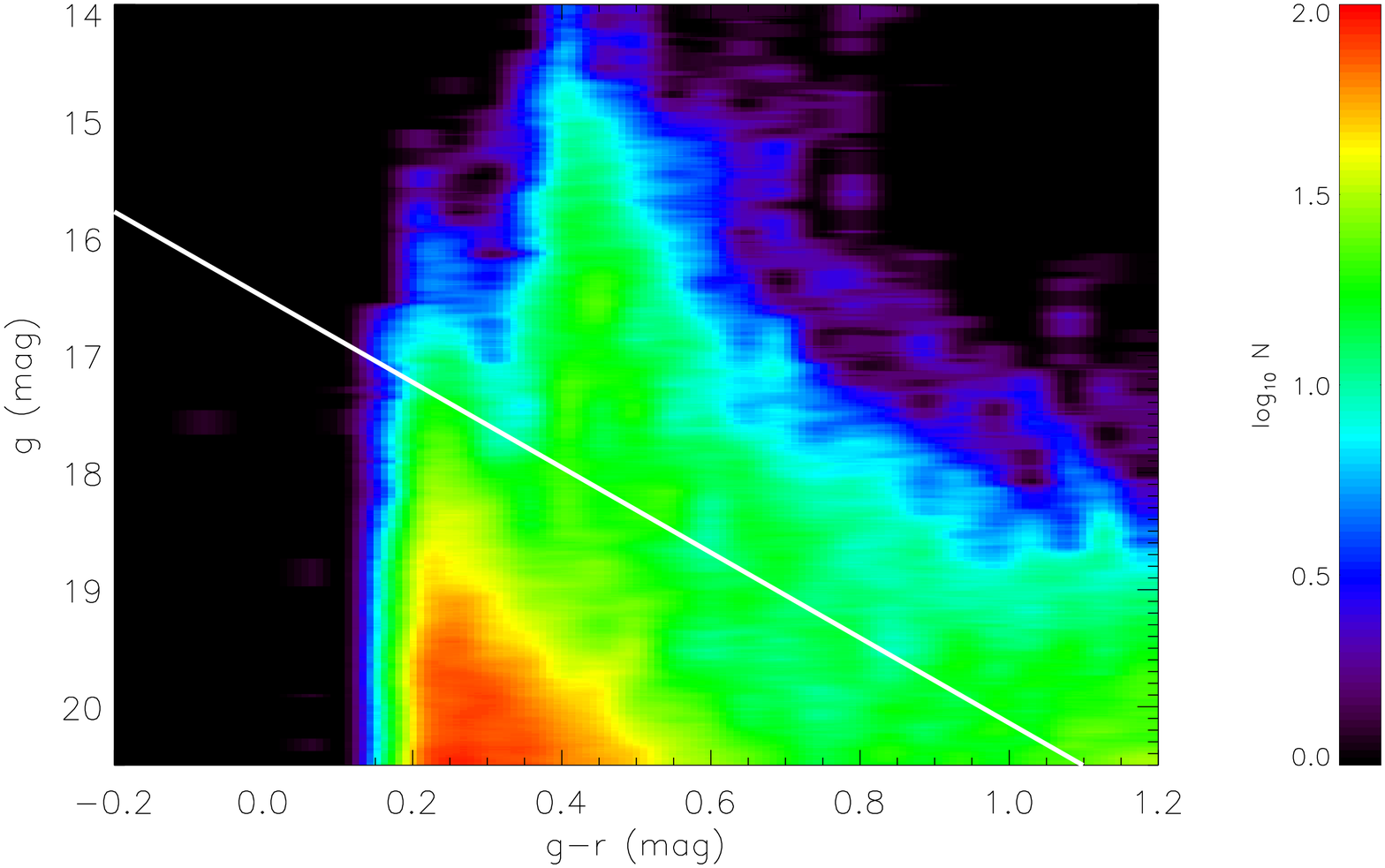}\\
   \includegraphics[width=0.5\textwidth]{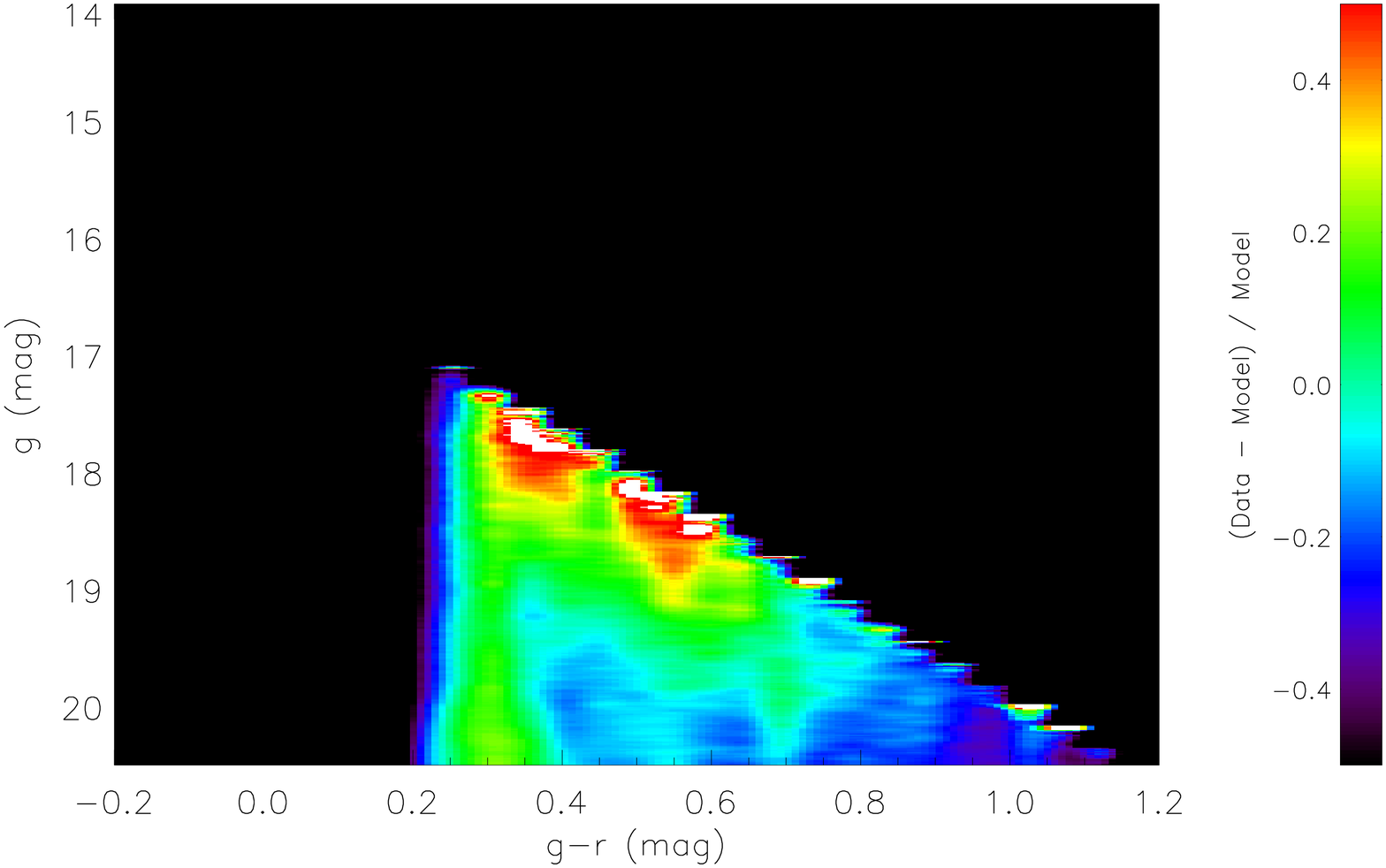}\\
   \includegraphics[width=0.5\textwidth]{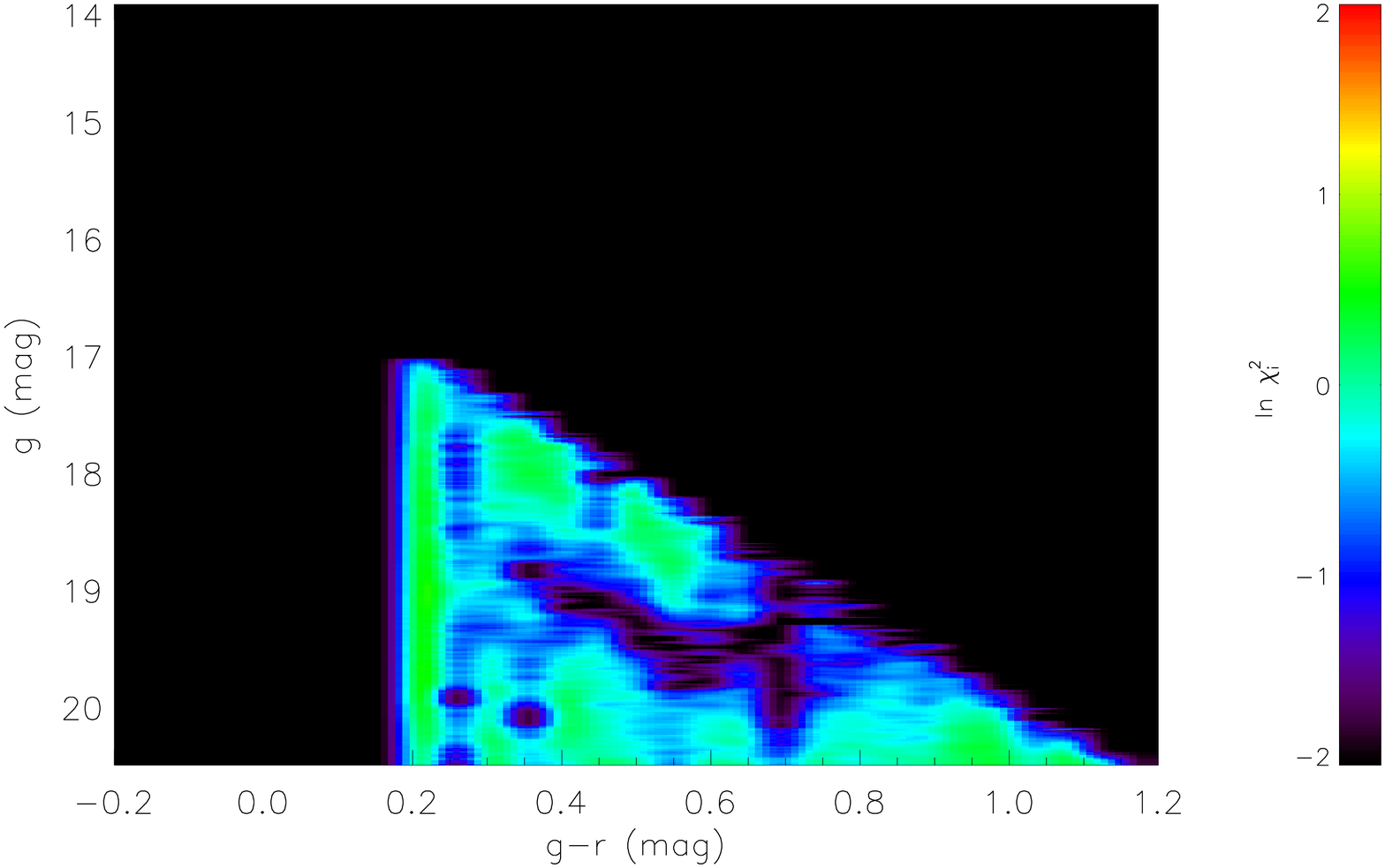}\\
   \caption{Same as in Fig.\ \ref{fig:set5:halo} but for thick disc plus halo stars and the parameter set 5. The upper panel is the Hess diagram of the halo plus the thick disc of Set 5. The white thick line in Fig.\ \ref{fig:triangle} shows the area in which the thin disc is negligible. The two lower panels are the relative difference and $\chi^2$ distribution restricted in the thick disc area. (A colour version is available on-line.)}
   \label{fig:set5:thick}
\end{figure}

The $\mathrm{sech^2()}$ and $\exp()$ prescriptions are tried to fit the vertical profile of thick disc with the parameters scale height $h_{\rm{t}}$ and local calibration $\rho_{\rm{t}}({\sun})$ within the following ranges:
\begin{description}
    \item[] $0.0010 ~\rm{M_{\sun}/pc^3} \leqslant \rho_{\rm{t}}({\sun}) \leqslant 0.00035 \rm{M_{\sun}/pc^3}$;
    \item[] $400 ~\rm{pc} \leqslant h_{t} \leqslant 900 ~\rm{pc}$.
\end{description}
The minimum $\langle\chi^2\rangle=16.22$ was found for the $\mathrm{sech^2()}$ profile with the best-fitting parameters listed in Col.\ 7 of Table \ref{tab:tri:def}. The minimum $\langle\chi^2\rangle$ of the fit using an exponential function is 0.312.

The fit result, shown in Fig.\ \ref{fig:set5:thick}, shows on the one hand the homogeneous quality of the fit, but on the other hand a systematic trend along the MS of the thick disc with too few stars in the G-dwarf regime and too many K and M dwarfs.

%
\subsubsection{Thin disc choice}

For the thin disc fitting, we now fix the halo and thick disc as determined above. For the thin disc, we cannot get a best fit for all the free parameters. On the basis of the analysis of the advantages and disadvantages of the thin disc models of for the parameter sets 1, 2, 3, and 4, we adjust the parameters iteratively to get a better result. All iterations to improve the parameters in order to find an optimal Milky Way model must be done by hand. The power-law index $\alpha$ in the scale-height function changes the fraction of different sub-populations as a function of height above the mid-plane. In the Hess diagram, this affects the balance of densities between $g-r>1$ (faint, nearby) and $0.2<g-r<0.4$ (bright, more distant). The values for $z_0$ and $\rho_{\rm{d,0}}$ affect the total density of the thin disc. These are the general directions for adjusting the parameters. The value of $t_0$ was not changed.

We performed these investigations for both vertical profile shapes. The sech$^2$ profile results in a better fit with a minimum $\langle\chi^2\rangle=1.33$. (The exponential profile leads to a best fit with $\langle\chi^2\rangle=2.95$). 

\subsubsection{Model properties}

\begin{figure} 
    \includegraphics[width=0.5\textwidth]{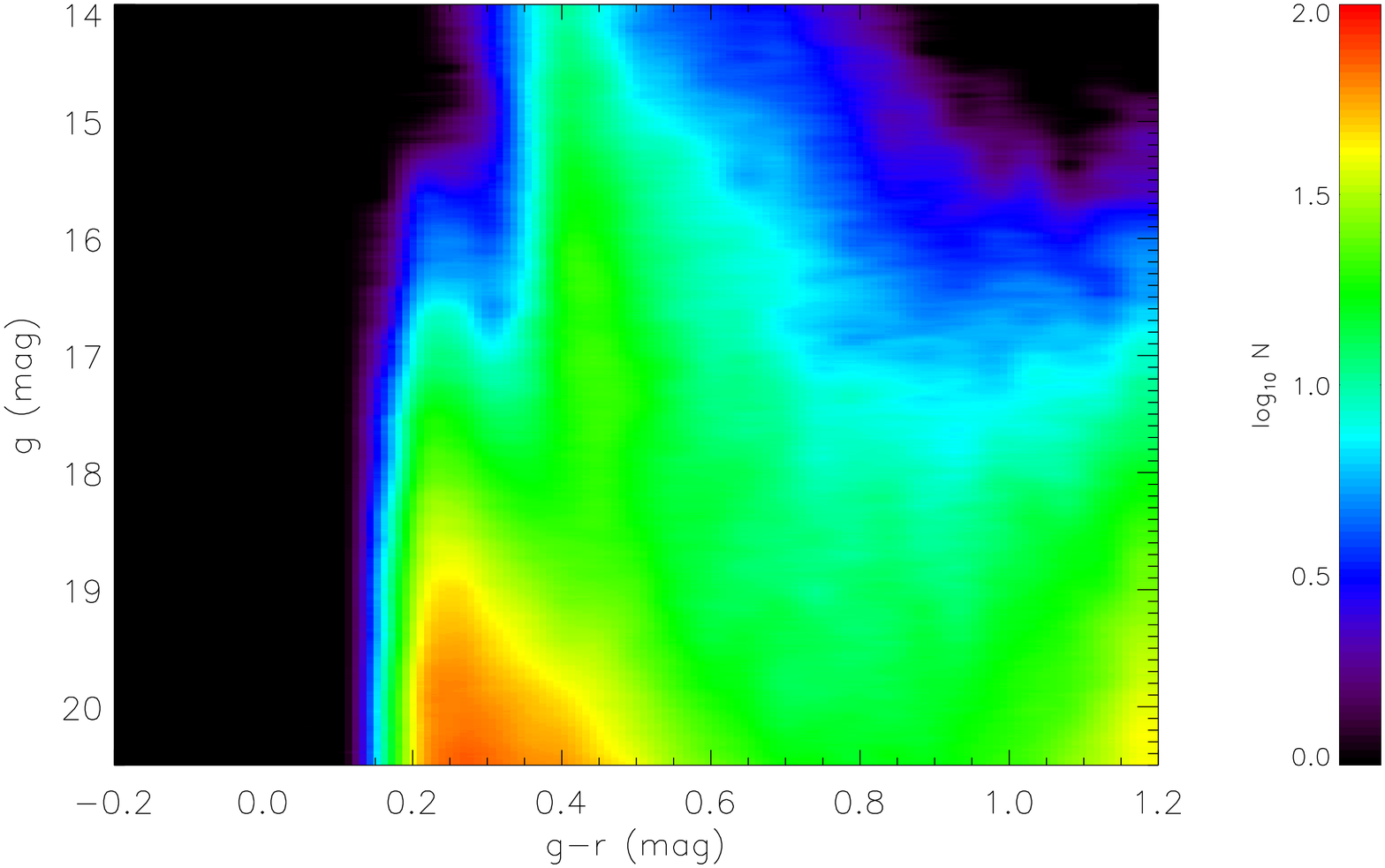}\\
    \includegraphics[width=0.5\textwidth]{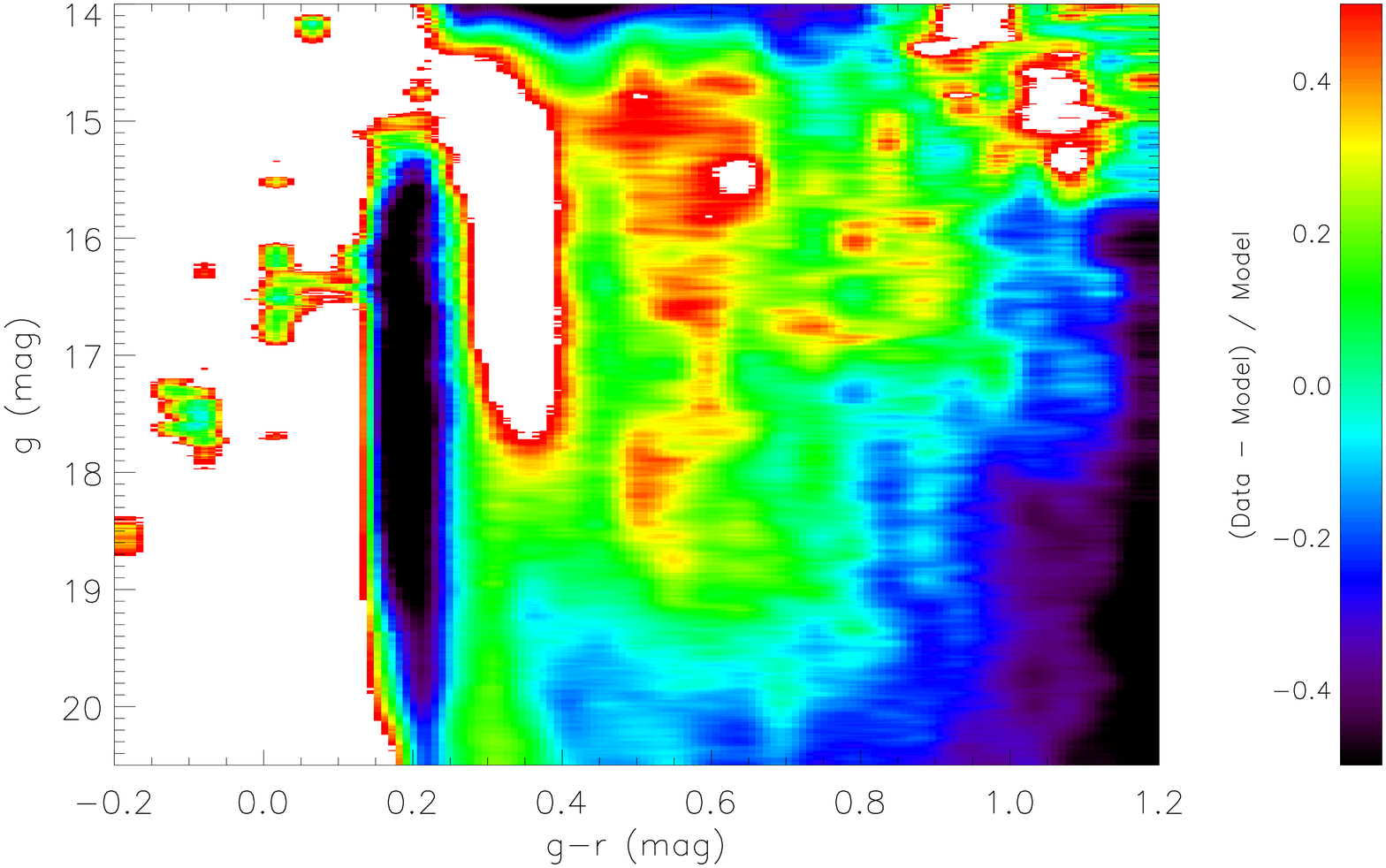}\\
    \includegraphics[width=0.5\textwidth]{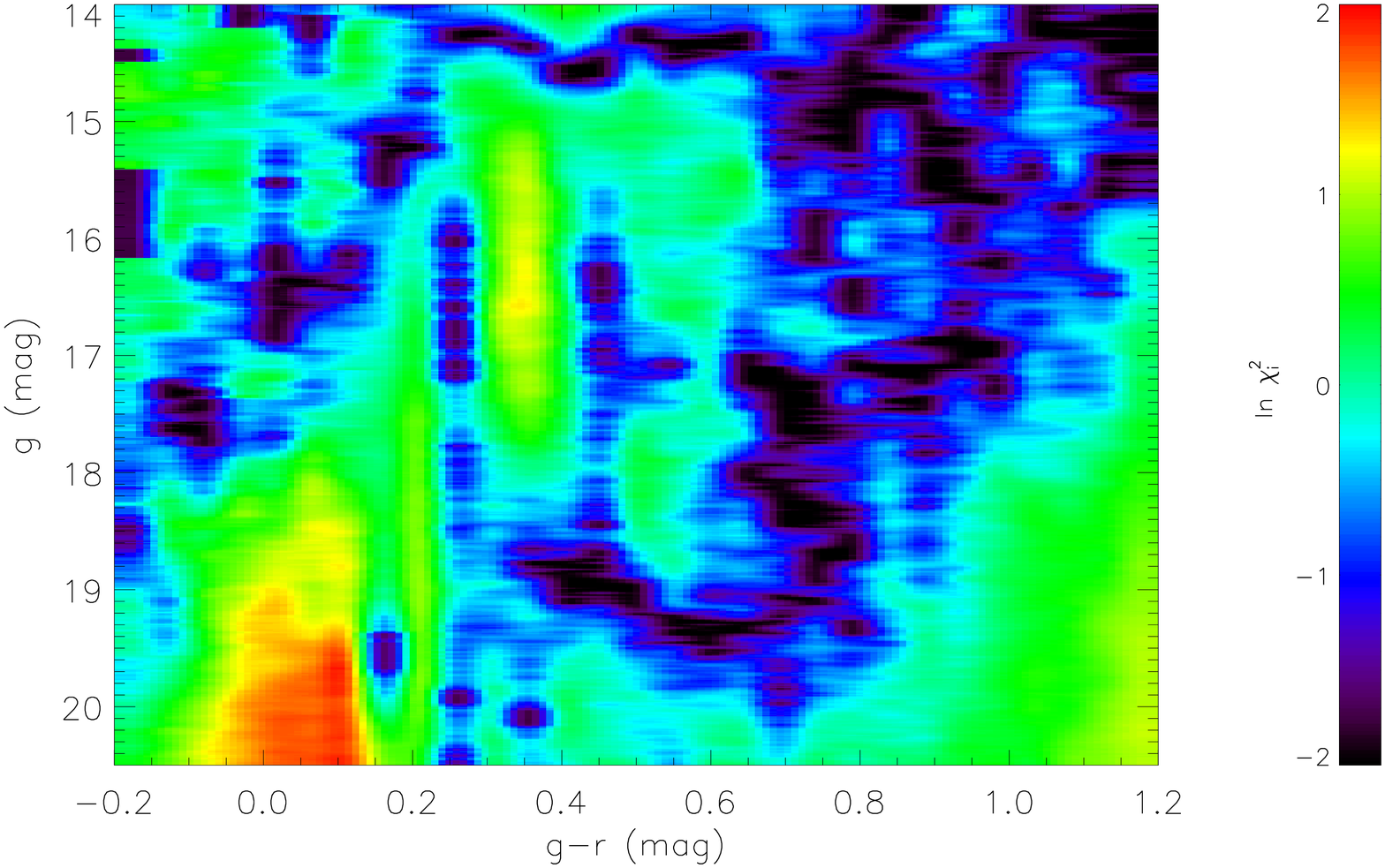}\\
    \caption{Hess diagrams (top), relative differences (middle)
between the Hess diagrams of the parameter set 5 and the SDSS observations, and
the $\chi^2$ diagrams (bottom). The notation is the same as in Fig.\
\ref{fig:NGP_SDSS_hess_set}. (A colour version is available on-line.)}
    \label{fig:hess:set5}
\end{figure}

\begin{figure}
\includegraphics[width=0.5\textwidth,height=0.33\textwidth]{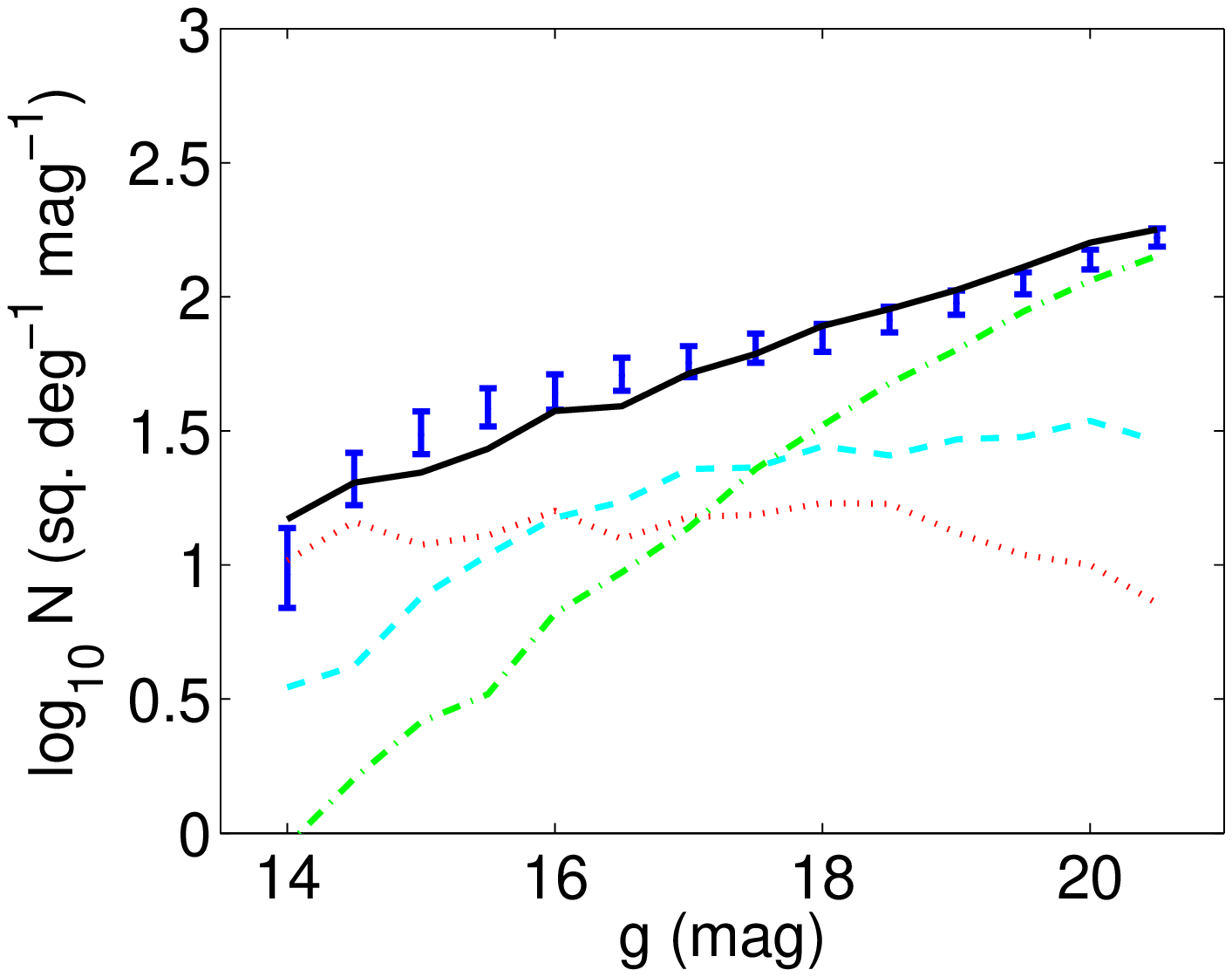}\\
\includegraphics[width=0.5\textwidth,height=0.33\textwidth]{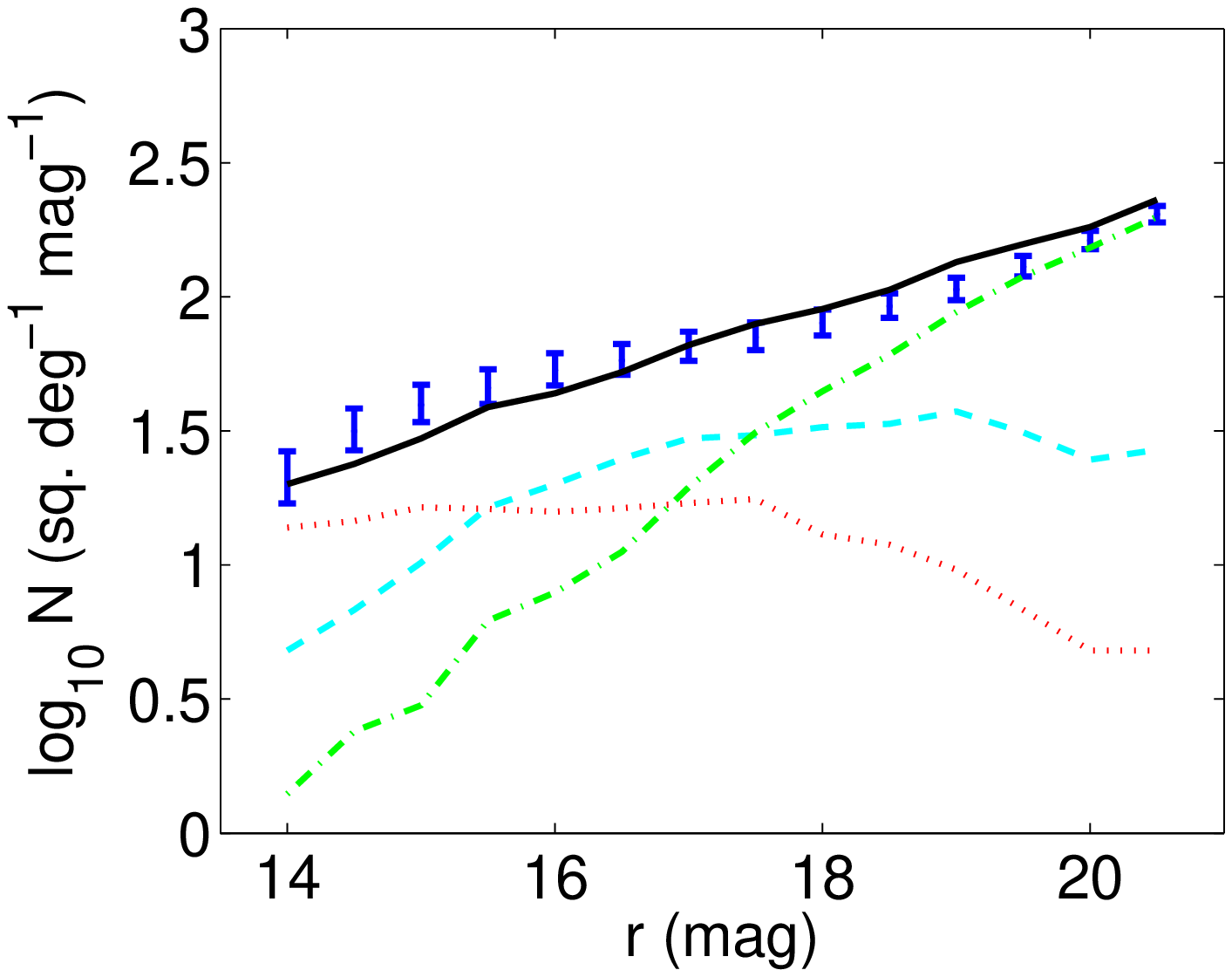}
    \caption{The magnitude distributions of the SDSS observations and
the \tri simulations generated with parameter set 5 in the $g$ (top) and $r$
(bottom) band. The x- and y-axes are similar to the bottom panel of
Fig. \ref{fig:NGP_SDSS_only_hess}. The error bars (blue) are from the SDSS
DR7 data with a colour limit of $0.2\leq(g-r)<1.2$. The black solid
line represents the model star counts from \tri set 5. The dotted
(red), dashed (cyan), and dot-dashed (green) line stand for the thin
disc, thick disc, and halo of \tri set 5, respectively. (A colour version is available on-line.)}
    \label{fig:mag:set5}
\end{figure}

\begin{figure} 
\includegraphics[width=0.5\textwidth,height=0.33\textwidth]{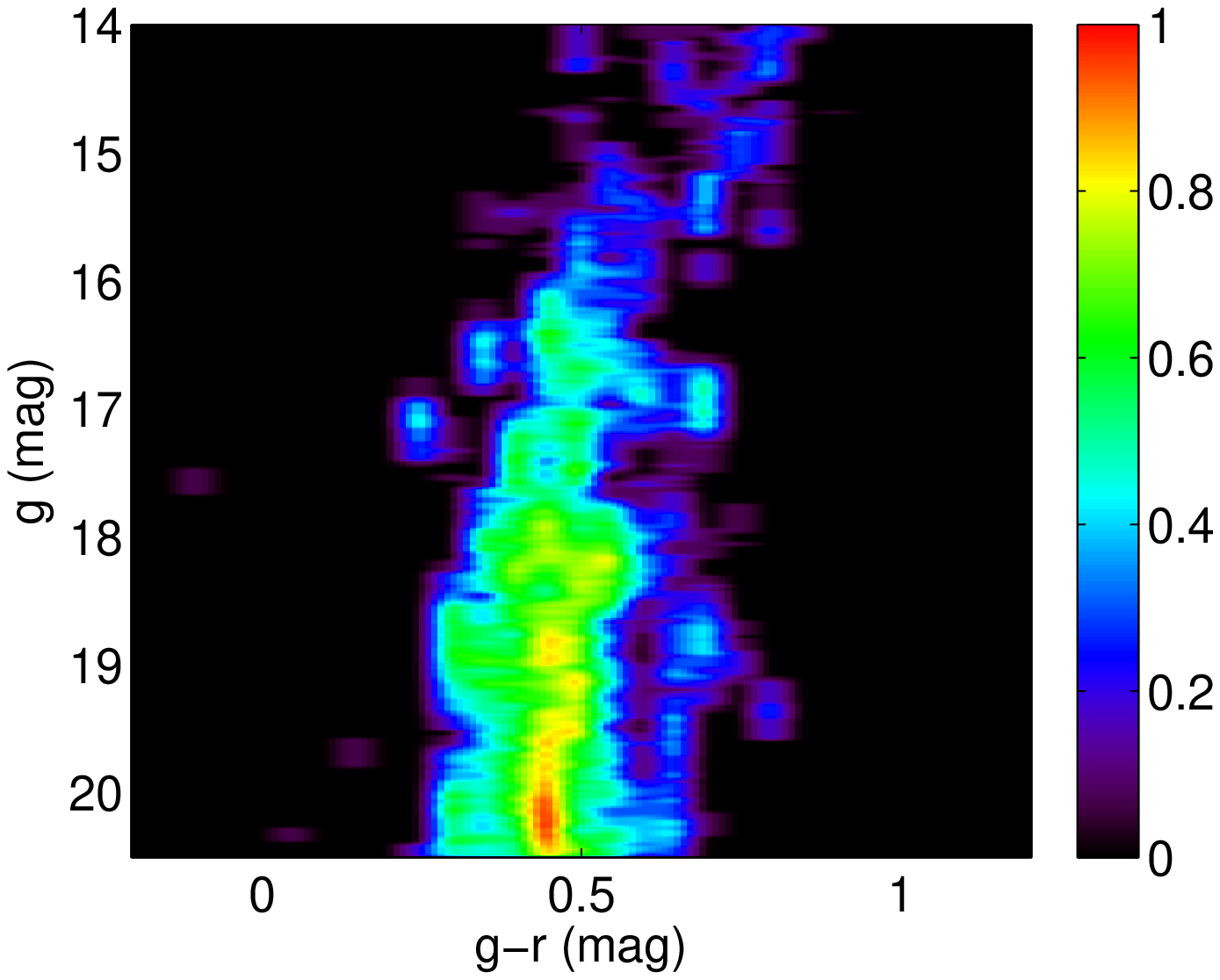}\\
\includegraphics[width=0.5\textwidth,height=0.33\textwidth]{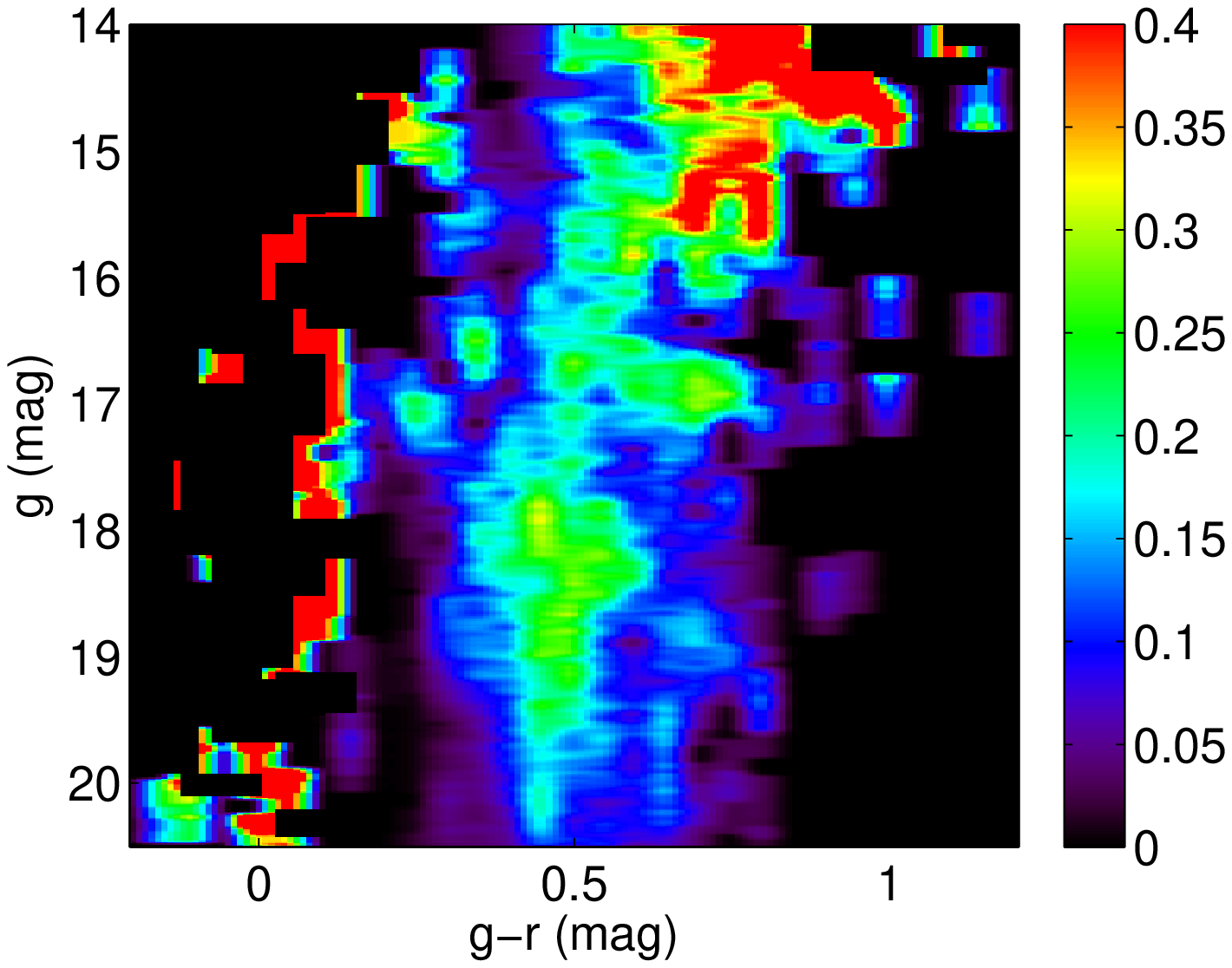}
    \caption{The fraction of the contribution by modelling giants of parameter set 5
in the \tri model. The upper panel shows the giant counts simulated by
set 5 of the \tri model in the NGP field. The colour code covers the
star number density from 1 to 10 per square degree, 1 mag in $g$, and 0.1 mag
in $g-r$. Lower panel: The relative fraction of giants to all stars in
the model. The colour code is from $0\%$ to $40\%$. (A colour version is available on-line.)}
    \label{fig:set5:giants}
\end{figure}

The Hess diagram analysis for input parameter set 5 is plotted in Fig.\ \ref{fig:hess:set5}.  The $\chi^2$ distribution is smoother than for the other four sets, but the systematic features in the relative deviations of data from model do not vanish. These are as follows(see middle panel of Fig. \ref{fig:hess:set5}):
\begin{enumerate}
  \item The dark vertical stripe at $g-r=0.2$ is a sign of too many blue thin-disc stars (A stars), which do not show up in Figures \ref{fig:set5:thick} and \ref{fig:set5:halo}. The constant SFR overestimates the fraction of young stars.
  \item The white island at $g-r=0.35$ and $15<g<17$\,mag points to a mismatch of the thick disc turn-off stars. There are too few nearby stars owing to the flat sech$^2$ profile and the turn-off is too red.
  \item The faint red end with a colour-magnitude range of $g-r>1.0$ and $g>18$ mag shows too many M dwarfs. The reason may be that the $g-r$ colour of the thin disc M dwarfs is slightly too blue or that the IMF allows too many stars at the low mass end.
  \item The white islands at the bright red end are another hint that the red giants of the thick disc and halo are too blue (see also Figs.\ \ref{fig:iso:tri} and \ref{fig:set5:giants}).
\end{enumerate}

The magnitude distribution in the $g$ and $r$ bands (see Fig.\ \ref{fig:mag:set5}) shows larger discrepancies than for the default Set 1 (top row of Fig.\ \ref{fig:mag_distri_SDSS_set1234}). The shoulder at $17\mathrm{mag}<g<19\mathrm{mag}$ is still missing. This is again a hint that the Hess diagram analysis is a much more powerful constraint of the stellar population than luminosity function fitting.

To quantify the impact of giants on the Hess diagram, we select the giants using the surface gravity $\log{g}<3.8$. The absolute and relative contributions of giants simulated by parameter set 5 are shown in Figure \ref{fig:set5:giants}. The giants of $\sim 30\%$ stars exist in the Hess diagram. The rigid line of giant counts is close to $g-r=0.5\sim 0.7$. The giants in the thin disc have a higher relative contribution, though the absolute counts are too low for the thin disc.

The parameters of the optimized model are listed in Table \ref{tab:tri:def} (set 5). Compared to the default model (Set 1), the final Set 5 has a lower local surface density, a smaller maximum scale height $h_\mathrm{z,max}$ of the thin disc, and a shallower nearby halo profile of lower local density, because the old, extended thin disc and the halo do not need to account also for the thick disc stars.

Compared to the \jj model with $\langle\chi^2\rangle=4.31$ \citep{jgv}, the fit of Set 5 is much worse. This is also obvious from the median relative deviations $\Delta_\mathrm{med}=0.2576$. The mismatch of the isochrones in the $ugriz$ filters alone cannot account for these deviations. The simple structural forms of the thin and thick discs are the main reasons for the large discrepancies.

\section{Simulations and results of the \besan model}

The \besan model allows no variations in the choice of free parameters. We performed a multi-colour analysis of the predictions in different colours, as given in Table \ref{tab:range}.  We selected all stars in the NGP field from DR7 within the colour and apparent magnitude ranges listed in Table \ref{tab:range} \footnote{The mock catalogues of the thin and thick discs and the halo and of their sum are generated by Monte Carlo procedures in different procedures. The number of stars in the sum of the three components are therefore not exactly the same as the number of stars in the general simulation.} and with magnitude errors smaller than 0.2 mag in each filter. The selection of the field and objects is similar to that used in \citet{jgv} and Section 4.

\begin{table*}
\caption{Properties of the observed and simulated CMDs.}
\label{tab:range} \centering
\begin{tabular}{ccccrrrrrcc}
  \hline
  diagram & colour & magnitude & $N_{obs}$ & $N_{mod}$ & ${\approx}~~(N_{thin}$ & $+N_{thick}$ & $+N_{halo})$ & ${N_{mod}}/{N_{obs}}$ & $<\chi^2>$ &$\Delta_{\rm{med}}$\\
  \hline
  $(u-g, g)$ & $[-0.2, 2.0)$ & $[14.0, 20.5)$ & 238\,724 & 261\,205 &  43\,285 & 133\,603 &  84\,550 & 1.09 & 555.43 &0.5256 \\
  $(g-r, g)$ & $[-0.2, 1.2)$ & $[14.0, 20.5)$ & 279\,174 & 312\,093 &  94\,590 & 134\,209 &  83\,587 & 1.12 & 408.15 &0.2724 \\
  $(r-i, r)$ & $[-0.2, 1.2)$ & $[14.0, 20.5)$ & 479\,913 & 433\,533 & 145\,784 & 183\,777 & 103\,955 &  0.90 & 222.62 &0.2595\\
  $(i-z, i)$ & $[-0.2, 0.5)$ & $[14.0, 18.5)$ & 220\,144 & 237\,249 &  97\,560 & 114\,468 &  25\,236 & 1.08 & 160.60 &0.2029 \\
  \hline
\end{tabular}
\tablefoot{Col.\ 1 gives the colour and magnitude of the CMD. Cols.\ 2 and 3 list the ranges in colour and apparent magnitude, and Cols.\ 4 and 5 show the total number of stars in the data ($N_{obs}$) and model ($N_{mod}$) for each CMD. The stellar numbers of the three components are listed as Cols.\ 6 to 8 ($N_{thin}$, $N_{thick}$ and $N_{halo}$). Col.\ 9 shows the ratio of star numbers of the model to the observations. Col.\ 10 lists the $<\chi^2>$ of the Hess diagrams. The last column lists the median absolute relative difference $\Delta_{rm{med}}$ of each Hess diagram for the region redder than the turn-off.}
\end{table*}

\subsection{Isochrones}

Since the thick disc and the halo are modelled in the \besan model by simple stellar populations with a single age and a relatively narrow metallicity range (11 Gyr and $\mathrm{[Fe/H]}=-0.78\pm0.30$ for the thick disc, 14 Gyr and $\mathrm{[Fe/H]}=-1.78\pm0.50$ for the halo), we can compare the isochrones with the fiducial sequences of star clusters with corresponding properties (see Table \ref{tab:gcs}). In \citet{An09}, a detailed analysis of isochrone fitting to fiducial sequences in the $ugriz$ filters of globular and open clusters from \citet{An08} is presented. The authors included a comparison with clusters observed with the CFHT-MegaCam by \citet{Clem08}. \citet{An09} found that the clusters show a systematic offset relative to the Yale Rotating Evolutionary Code (YREC) and MARCS models owing to a zero-point problem. We use five old globular clusters to cover the metallicity range of the thick disc and halo. 

\begin{figure*}\centering 
  \includegraphics[width=0.45\textwidth]{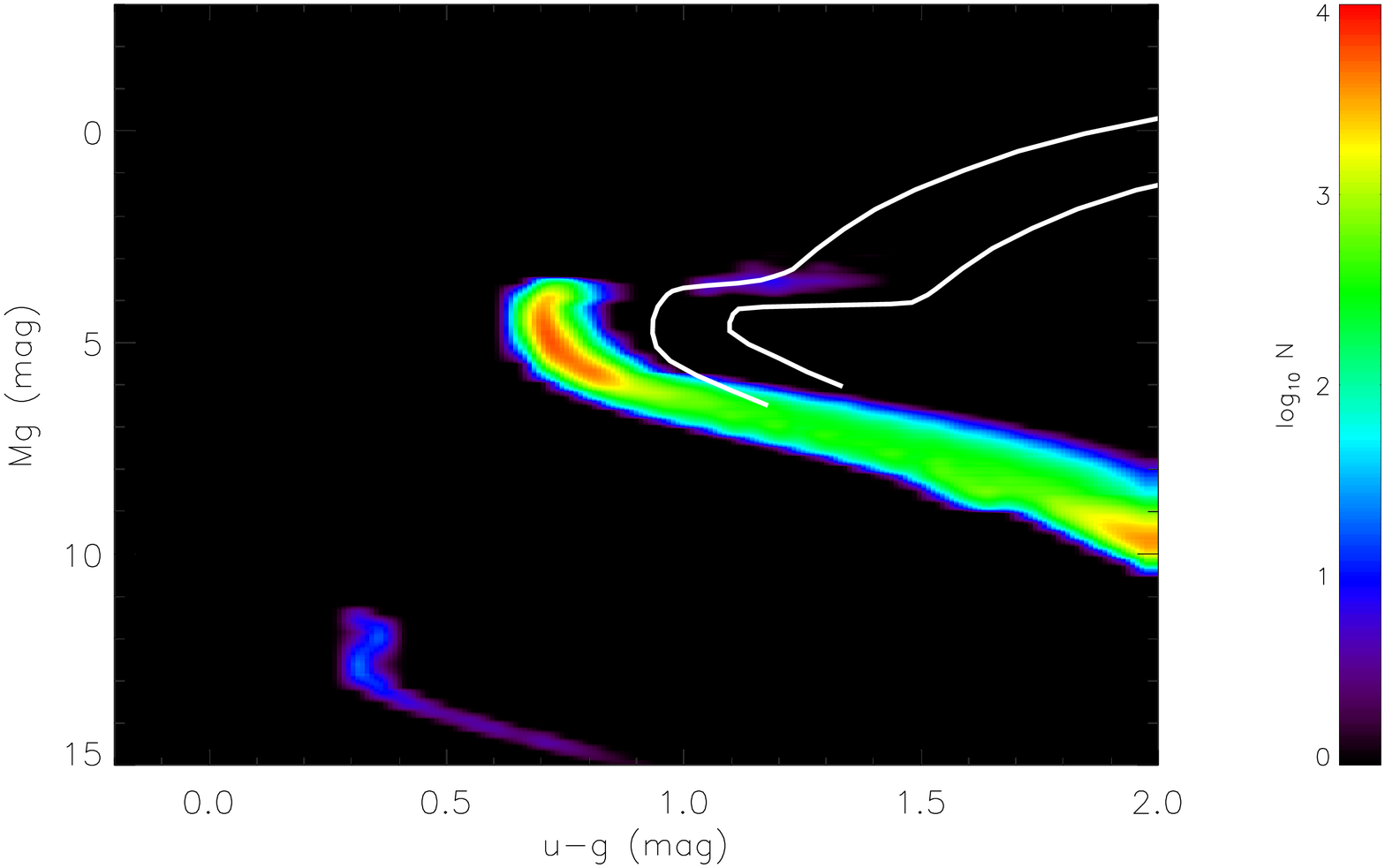}
  \includegraphics[width=0.45\textwidth]{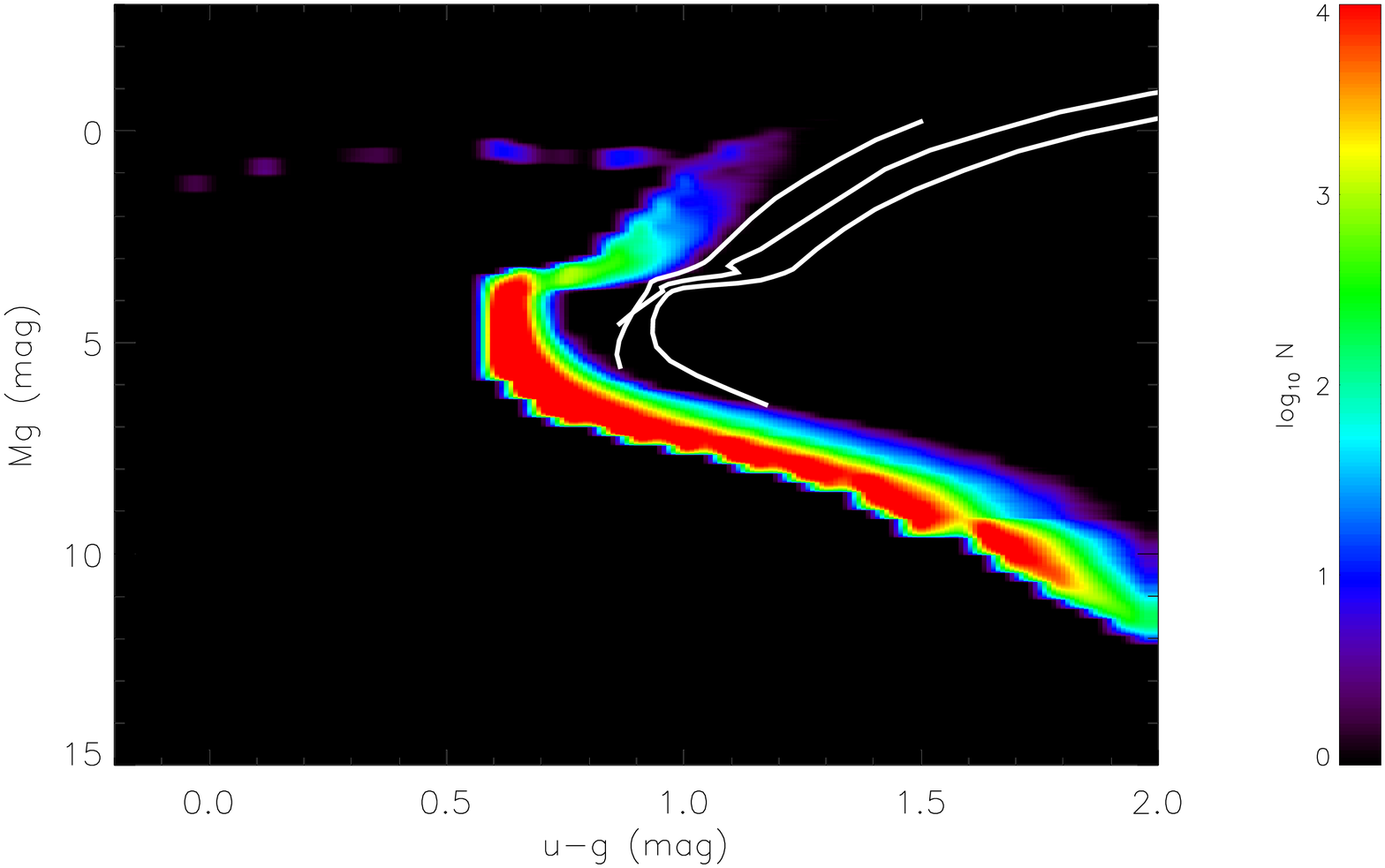}\\
  \includegraphics[width=0.45\textwidth]{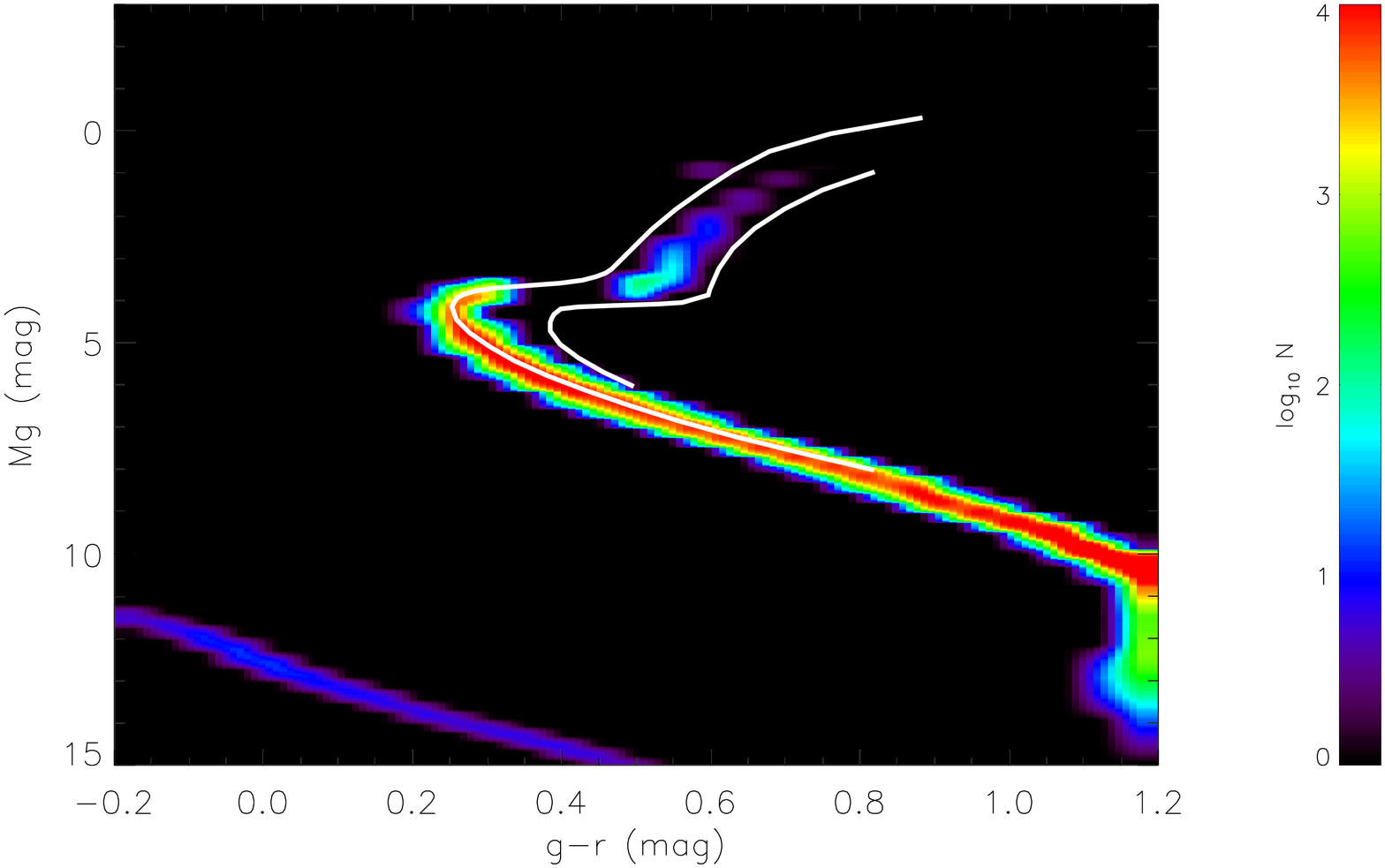}
  \includegraphics[width=0.45\textwidth]{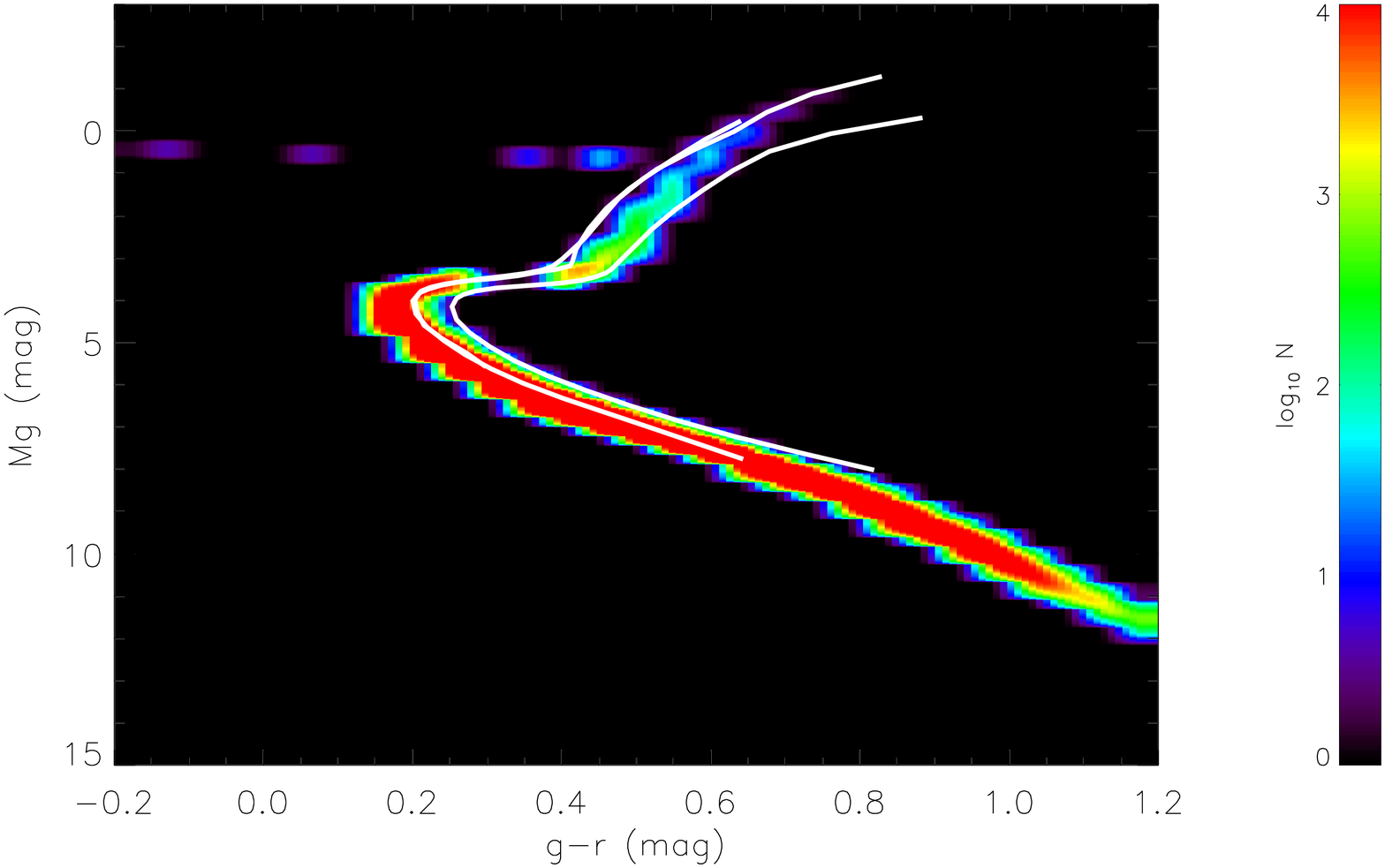}\\
  \includegraphics[width=0.45\textwidth]{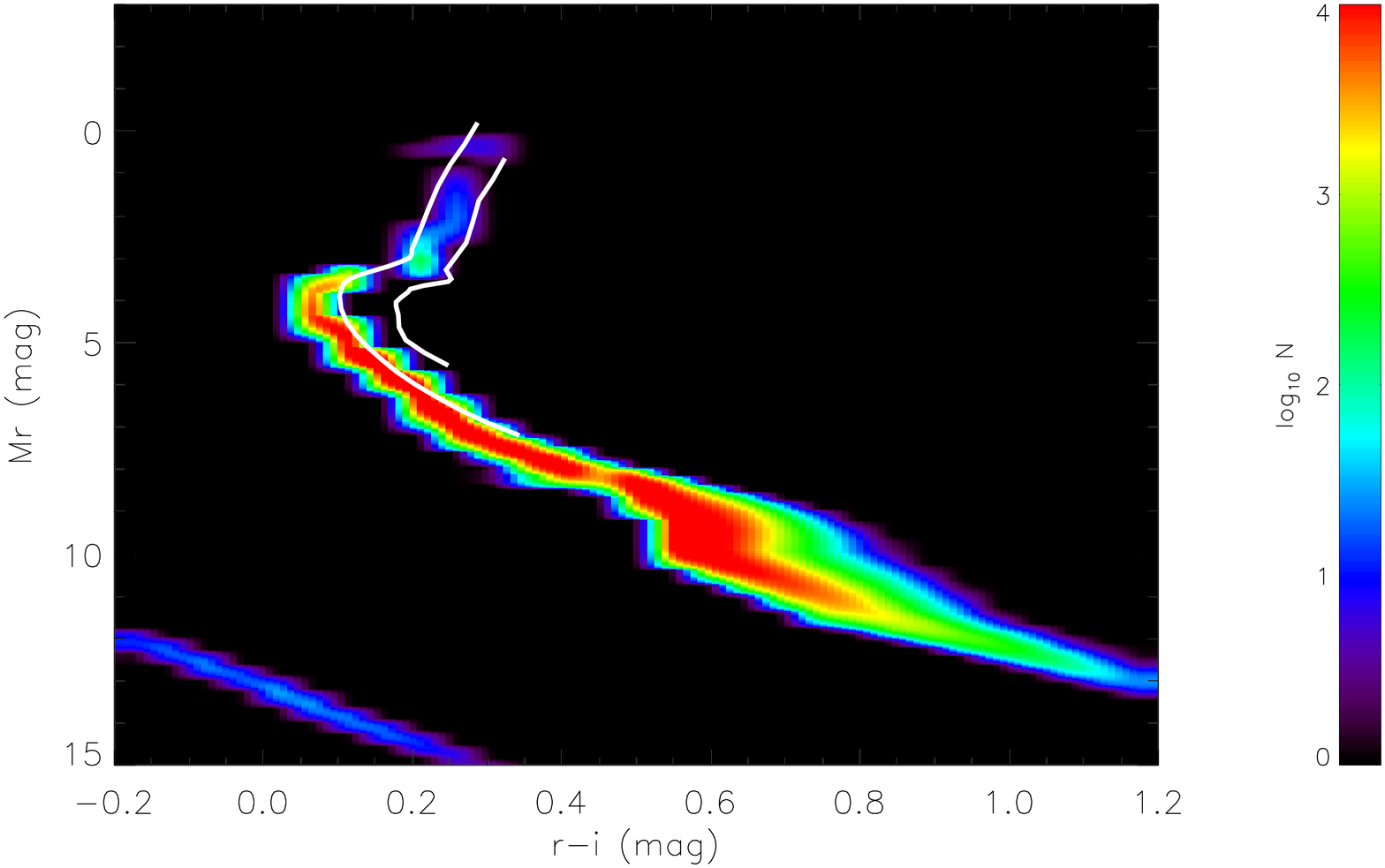}
  \includegraphics[width=0.45\textwidth]{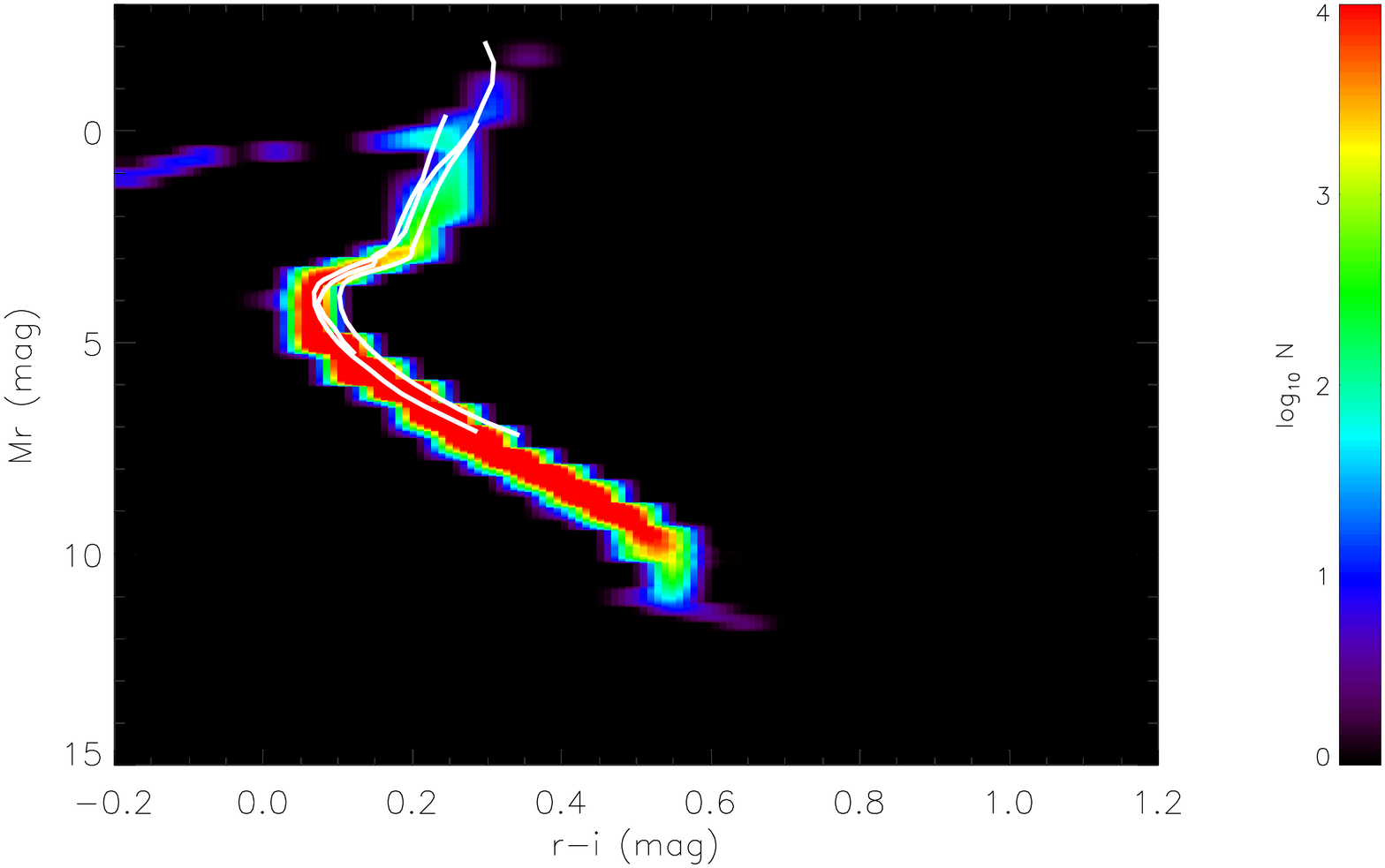}\\
  \includegraphics[width=0.45\textwidth]{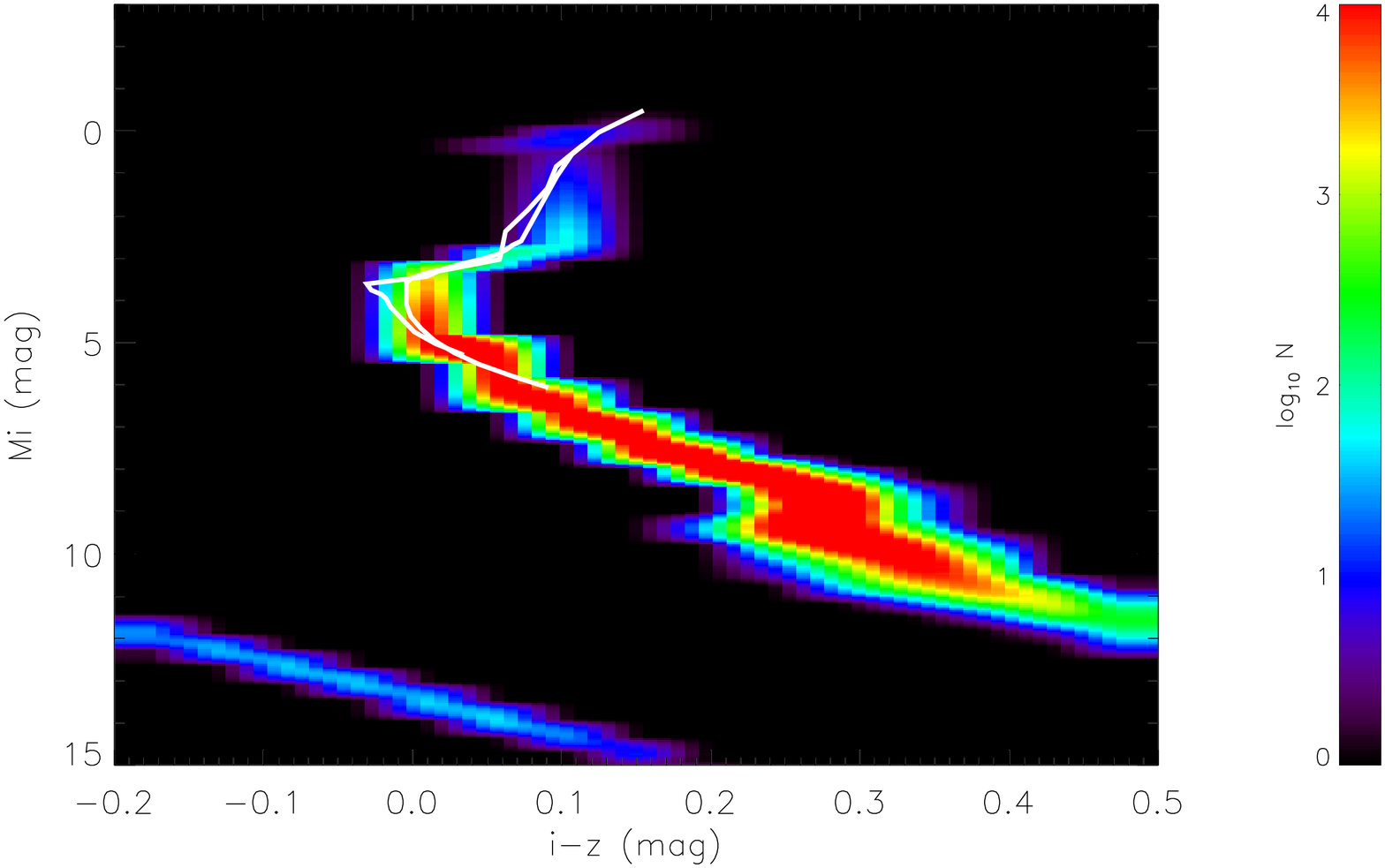}
  \includegraphics[width=0.45\textwidth]{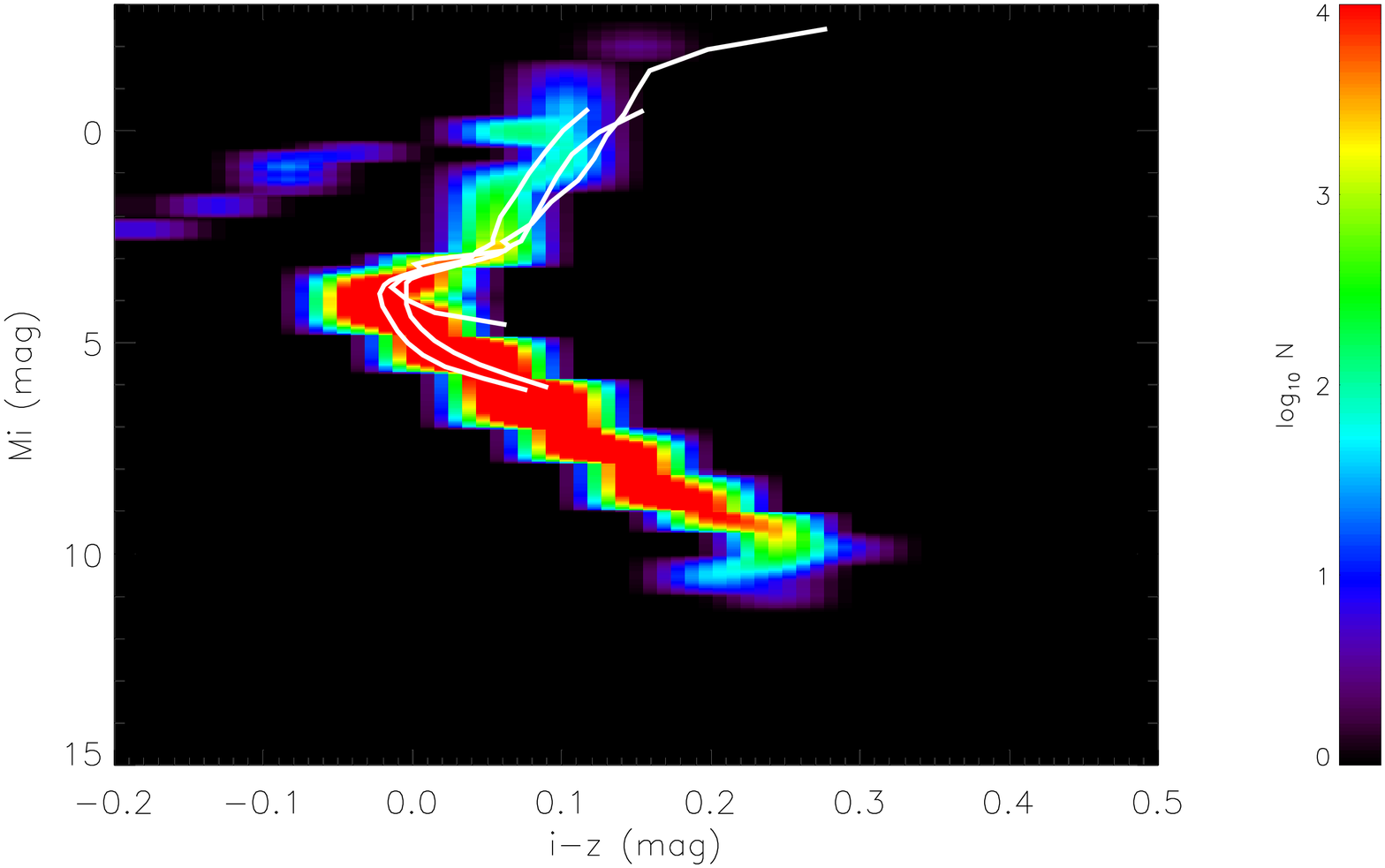}\\
\caption{The colour-absolute magnitude diagrams of the thick disc and halo of the \besan model and the fiducial isochrones of globular clusters in $(g-r, M_g)$. The Hess diagrams show the stellar distributions in $(u-g, M_g)$, $(g-r, M_g)$, $(r-i, M_r)$, and $(i-z, M_i)$ from top to bottom for the thick disc (left column) and halo (right column) simulated by the \besan model.  The overplotted thick lines (white) are the \citet{An08}'s fiducial isochrones based on SDSS observations of globular clusters. M5 and M71 are overplotted on the thick disc panels. M5, NGC 4147, and M92 are overplotted on the halo panels. (A colour version is available on-line.)}\label{fig:iso:besan}
\end{figure*}

We plot CMDs using the absolute magnitudes $M_g$, $M_r$, and $M_i$ from $-3$ mag to 15 mag and compare the CMDs to the fiducial isochrones of the globular clusters presented in Table \ref{tab:gcs}. There is a considerable offset of 0.3\,mag in $(u-g)$ between the fiducial isochrones and the simulated CMDs (top row of Fig.\ \ref{fig:iso:besan}), which is a hint of a zero-point problem in the $u$ band. This large offset explains why the Hess diagram of the $u-g$ completely fail to the actual data (see below).

In Fig.\ \ref{fig:iso:besan}, the left panels show the comparison for the thick disc, whereas the right-hand panels are for the halo. In the different CMDs, the fiducial isochrones of the globular clusters are overplotted: M5 and M71 are used to represent the thick disc; and M5, NGC 4147 and M92 are used for the halo (plotted in sequence from the turn-off point from the left to right in the right panels).

In the other CMDs of the thick disc, there is a smaller, but still significant, offset in the blue colour for the MS in the \besan model compared to the fiducial isochrones. The metallicity of the cluster M71 is closest to the metallicity of the thick disc adopted in \citet{Robin}. The matches of M5 with the lower metallicity are obviously better than for M71, though there is still an offset($\sim 0.2$ mag in the $u-g$) between the fiducial isochrones of M5 and the sequence of the thick disc of the \besan model. The isochrones of the thick disc correspond to fiducial isochrones with a metallicity below $-1.0$.

For the halo, the offsets are similar but slightly larger. The isochrone with $\mathrm{[Fe/H]}=-2.28$ is closest to the model but corresponds to the lower boundary of the metallicity range in the \besan model.

\subsection{Luminosity functions}

\begin{figure*}
  \includegraphics[width=0.5\textwidth,height=56mm]{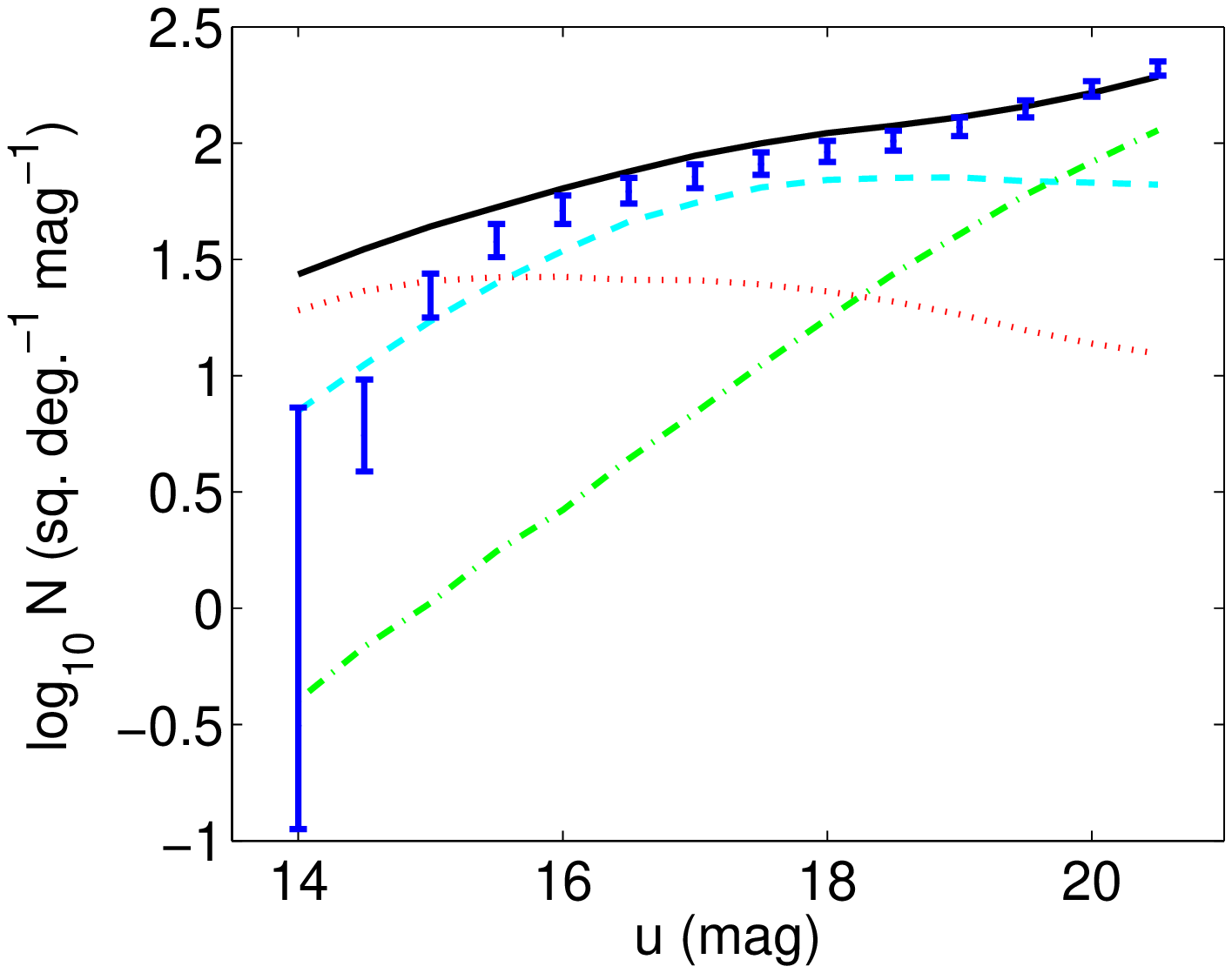}
  \includegraphics[width=0.5\textwidth,height=56mm]{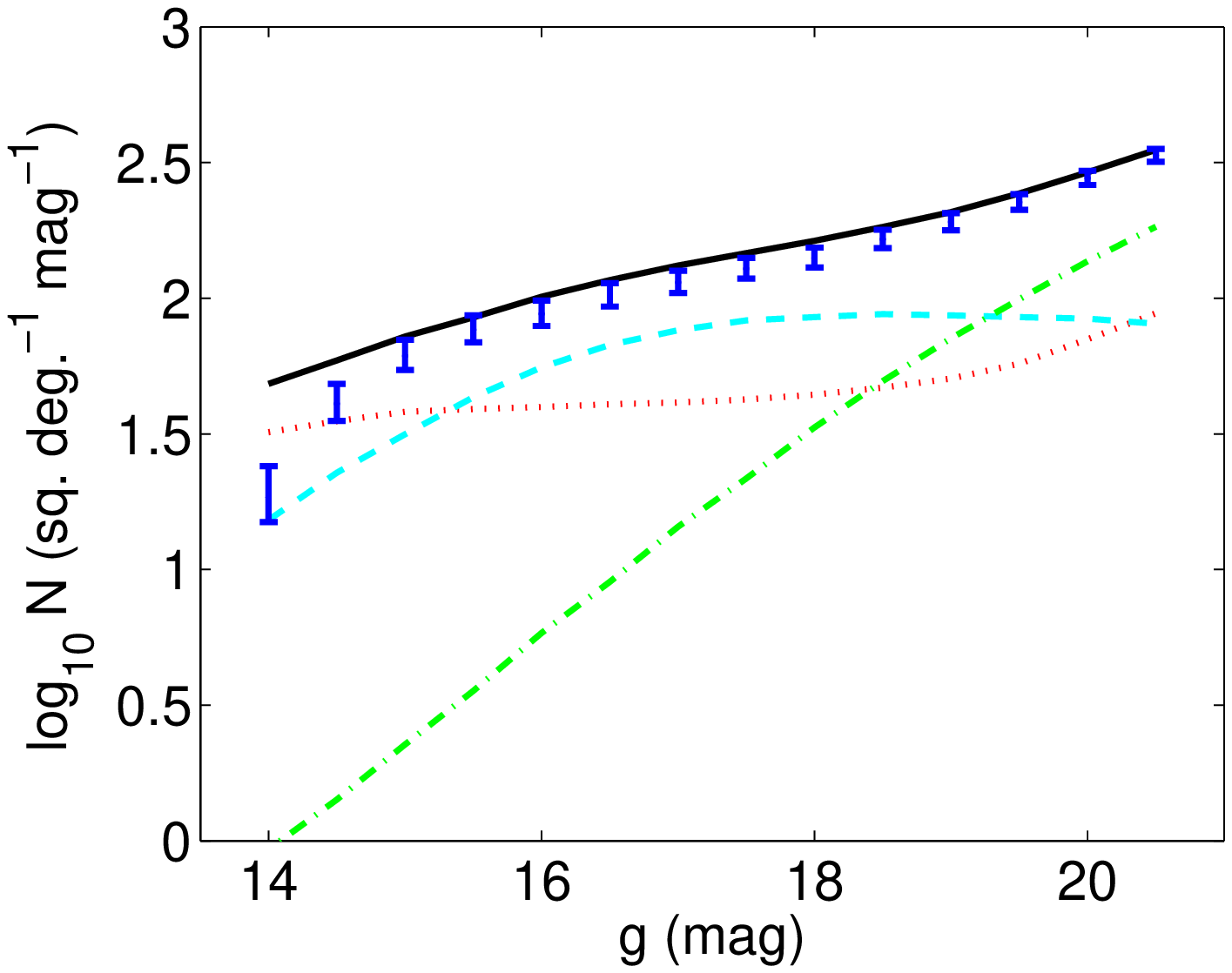}\\
  \includegraphics[width=0.5\textwidth,height=56mm]{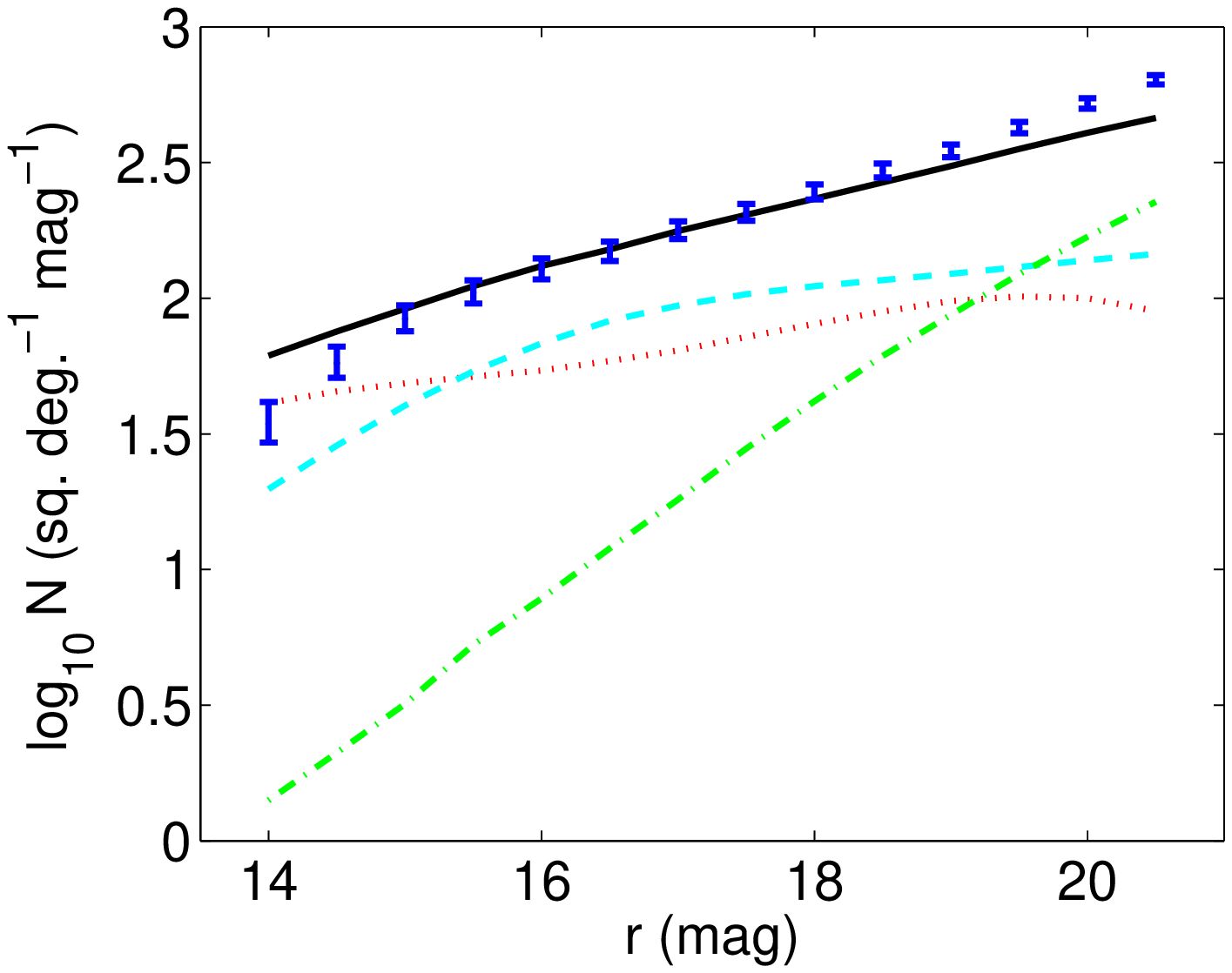}
  \includegraphics[width=0.5\textwidth,height=56mm]{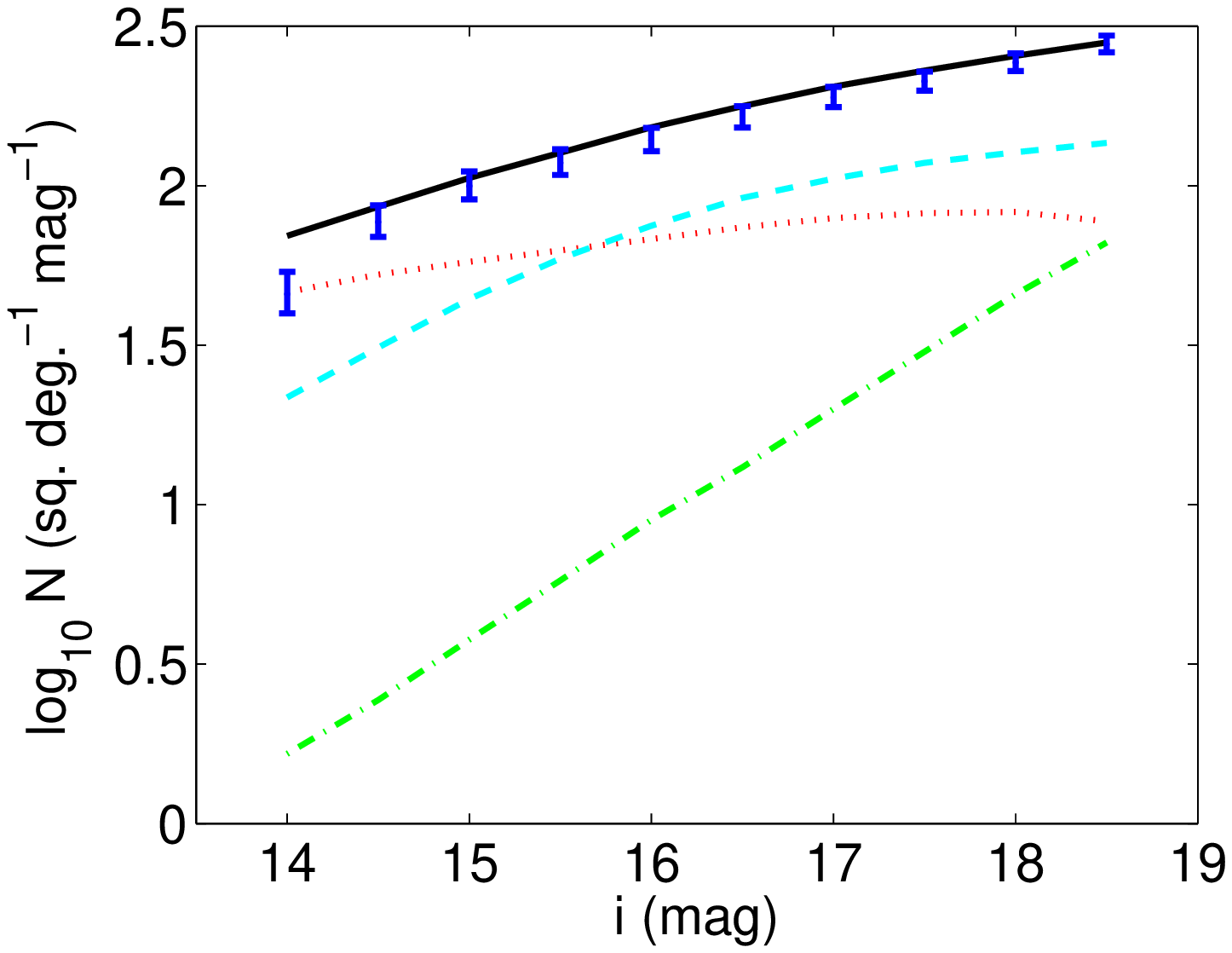}\\
  \includegraphics[width=0.5\textwidth,height=56mm]{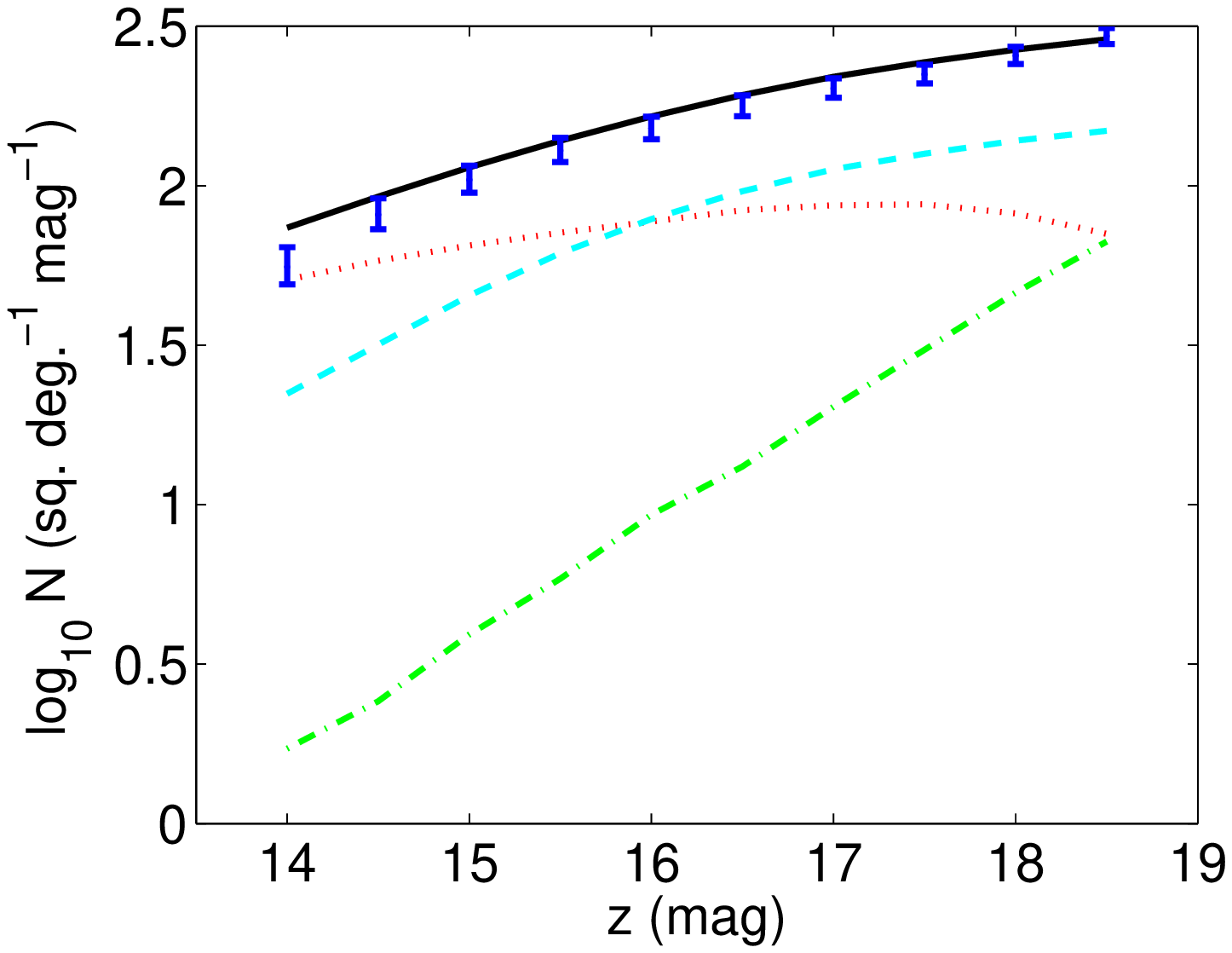}
  \caption{The distributions of apparent magnitude of the \besan model in the $u$, $g$,
$r$, $i$, and $z$ bands around the NGP. The thin disc, thick disc, and
halo are plotted as dotted lines (red), dashed lines (cyan), and
dot-dashed lines (green), respectively. The solid lines (black)
represent their sums.  The error bars (blue) show data from SDSS DR7
for comparison. The samples of the $u$ and $z$ band data are selected
in colour intervals of $-0.2\leq(u-g)<2.0$ and $-0.2\leq(i-z)<0.5$,
respectively. The other colour ranges extend from $-0.2$ to 1.2 (see
second column in Table \ref{tab:range}). (A colour version is available on-line.)}\label{fig:lf}
\end{figure*}

In Fig.\ \ref{fig:lf}, the luminosity functions of the SDSS data (with
error bars) and the simulated data are compared in all five filters.
The full (black) step-function shows all stars of the \besan model
with contributions from the thin disc (dotted red lines), thick disc
(dashed cyan lines), and halo (dot-dashed green lines). The model fits
the data well over a wide magnitude range in all filters. A few
systematic features can be seen:

\begin{description}
  \item[-] In the $u$, $g$, $i$, and $z$ bands, the predicted number
densities are slightly higher than in the actual data. At the brighter
magnitudes ($g<15$\,mag), these differences are significant in all
five filters. The thin disc dominates at the bright end where the
differences between the predictions and the data show the largest
deviations (0.2 dex to 0.5 dex). These stars correspond to F and early G
stars ($M_\mathrm{g}=4$ to 5 mag) of the MS of the thin disc with
distances above $\sim 500$ pc (see Sect.\ \ref{sec-hess}). The reason
for the discrepancy may be a higher mass density of the thin disc at
large $z$ or a higher fraction of early-type stars caused by the IMF
or the SFH.
   \item[-] Only in the $r$ band, the star counts are underestimated at
$r>19^{th}$ mag.
  \item[-] The thin disc dominates only at the bright end, but
contributes significantly up to the faint magnitude limit in the $r$,
$i$, and $z$ bands.
  \item[-] The contribution of the thick disc dominates over wide
magnitude ranges showing that the thick disc model parameters are very
important in attempting to reproduce the luminosity functions.
\item[-] The halo contributes significantly only at the faint end of the $u$, $g$, and $r$ band diagrams.
\end{description}

\subsection{Hess diagrams}\label{sec-hess}

We calculate the Hess diagrams of the SDSS data and the \besan
model in the same way. The star counts are binned in each CMD in intervals of 0.05
mag in colour and 0.5 mag in magnitude and smoothed in steps of 0.01
mag. Numerical counts are given on a logarithmic scale covering a range
of 1 to 100 per deg$^2$, 0.1 mag in colour, and 1 mag in luminosity.
The relative differences between the model and the data (i.e.,
$(\mathrm{data}-\mathrm{model})/\mathrm{model}$) are smoothed only in
luminosity and cover a linear range of $-1.0$ to $+1.0$. We
investigate star counts in the four CMDs ($u-g, g$), ($g-r, g$),
($r-i, r$), and ($i-z, i$). To show and compare each stellar
population in all the CMDs, the colour and magnitude ranges are different
in the different CMDs (see columns 2 and 3 of Table \ref{tab:range}).

\begin{figure*} 
\includegraphics[width=0.33\textwidth]{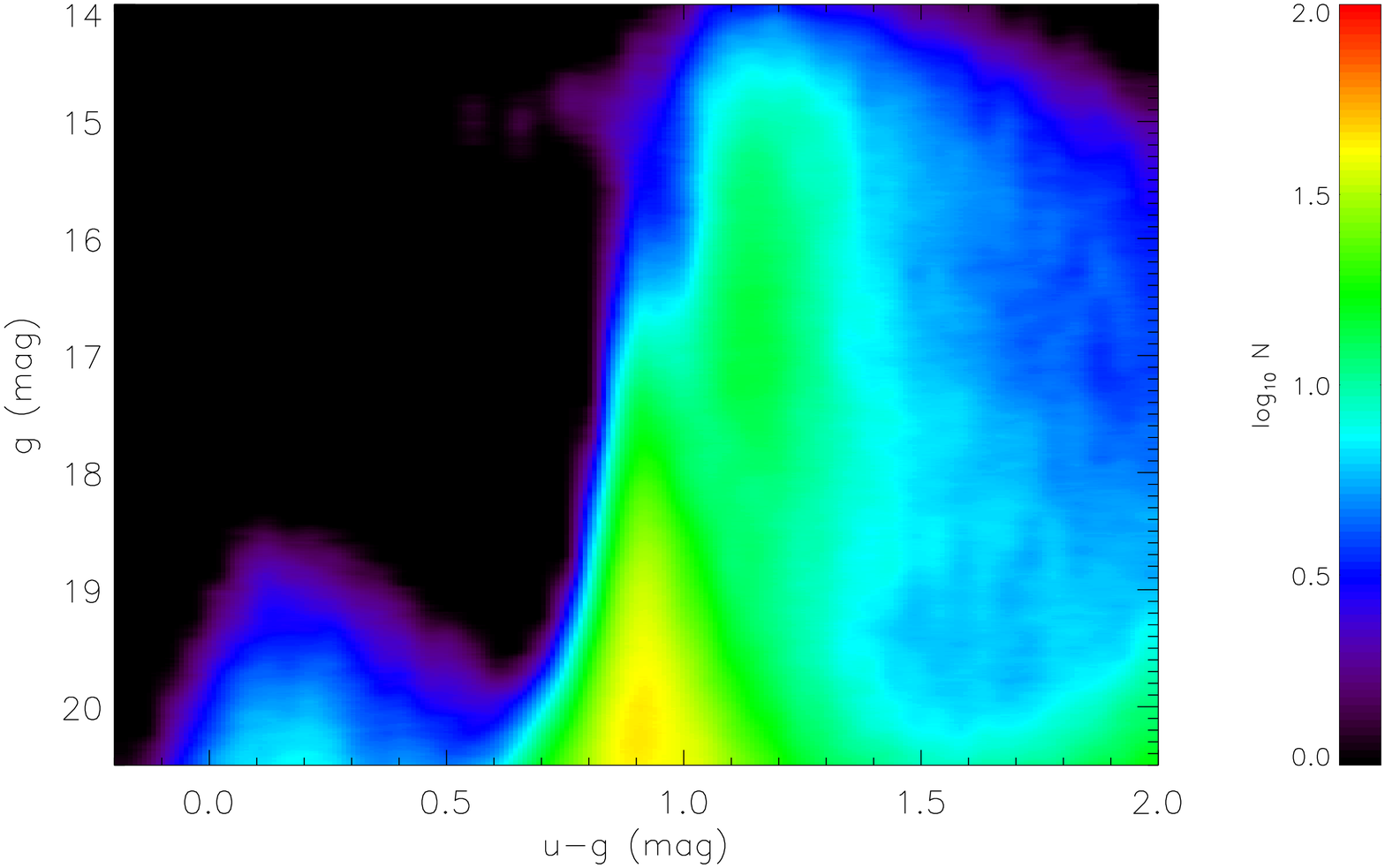}       
\includegraphics[width=0.33\textwidth]{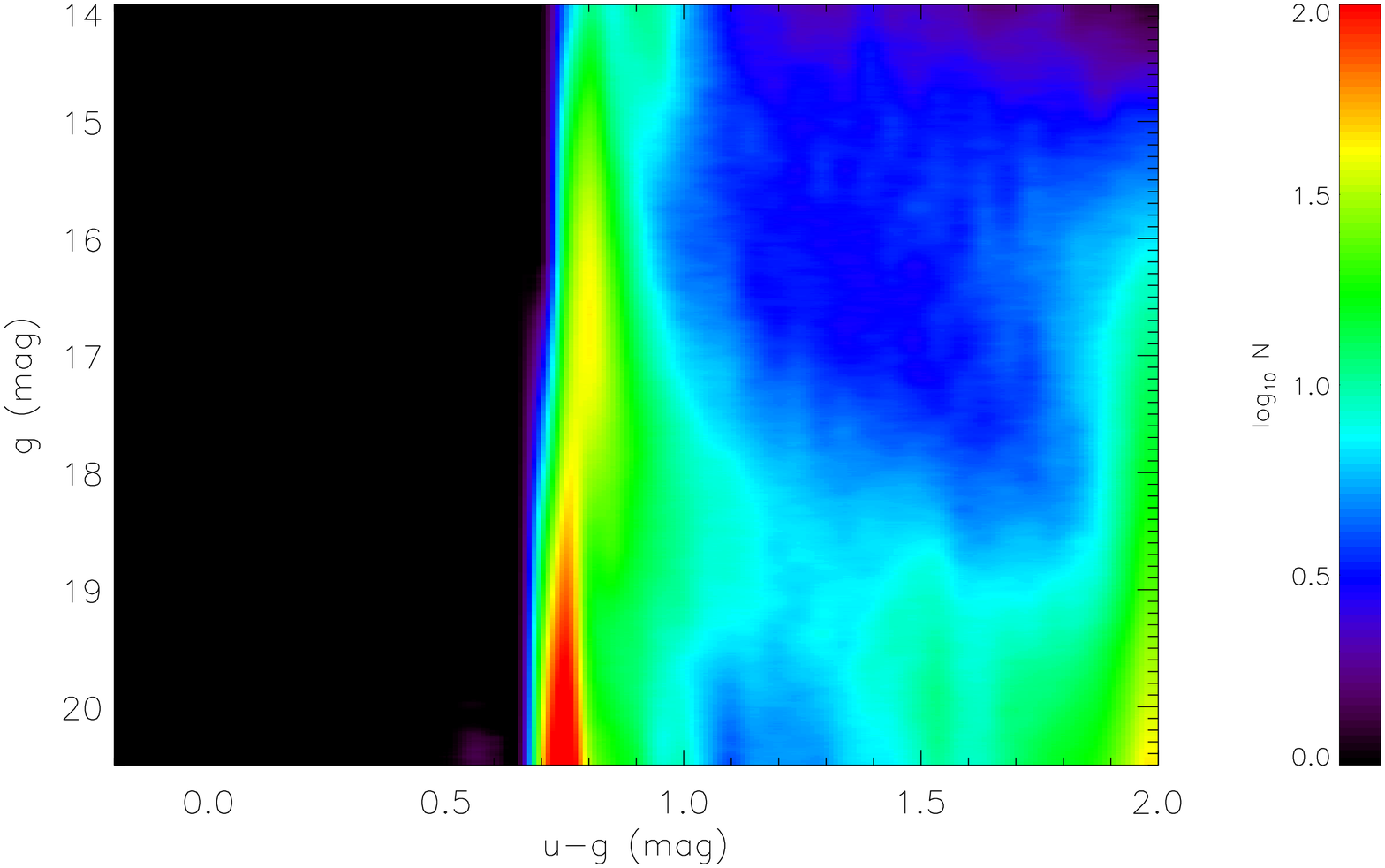}   
\includegraphics[width=0.33\textwidth]{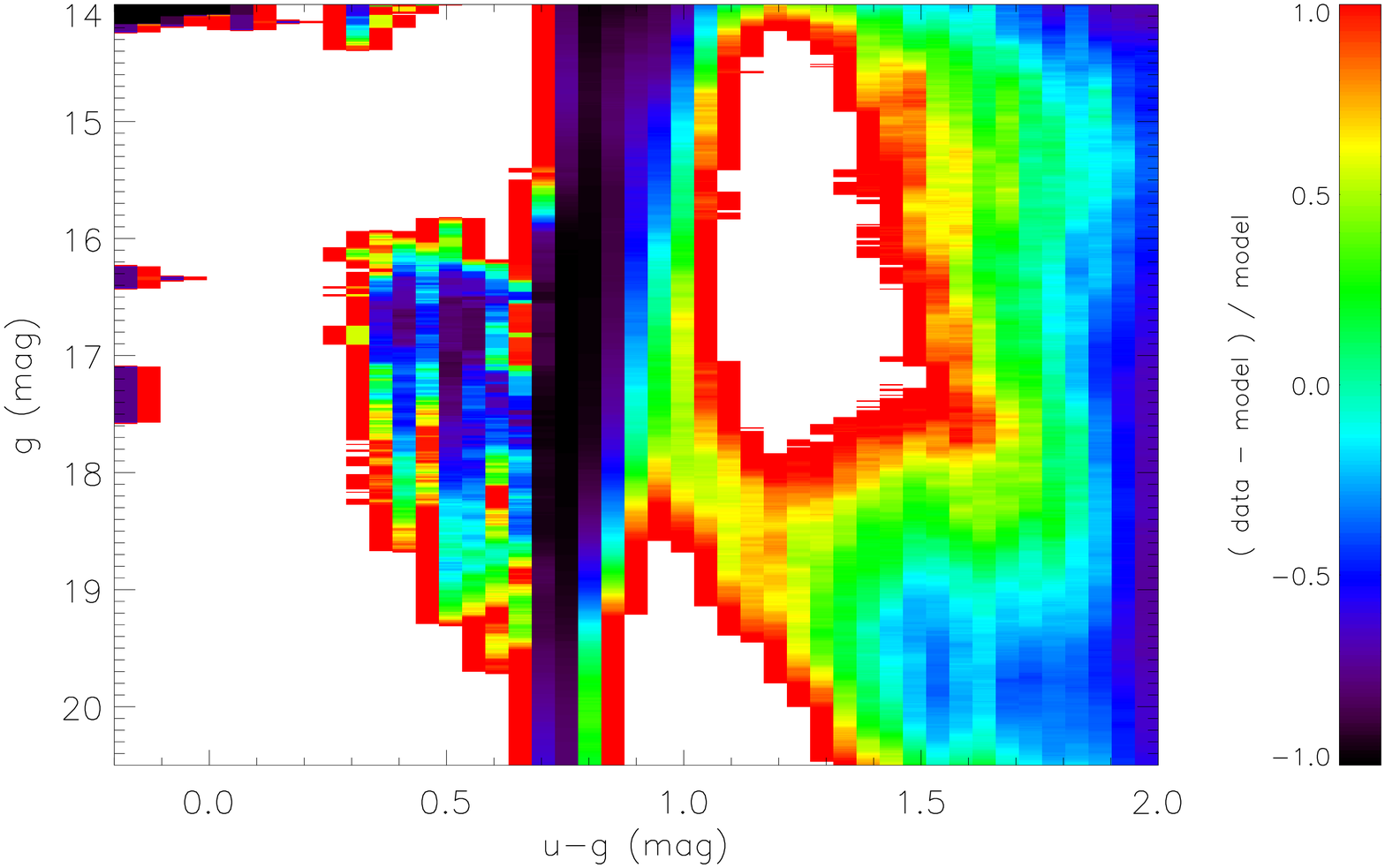}\\   
\includegraphics[width=0.33\textwidth]{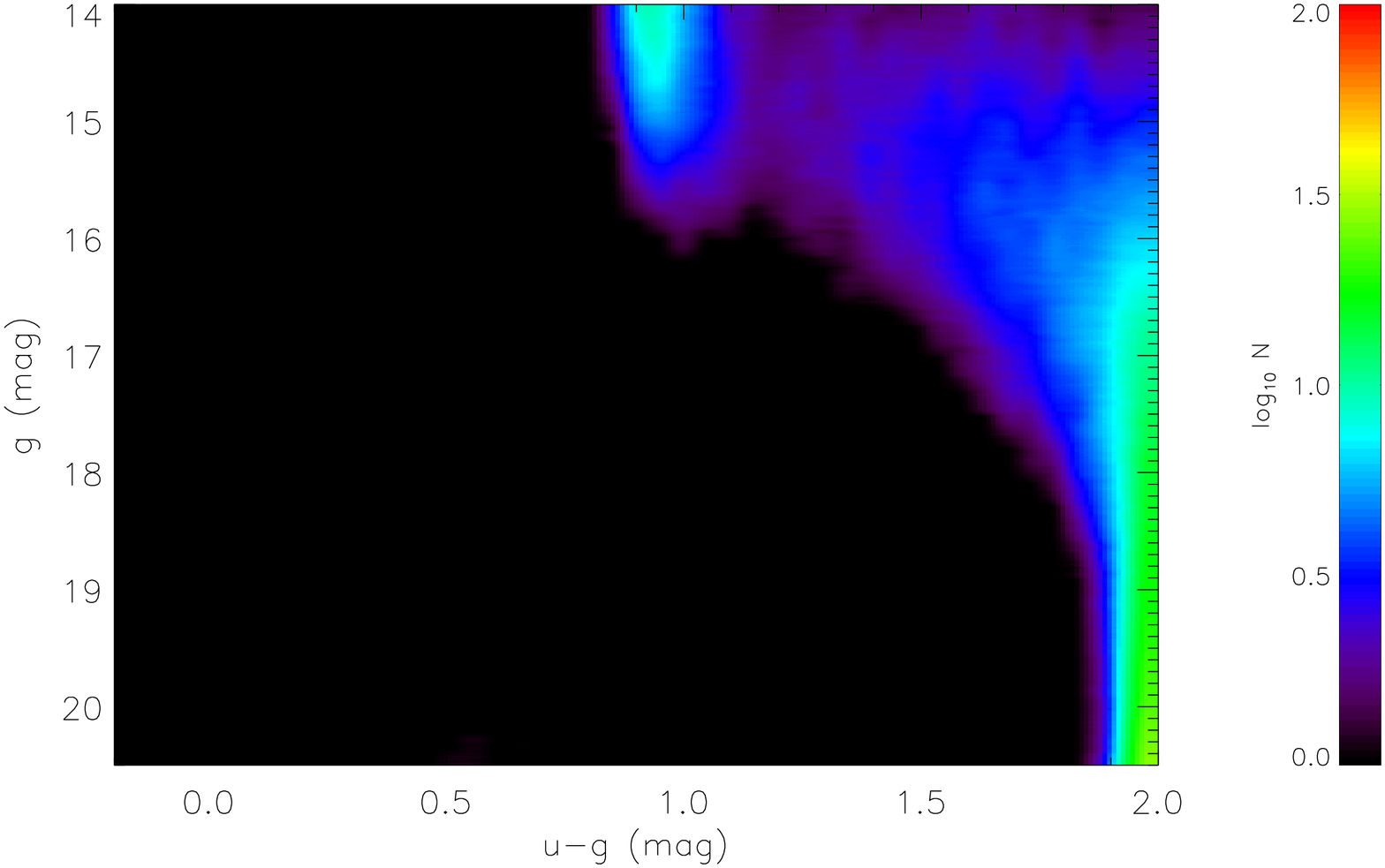}   
\includegraphics[width=0.33\textwidth]{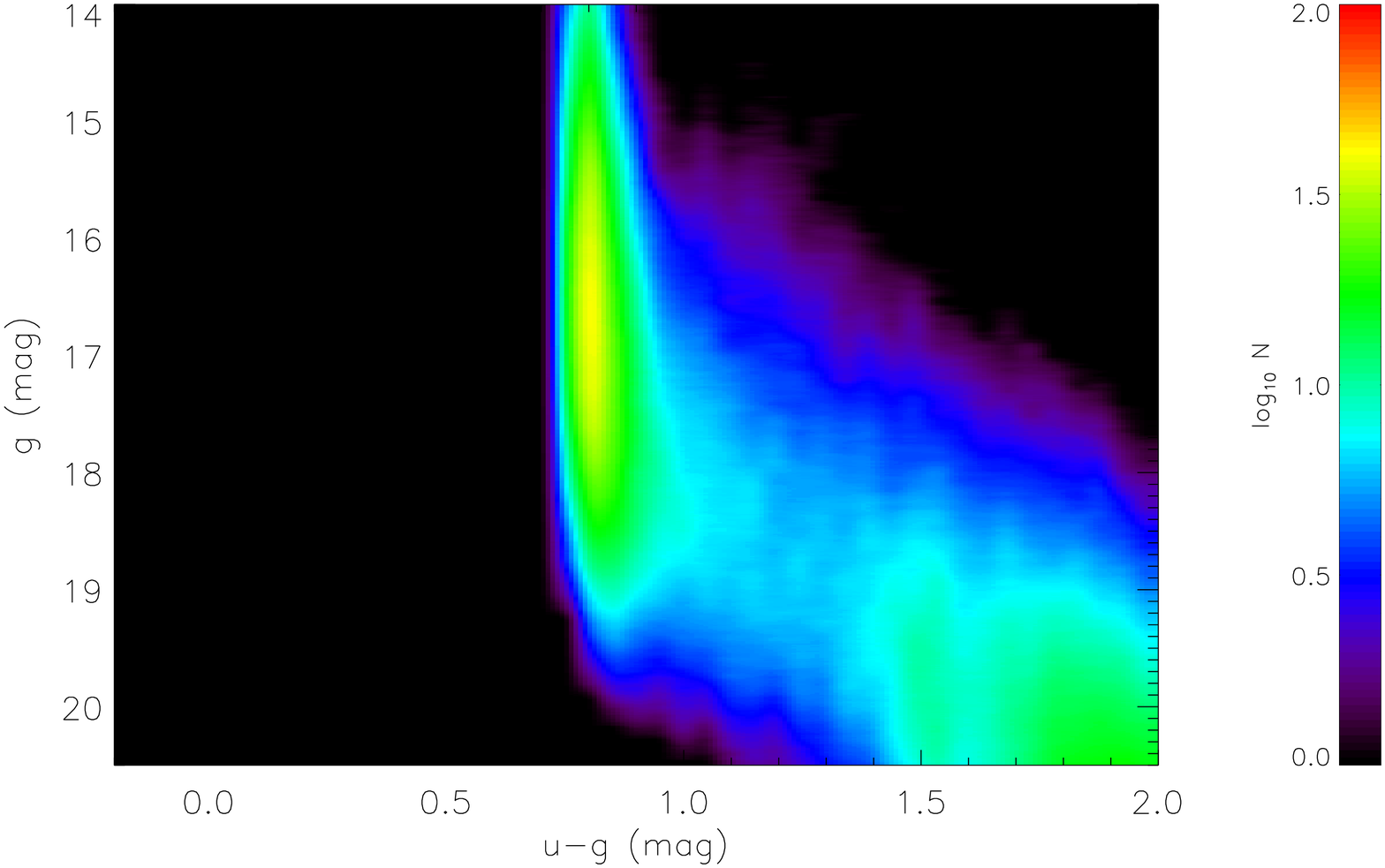}   
\includegraphics[width=0.33\textwidth]{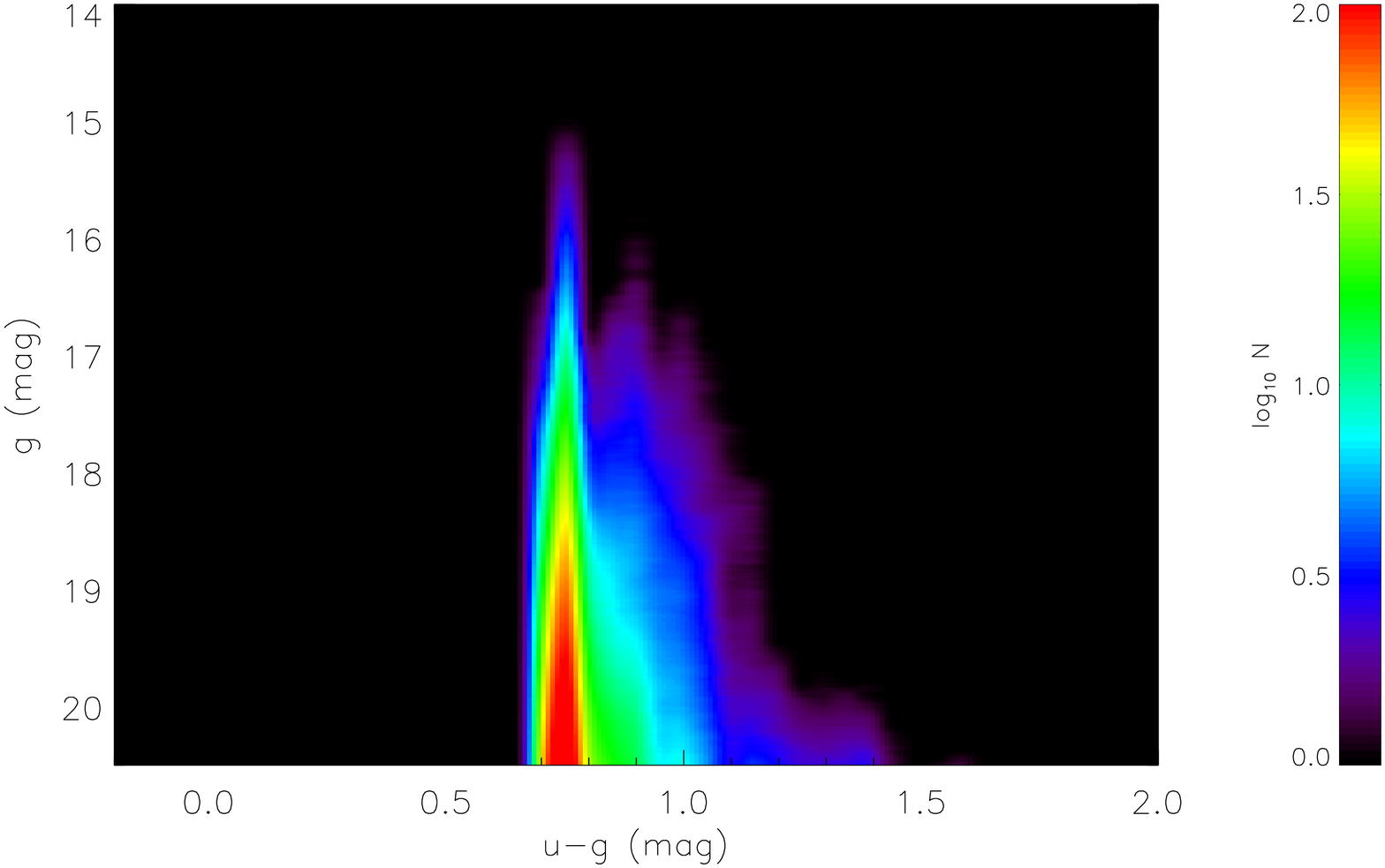}\\ 
\caption{Hess diagrams of the SDSS data and the \besan model in $(u-g,g)$. The upper three plots show the Hess diagram of the SDSS data, the model, and the relative difference between the data and model (from left to right). The lower panels show the Hess diagrams of the three simulated components of the \besan model: thin disc, thick disc, and stellar halo (from left to right). All Hess diagrams cover a range of stellar densities from 1 to 100 per square degree per magnitude in luminosity and 0.1 magnitude in colour (purple to red in the electronic version; black means yet lower number densities). The difference plot (top right) covers a range of relative differences from $-100\%$ to $+100\%$ (purple to red in the electronic version; black and white mean larger negative or positive differences, respectively). The ranges in apparent magnitude and colour of these six plots are listed in columns 2 and 3 of Table \ref{tab:range}).      (A colour version is available on-line.)}\label{fig:ug}
\end{figure*}

\begin{figure*} 
\includegraphics[width=0.33\textwidth]{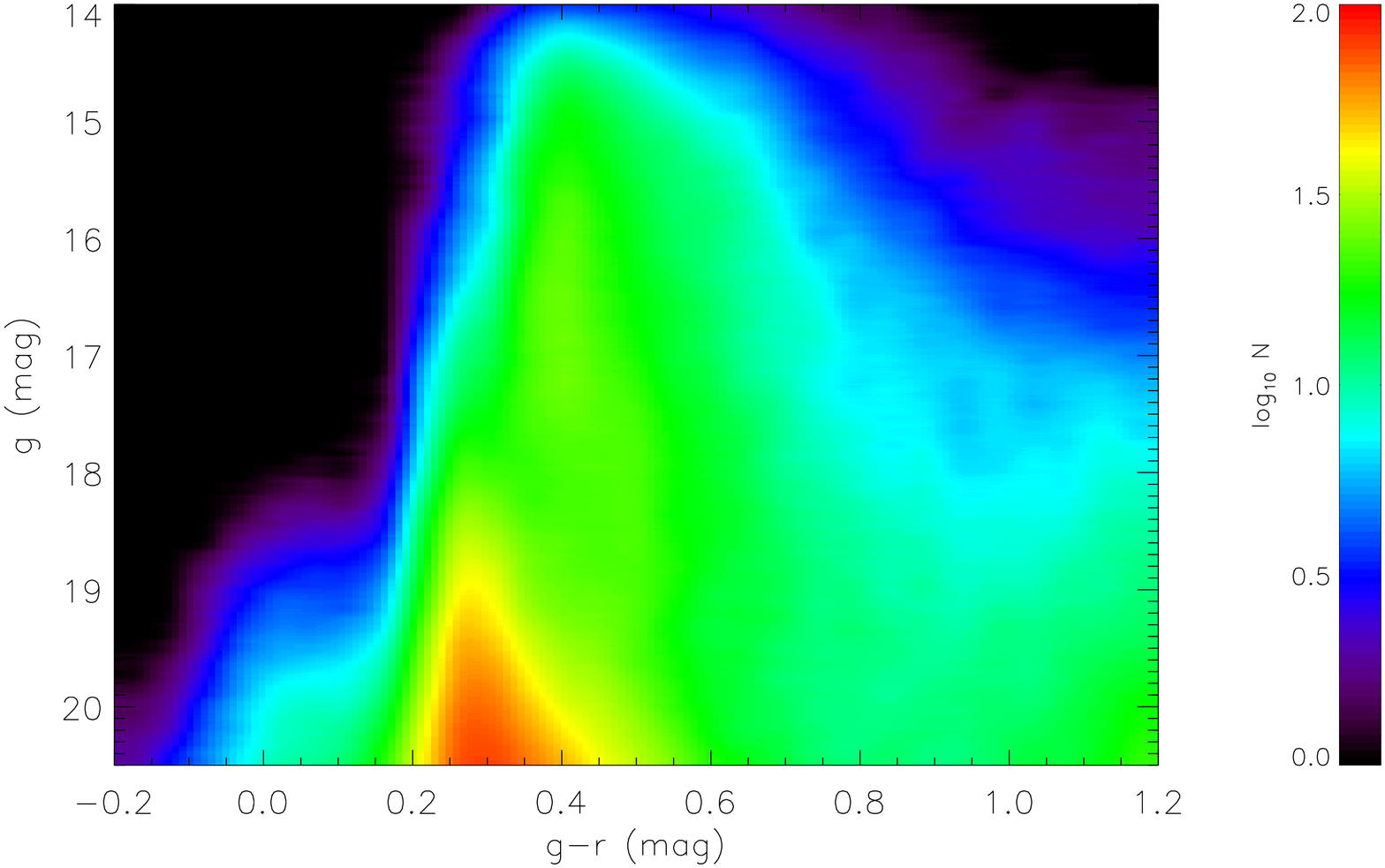}   
\includegraphics[width=0.33\textwidth]{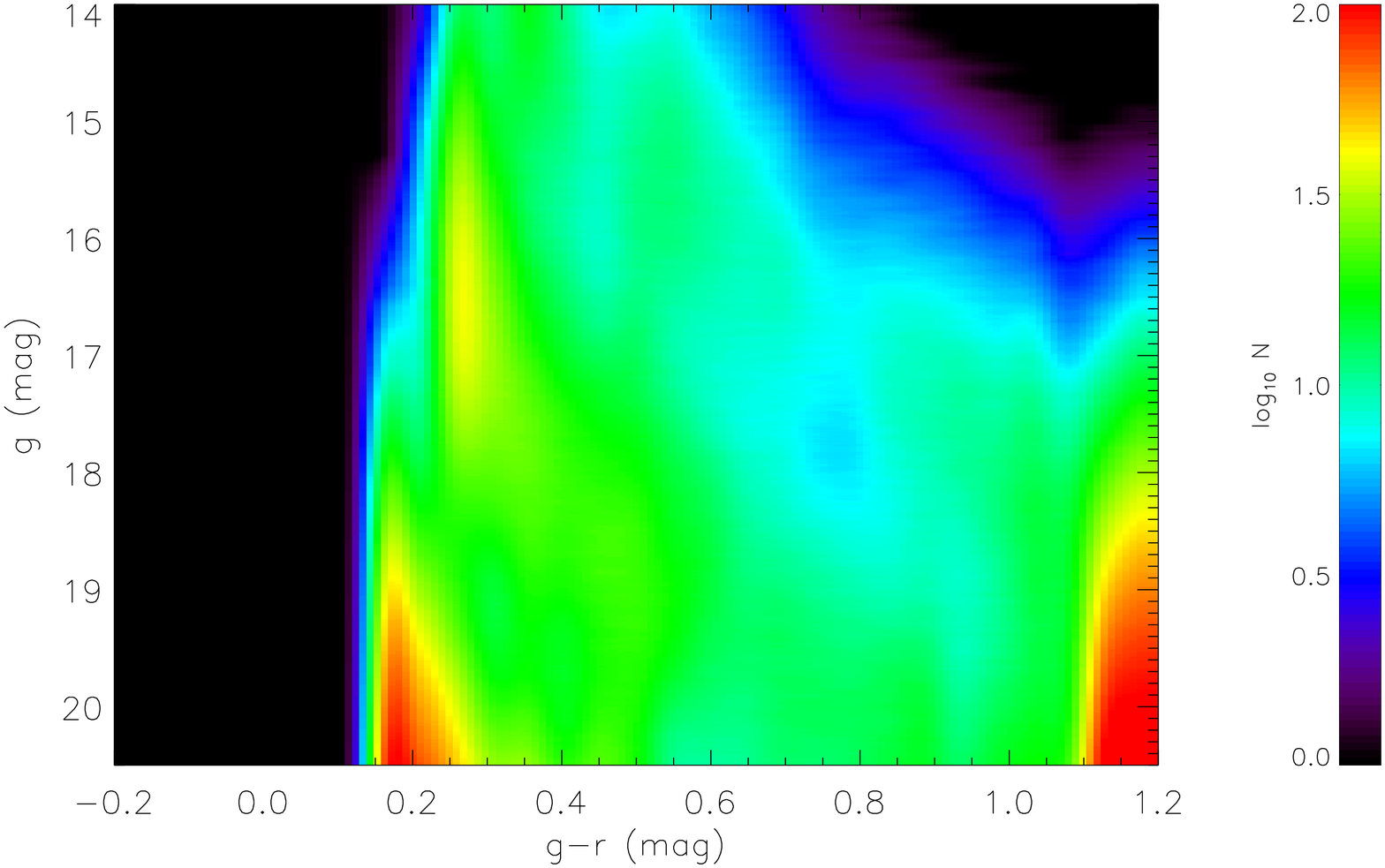}   
\includegraphics[width=0.33\textwidth]{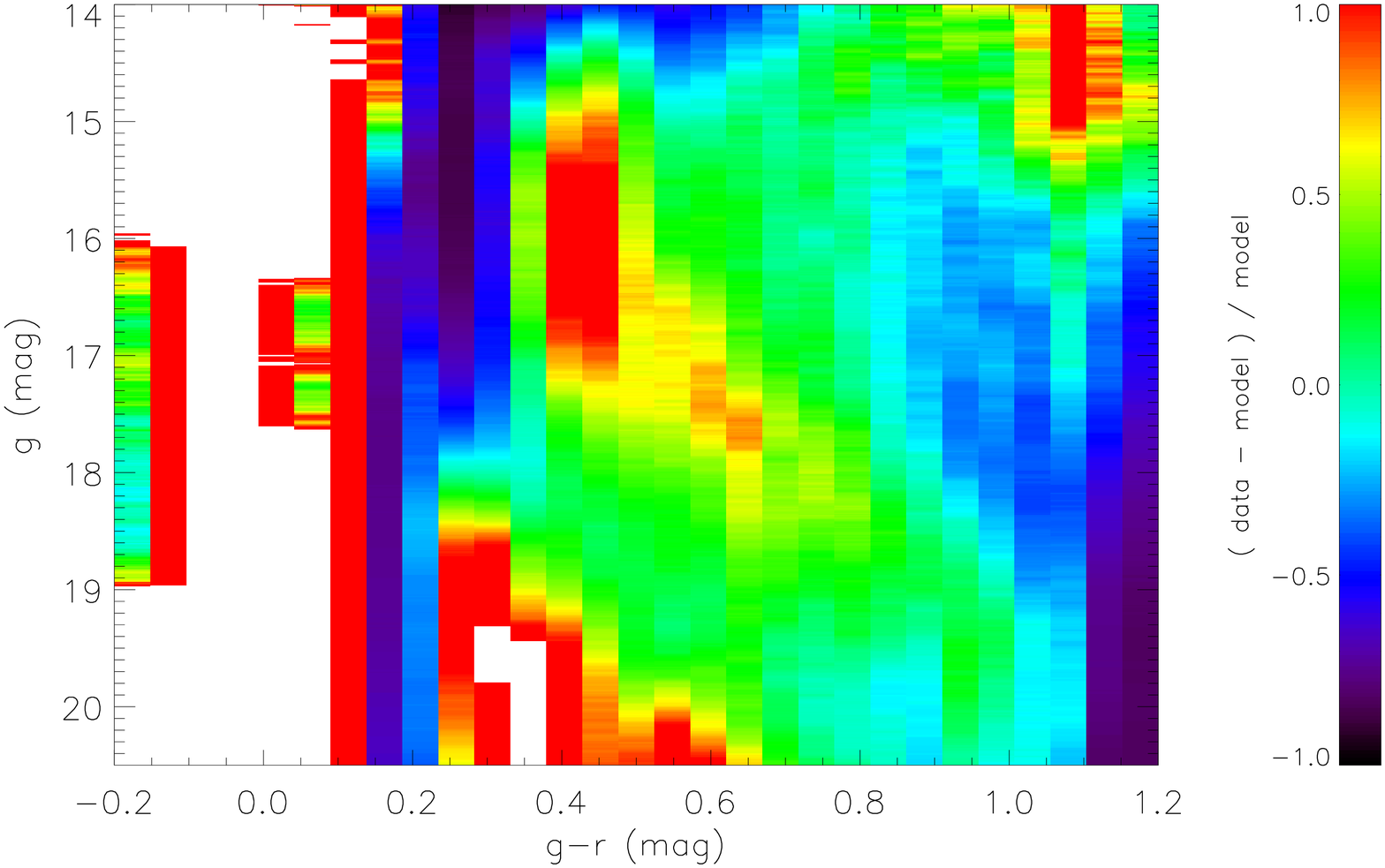}\\   
\includegraphics[width=0.33\textwidth]{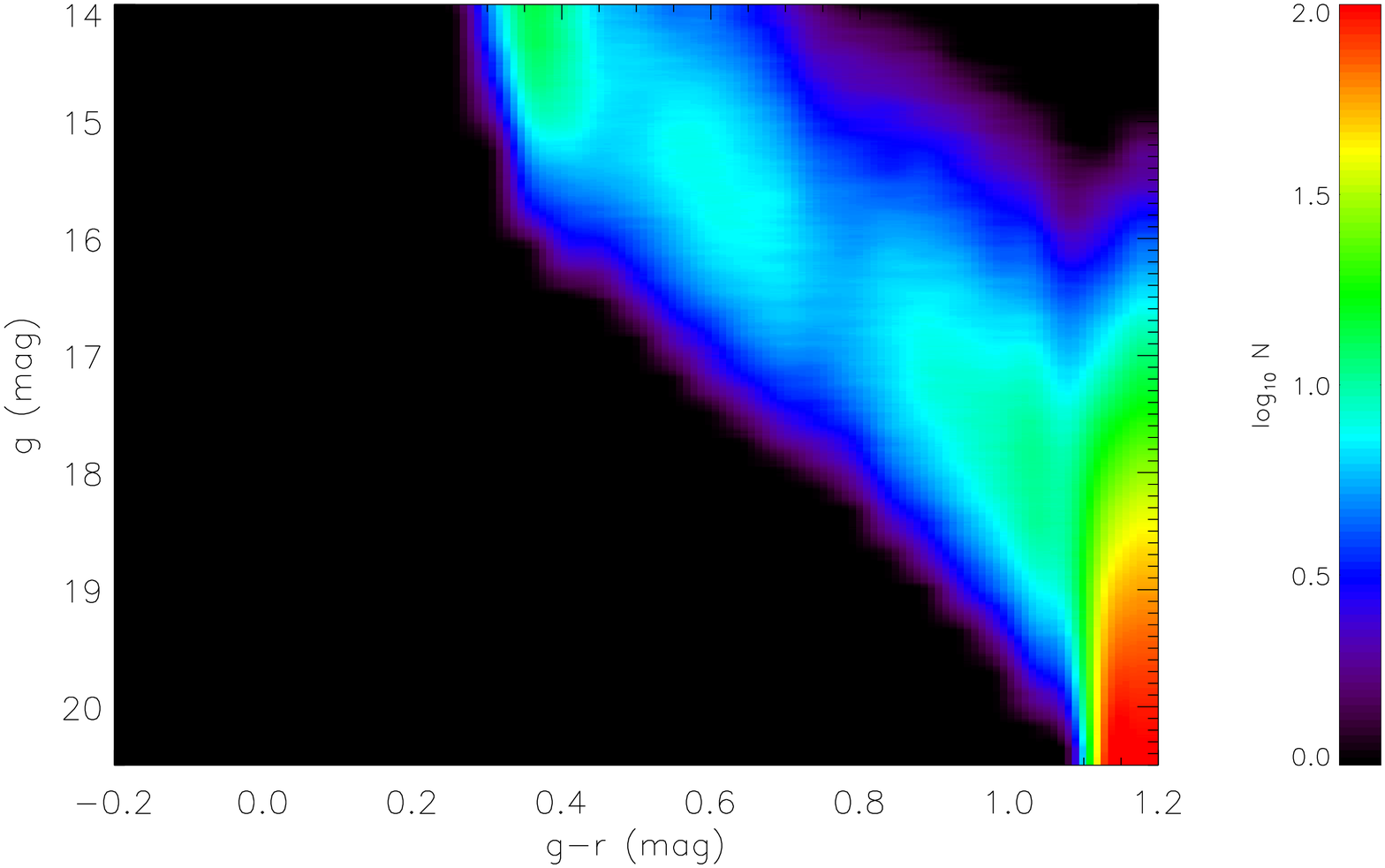}   
\includegraphics[width=0.33\textwidth]{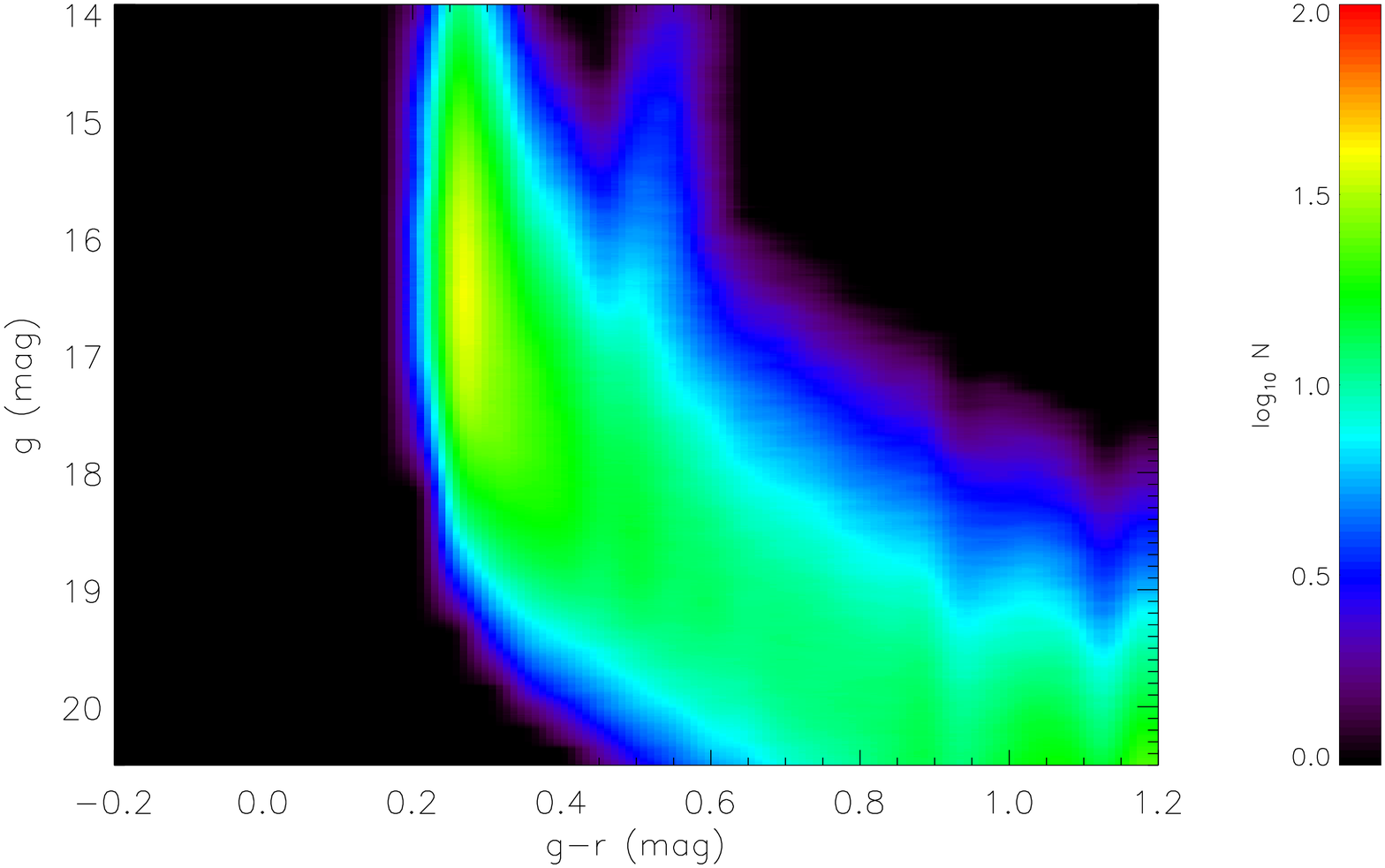}   
\includegraphics[width=0.33\textwidth]{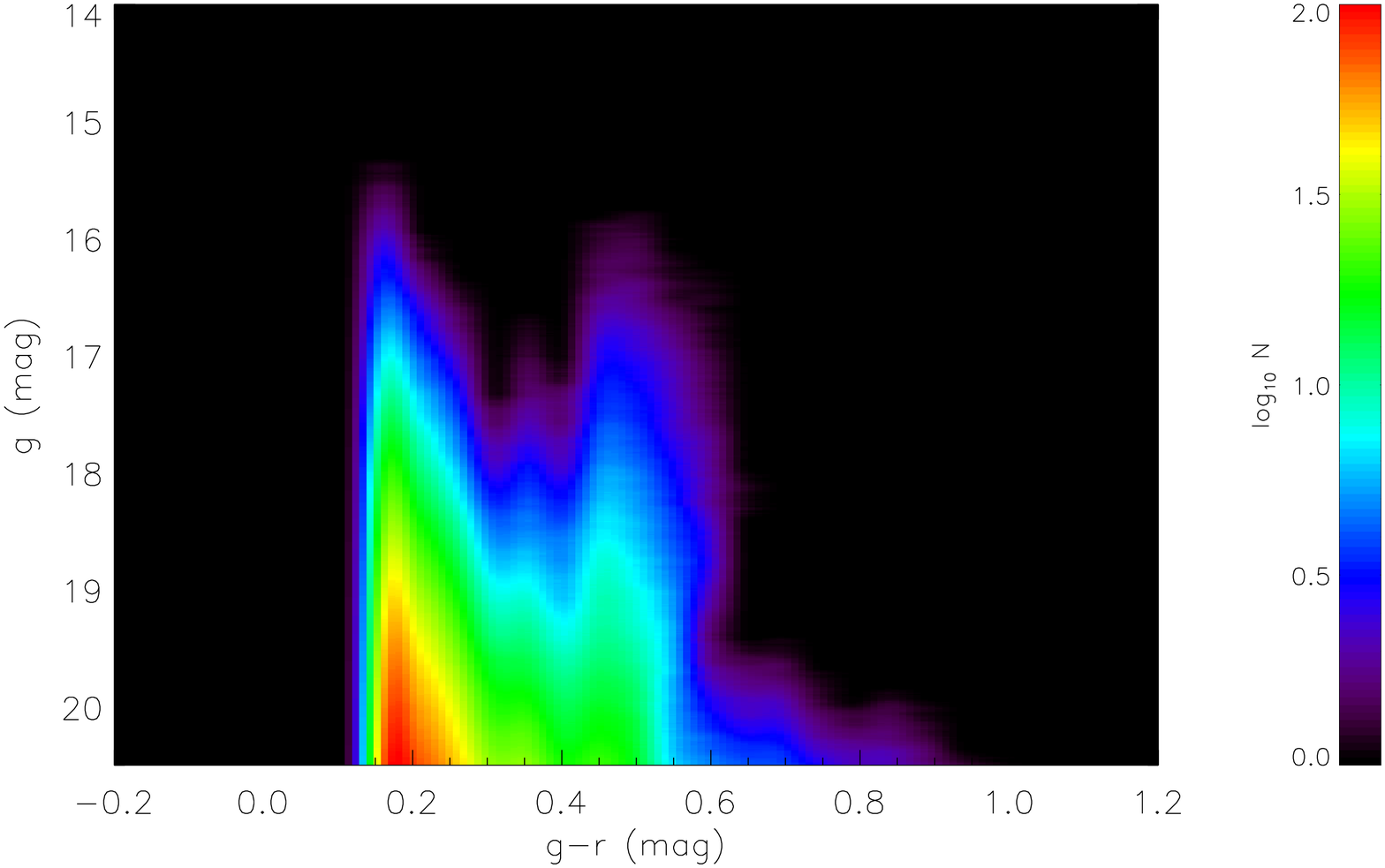}\\ 
  \caption{Same as Fig.\ \ref{fig:ug} but in $(g-r, g)$.}
  \label{fig:gr}
\end{figure*}

\begin{figure*} 
\includegraphics[width=0.33\textwidth]{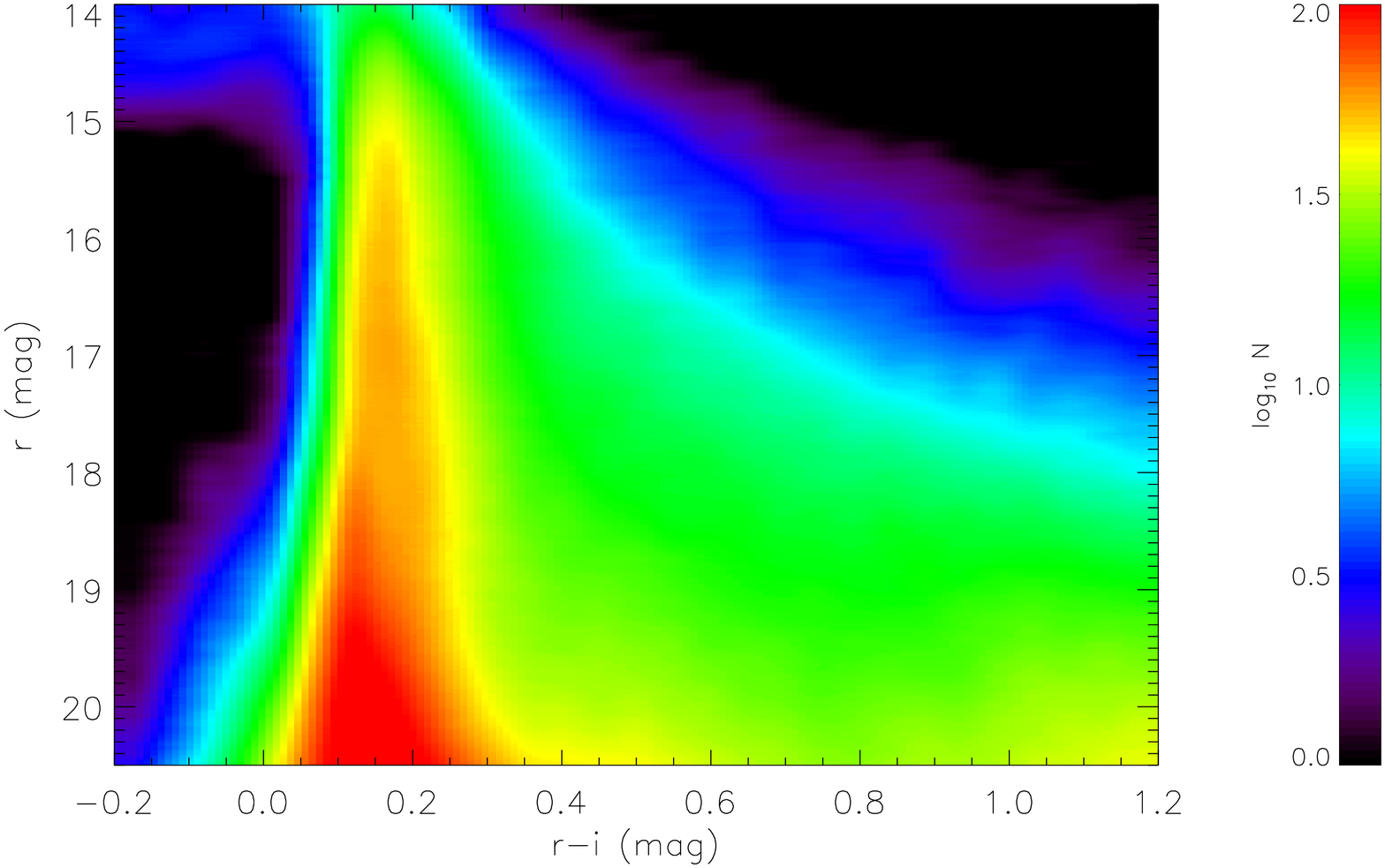}   
\includegraphics[width=0.33\textwidth]{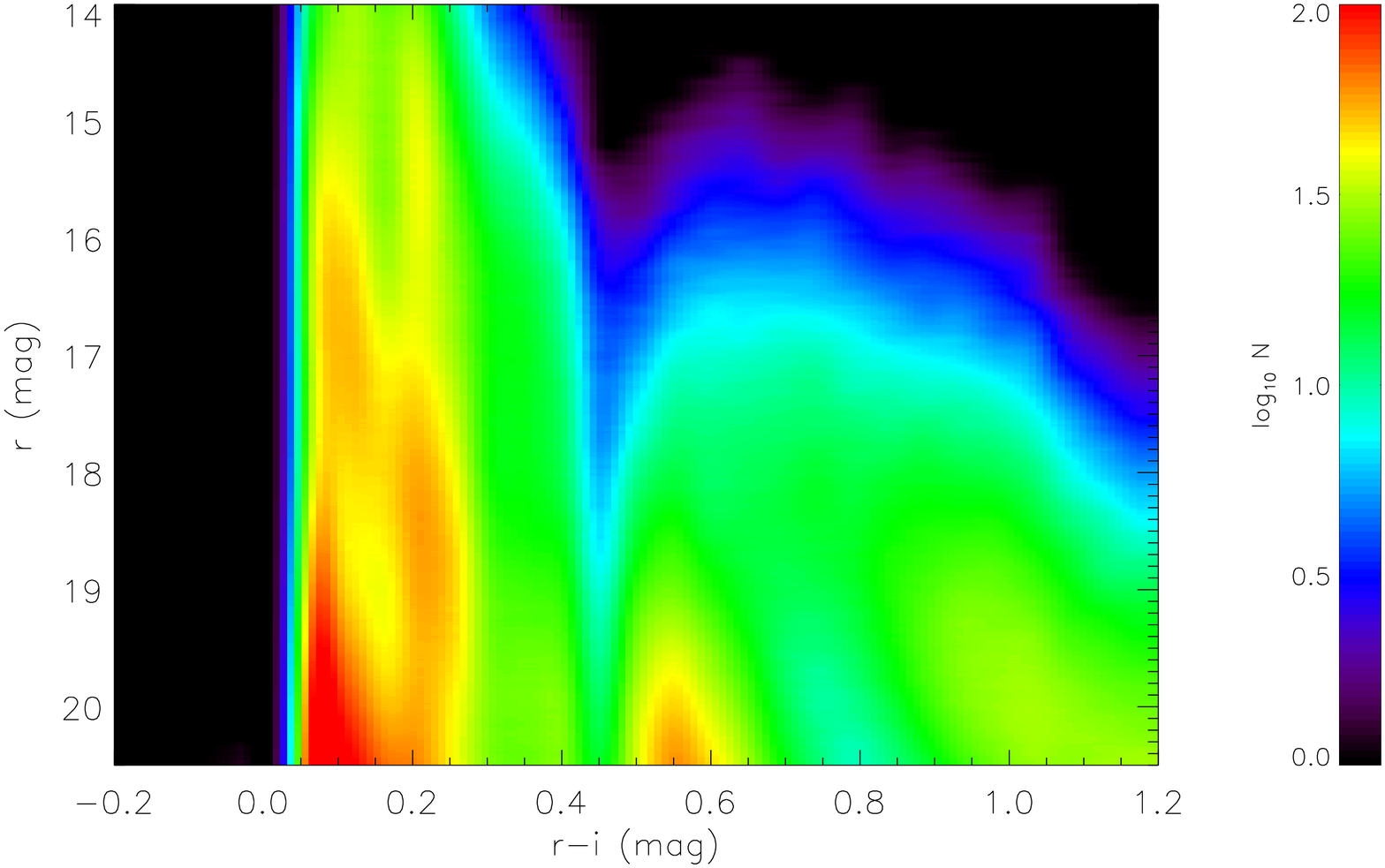}   
\includegraphics[width=0.33\textwidth]{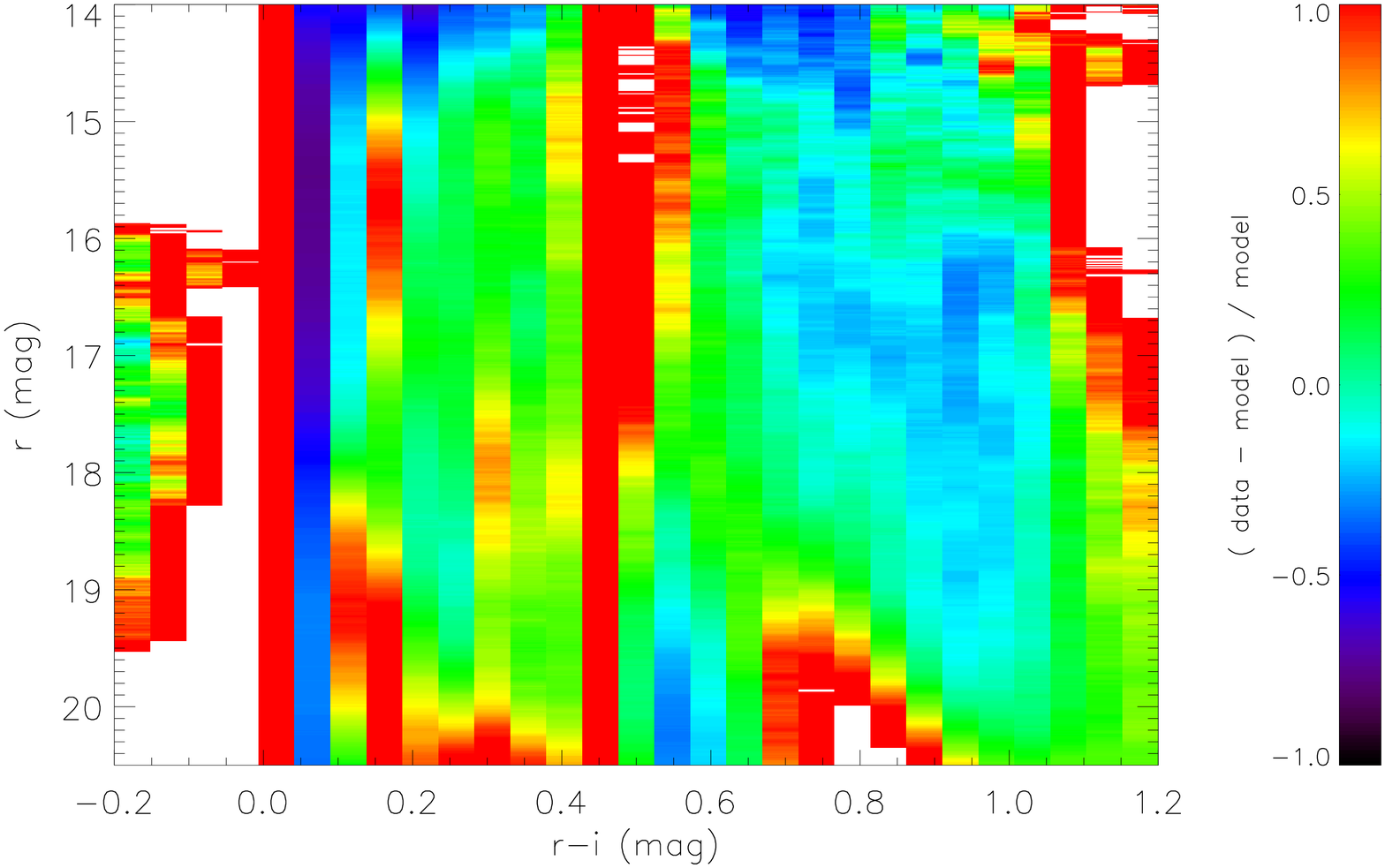}\\   
\includegraphics[width=0.33\textwidth]{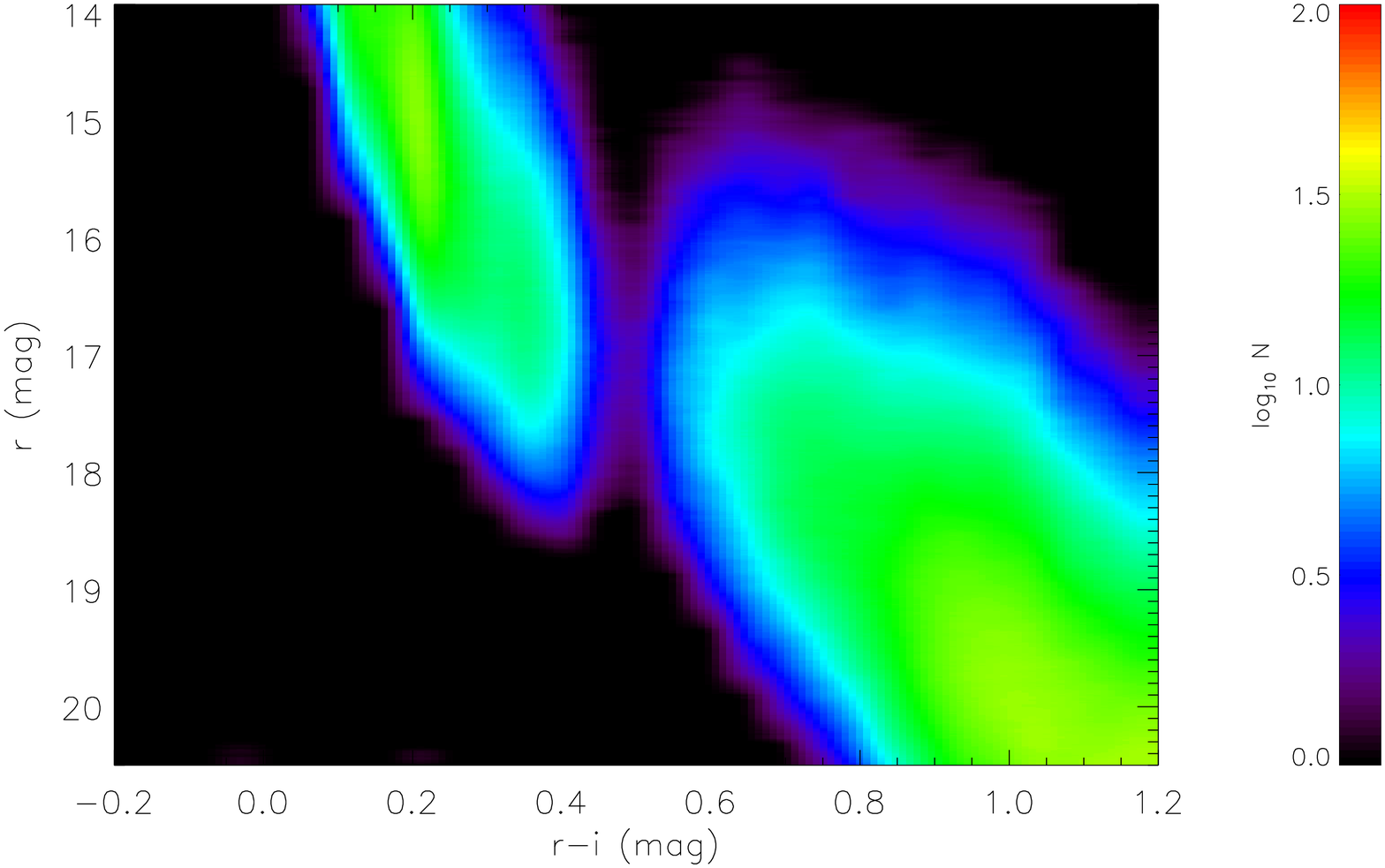}   
\includegraphics[width=0.33\textwidth]{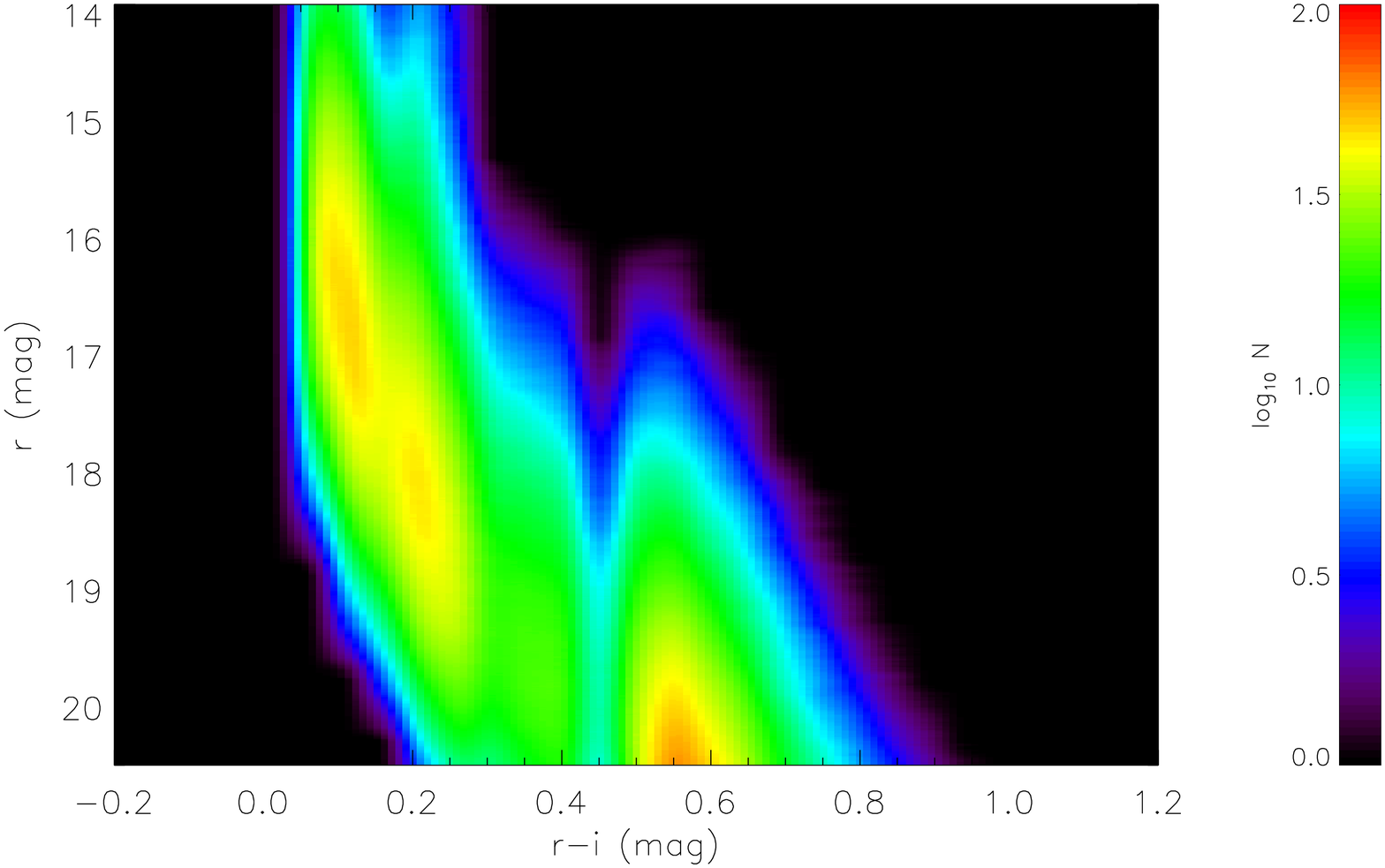}   
\includegraphics[width=0.33\textwidth]{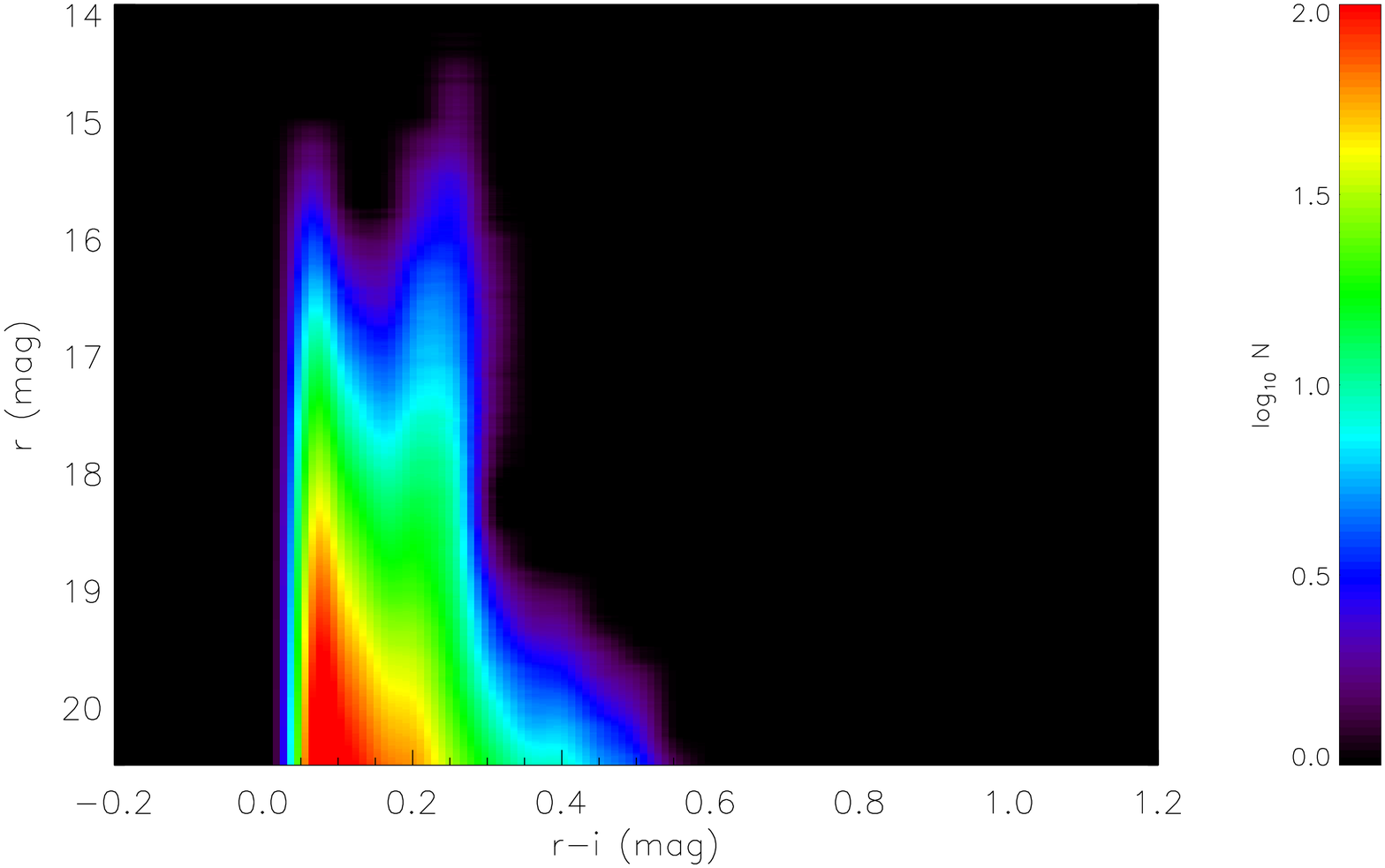}\\ 
  \caption{Same as Fig.\ \ref{fig:ug} but in $(r-i, r)$. }
  \label{fig:ri}
\end{figure*}

\begin{figure*} 
\includegraphics[width=0.33\textwidth]{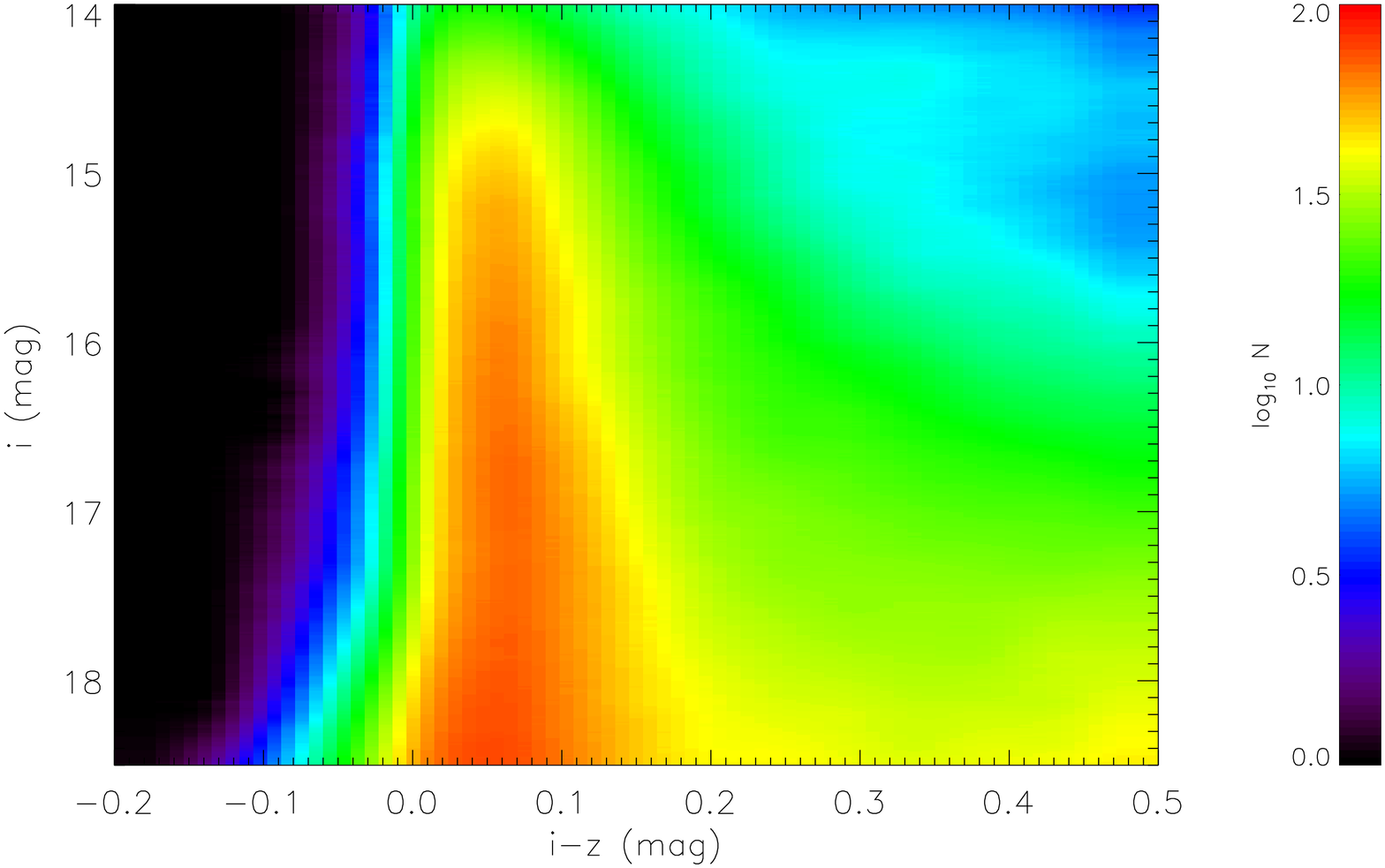}
\includegraphics[width=0.33\textwidth]{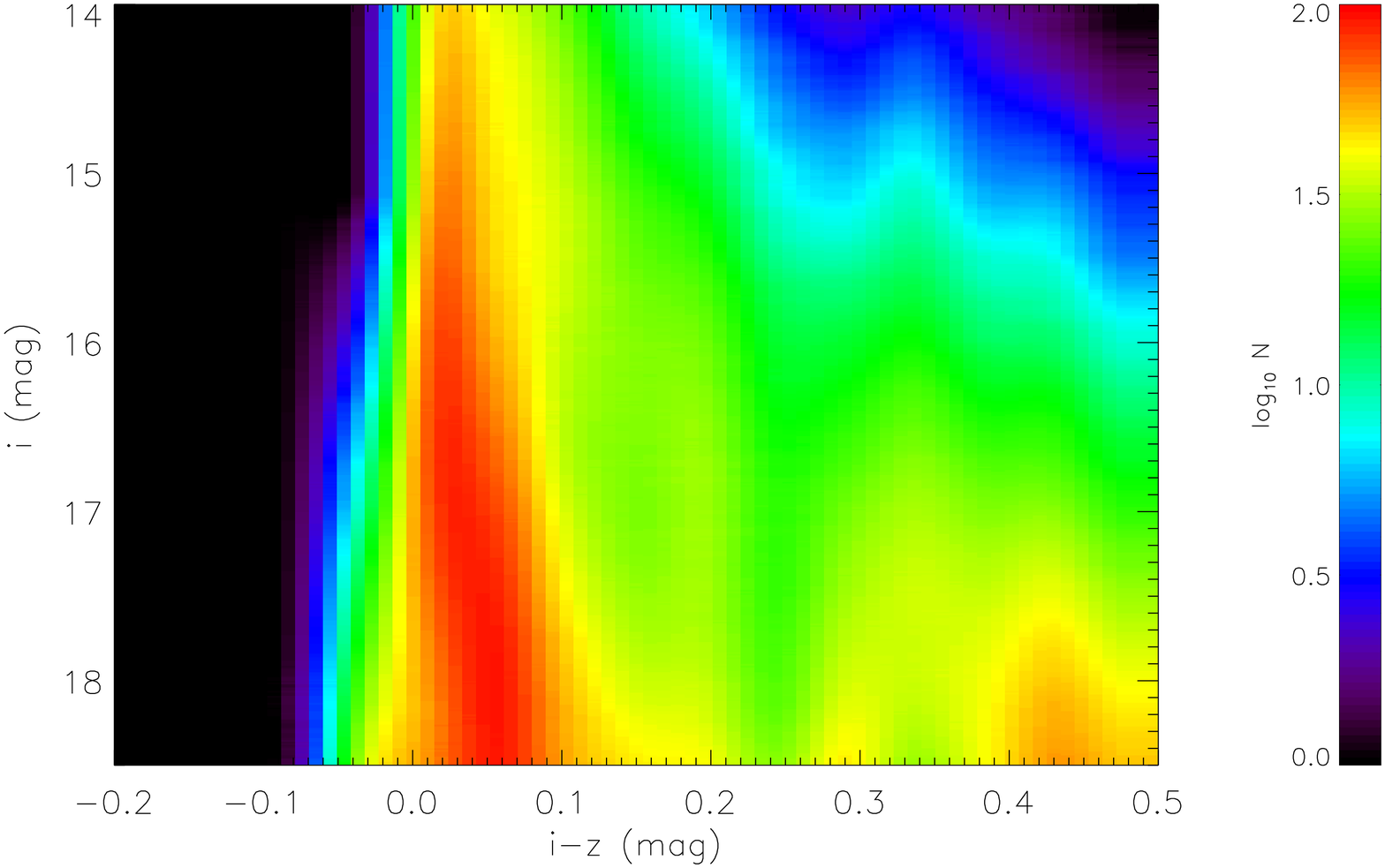}   
\includegraphics[width=0.33\textwidth]{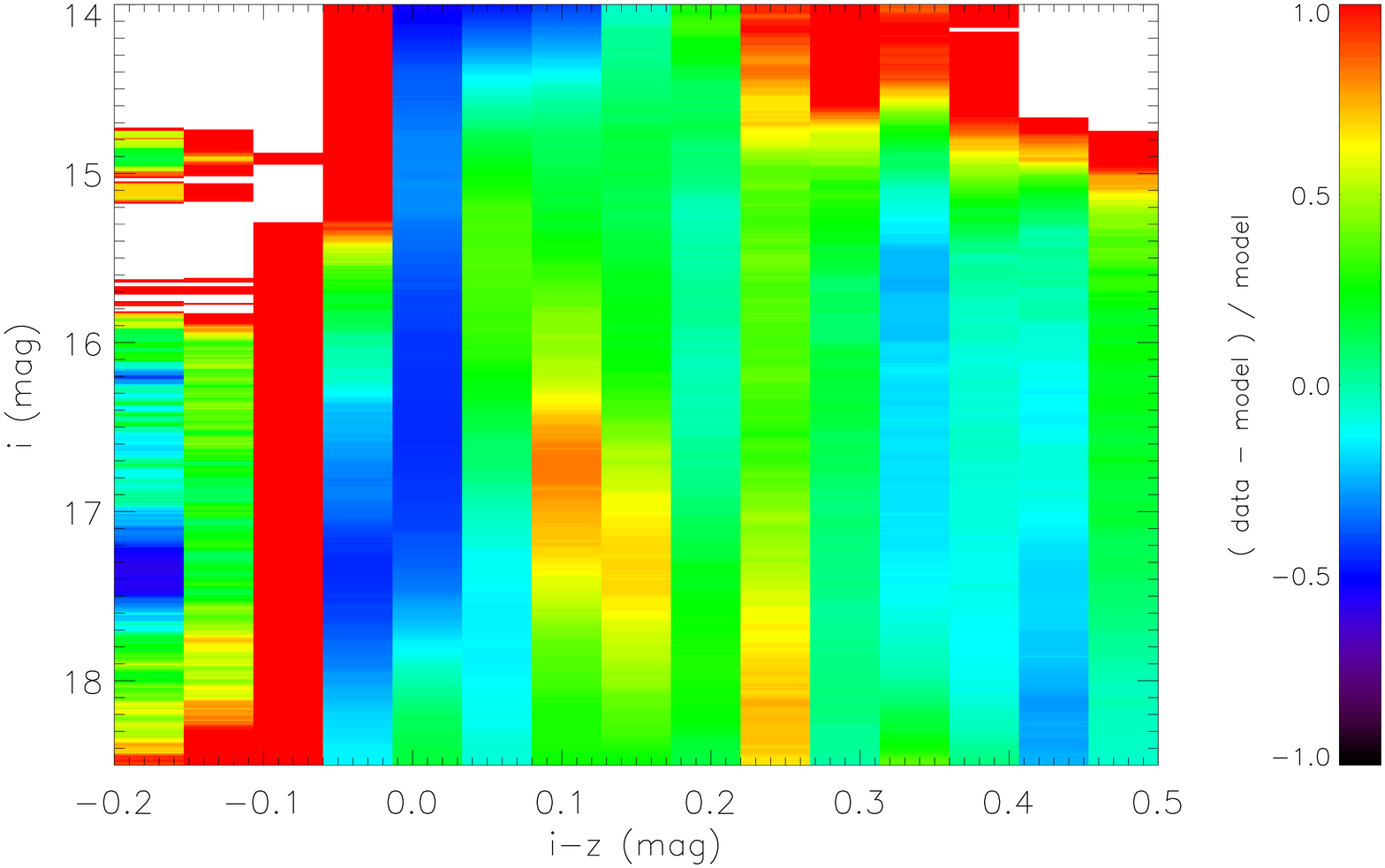}\\   
\includegraphics[width=0.33\textwidth]{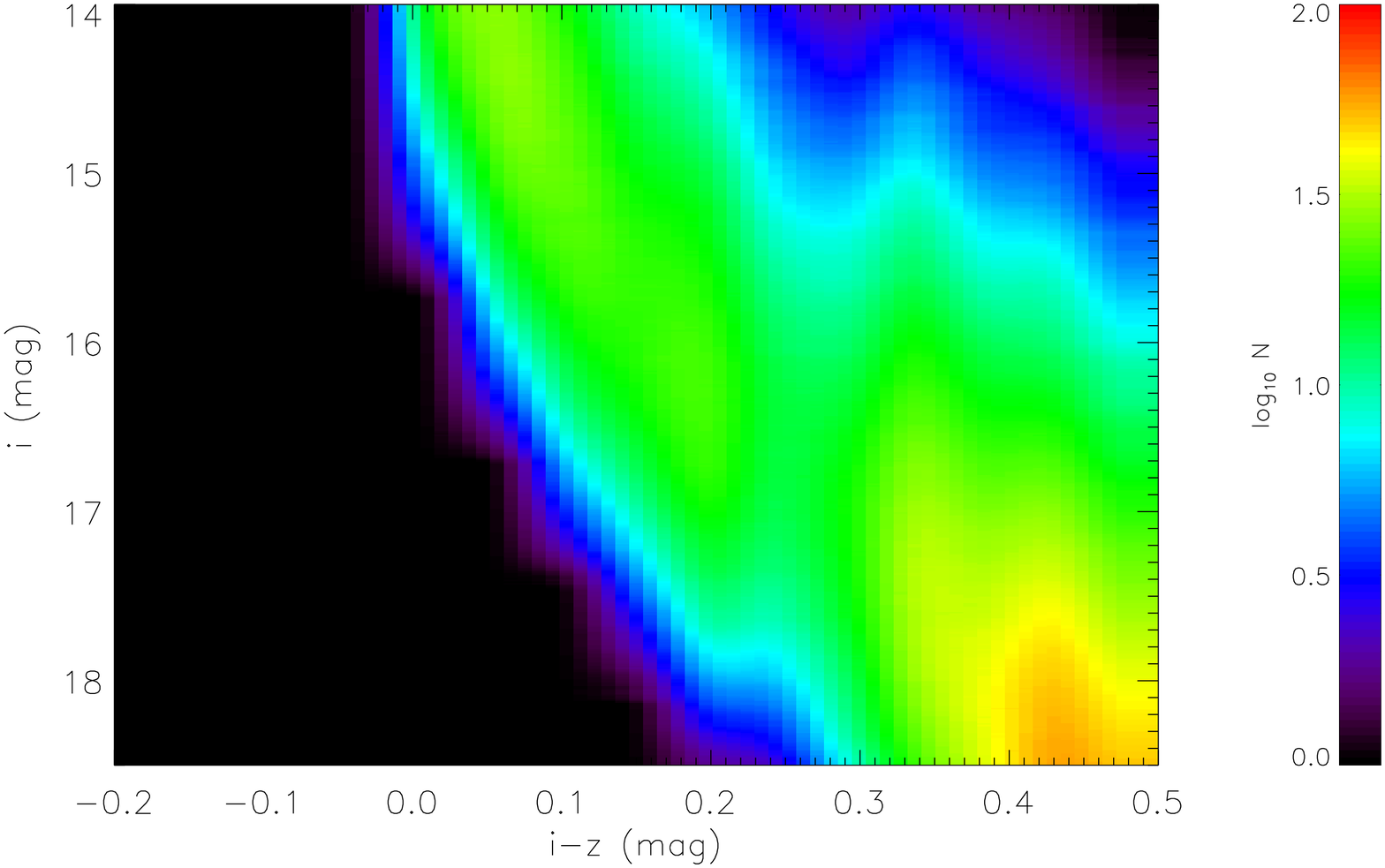}   
\includegraphics[width=0.33\textwidth]{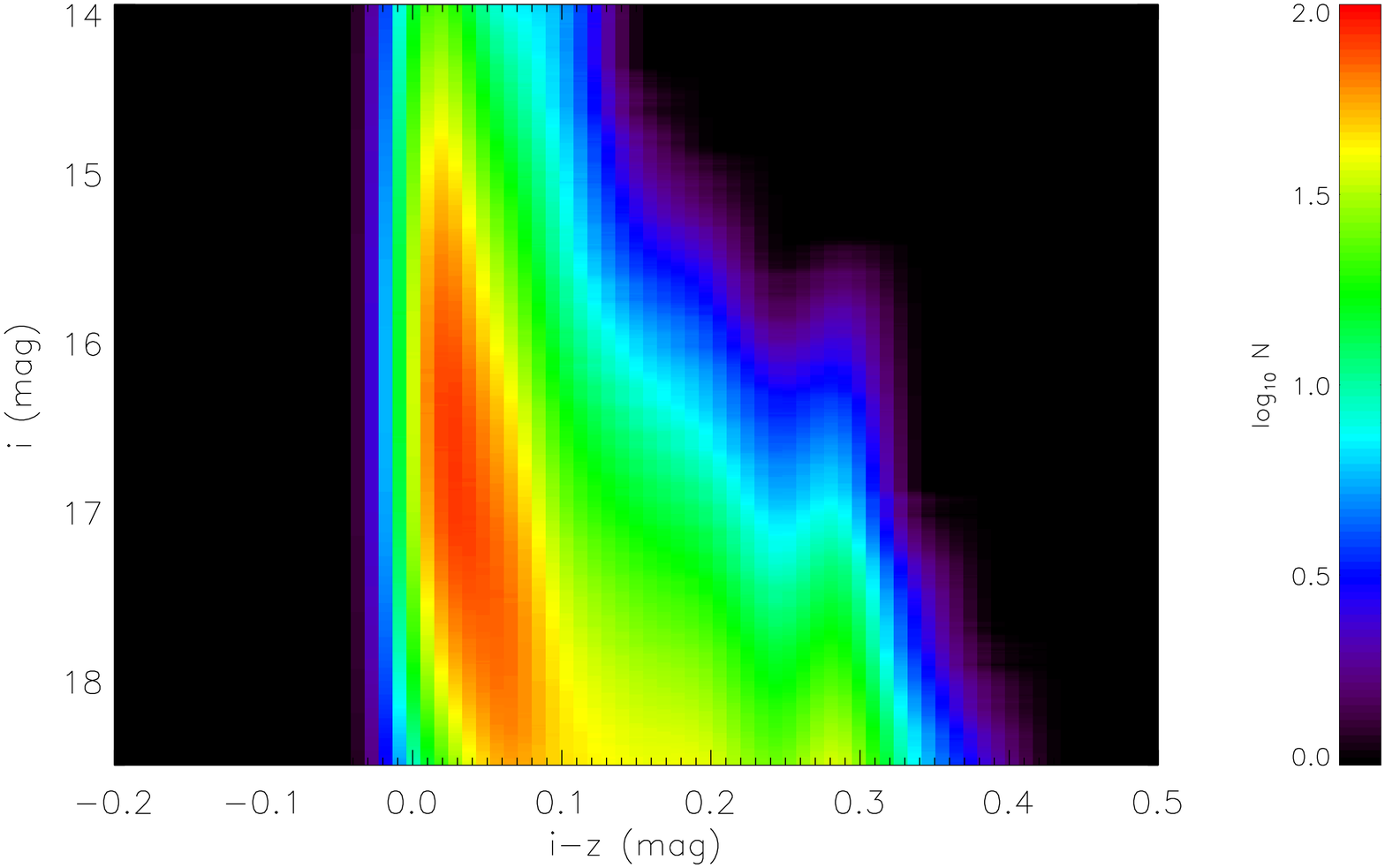}   
\includegraphics[width=0.33\textwidth]{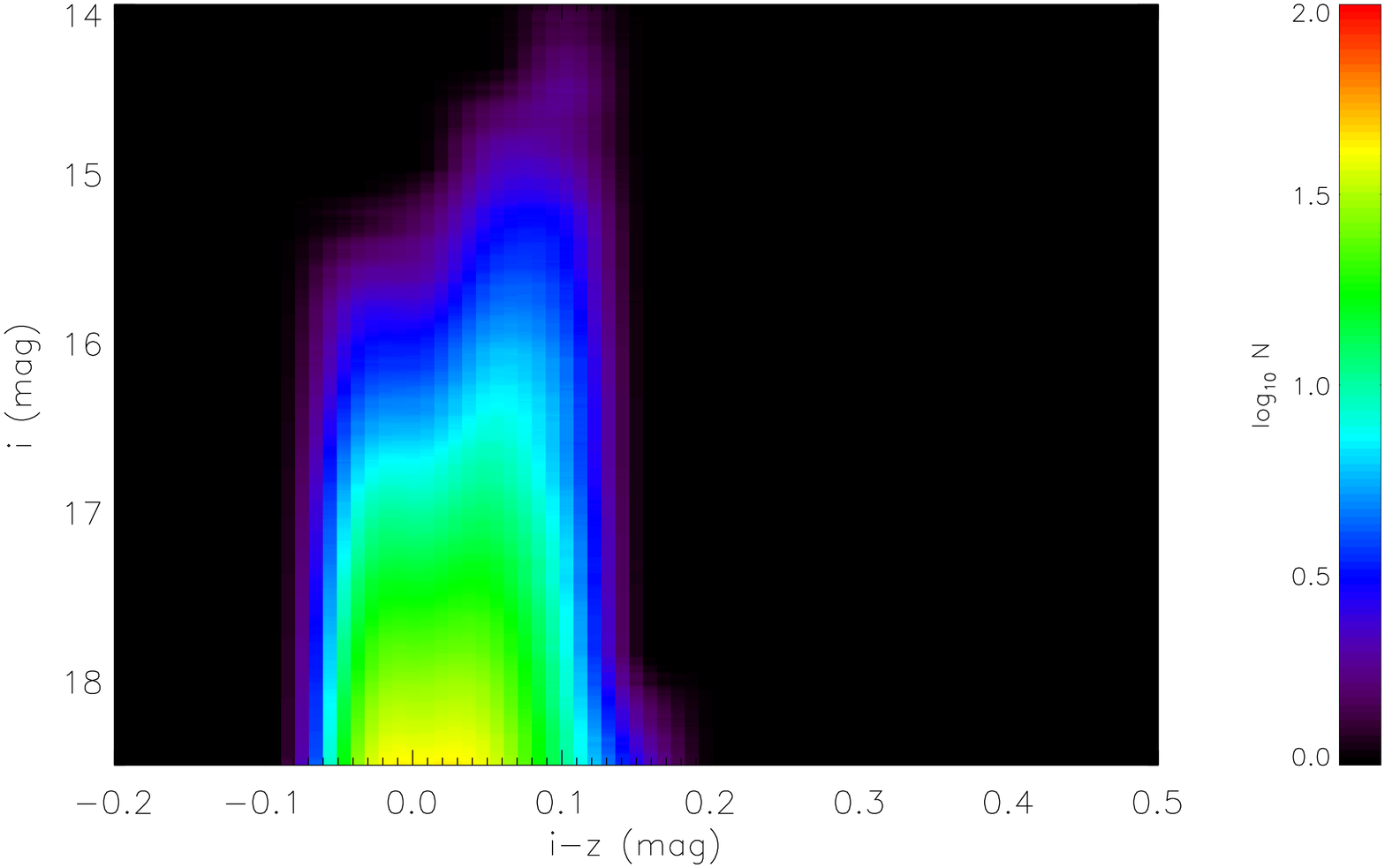}\\ 
  \caption{Same as Fig.\ \ref{fig:ug} but in $(i-z, i)$. }\label{fig:iz}
\end{figure*}

In Figures \ref{fig:ug} -- \ref{fig:iz}, the results are shown for all
CMDs. The upper panels show the Hess diagrams of the data, the model,
and their relative differences (from left to right). In the lower
panels, the contributions of the thin disc, the thick disc, and the
halo (from left to right) as given by the \besan model are shown with
the same colour coding.

We first discuss the results in $(g-r, g)$ for a direct comparison to
the \jj \citep{jj} and the TRILEGAL \citep{2005Girardi} models. The top
right panel in Fig.\ \ref{fig:gr} shows that the relative differences
for most regions are smaller than $50\%$. There are four regions in the
CMD exceeding that level:
\begin{description}
  \item[-] The deep minimum at the lower right corner, where the star count predictions exceed the observations by more than a factor of three (compare the middle and left panels in the top row) is due to M dwarfs in the thin disc model (lower left panel). It seems that the $(g-r)$ colour of the M dwarfs is too blue.
  \item[-] The F turn-off regions of the thick disc (lower middle panel) and the halo (lower right panel) are also blue-shifted by $\sim 0.1$\,mag in $(g-r)$, leading to the negative-positive features at $g=16$ and $g=20$ mag, respectively, in the difference plot.
  \item[-] At the bright end ($g=14$ to 15 mag), the star counts are determined by MS stars of the nearby thin disc, as well as bright F turn-off stars and giants in the thick disc. The thick-disc giants are responsible for the excess of stars on the blue side of the diagram. The maximum at the top right corner is a sign of missing thin-disc K dwarfs or red giants in the thick disc.
  \item[-] The faint blue part at $g>19$ mag and $g-r=0.2$ to 0.4 mag is dominated by the halo. The density of the halo is strongly under-represented by the model. This cannot be seen in the luminosity function (see Fig.\ \ref{fig:lf}), since it is compensated for by the over-estimated number of thin disc M dwarfs at $g-r<1.2$.
  \item[-] The \besan model confirms that the faint, very blue plume at $g-r<0.2$ cannot be reproduced by stellar populations of the Milky Way and is most probably dominated by mis-identified extragalactic sources \citep{Lu01} \citep[see also][]{Schultheis2006}.
\end{description}

In the \jj model, the median relative deviation in the Hess diagram is
$\Delta_\mathrm{med}=5.63\%$. This corresponds to a reduced $\chi^2$
value, measured by the star counts in logarithmic scale, of 4.31. In
the \besan model, typical deviations are five times larger with a
corresponding $\chi^2=408.15$. The median relative deviations of the
different colour diagrams are between $20\%$ and $53\%$. In the CMD (Hess
diagram), the star count distribution at a given colour is mainly
determined by the geometrical structure of the components, whereas the
distribution at a given magnitude is strongly influenced by the
colours of the underlying stellar populations. Since all strong
features in the relative differences between the data and model in  Fig.\
\ref{fig:gr} show a horizontal structure, the main reason for this
disagreement with in the \besan model are the mis-matching properties of the
synthetic stellar populations (see Fig. \ref{fig:iso:besan}). In the
\jj model, the MS properties of the stellar components fit by
construction in the solar neighbourhood.

In the \tri model (Section 4), the colour offsets in the synthetic
stellar populations are much smaller and do not completely dominate
the deviations of the data from the model. For features in the Hess
diagram other than the ones just discussed, the absolute relative difference
has a median value of $0.2576$ corresponding to $\chi^2=104.19$.

In case of the \besan model, the overall differences between the data and model arising
from deviations of the density profiles are large and cannot be
quantified here.

In $(u-g, g)$ (see Fig.\ \ref{fig:ug}) diagram the differences between the data
and the \besan model are much larger than in the $(g-r,g)$ plot. The main
reason is that the isochrones are even more blue-shifted by $\sim
0.3$\,mag in $(u-g)$ (see Fig.\ \ref{fig:iso:besan}). In addition, the
features in the number density distributions of the Hess diagrams are
much wider in colour in the observations than in the model.

In the $(r-i, r)$ (see Fig.\ \ref{fig:ri}) diagram, the thin disc dominates
for $r-i>0.8$, while the thick disc dominates for $0.3<r-i<0.7$. That the largest significant difference between
the model and the observations is interestingly located at the interface of the
populations of the thin and thick disc is caused by a gap at
$0.75<r-i<0.85$ and $20<r<20.5$. Another feature is a pronounced gap
at $r-i\approx0.45$ in the MS of the thin disc (lower left panel),
which is much weaker in the other colours. These minima correspond to the
relative number counts along the MS of the thin disc. At red colours
(low-mass end), the star numbers are determined by the IMF, whereas in
the blue (high-mass end) there is a strong impact of the SFH owing to
the finite MS lifetime of the stars.

The Hess diagram of $(i-z,i)$ (Fig.\ \ref{fig:iz}) represents the best
fit of the \besan model and SDSS data. In most regions of the CMD， the
median absolute difference is about $0.20$.

Altogether，our multi-colour analysis confirms that the synthetic
population properties first of all need to be improved in the \besan
model. In a second step, the IMF, SFH, and the spatial structure should
be investigated with higher precision.

%
\section{Summary}\label{sec:Sum}

We have compared star-count predictions generated by the
\tri and \besan models with the observed sample in the NGP field
($b>80{\degr}$) from the SDSS DR7 photometric stellar catalogue using
de-reddened magnitudes adopting the extinction values of
\citet{Schlegel}. We measured the quality of the agreement between
different models and the data by performing a $\chi^2$ determination of the Hess
diagrams, i.e., the star count distribution in the CMD. As a
comparison, we referred to the \jj model (model A in \citealt{jj,jgv})
with $\langle\chi^2\rangle=4.31$ corresponding to a median relative
deviation between the data and model of $\Delta_\mathrm{med}=5.63\%$ in
$(g-r,g)$. Additionally, fiducial isochrones of five globular clusters,
based on the SDSS observations of \citet{An08} combined with the cluster
parameters of \citet{Harris1996}, were used to test the synthetic
stellar populations of the thick disc and halo modelled by the \tri and
\besan models.

\subsection{The \tri model}

We studied with \tri five different models simulating the NGP region
and compared their output to observational photometric data of this
region from the SDSS.  The web interface of \tri allowed us a
restricted choice of input parameters and functions for the Galactic
components. We compared the default input set (lacking a thick disc
component) with best-fit adjustments for three other Galaxy models from
the literature, namely the \jj model \citep{jj}, the \besan model
\citep{Robin}, and the model of \citet{Juric}. We found that the
default set reproduces best the luminosity functions in the $g$ and
$r$ bands. The $\langle\chi^2\rangle=151.98$ from the Hess diagram
analysis is slightly worse than that based on the \juric model with
$\langle\chi^2\rangle=137.88$. All these models show a typical
discrepancy $\Delta_{\rm{med}}$ of $26\%$ to $46\%$ in the Hess diagram,
which is significantly larger than the disagreements from the \jj model.

A comparison of the colour-absolute magnitude diagrams of the thick disc
and halo with the observed fiducial isochrones of globular clusters
\citep{An08} shows that the MSs and the F-turnoff regions of the
stellar evolution library are roughly consistent in the $g$ and $r$
filters. The small colour offset of $g-r\sim 0.05$\,mag may be due to
age differences among the components. However, the giant branches show
large systematic disagreements. The red colour of the K giants observed
in the globular clusters are not reproduced.

Since the parameter space was too large to find the absolute minimum
$\langle\chi^2\rangle$ with \tri, we determined the parameters for
halo, thick disc, and thin disc sequentially by fitting smaller areas
of the CMD. Finally we found an optimised set 5 with
$\langle\chi^2\rangle=1.33$ and a median deviation $\Delta_{\rm{med}}$
of $26\%$, which is five times larger than in the \jj model. The parameters of
this model are listed in the third-last column of Table
\ref{tab:tri:def}. The reason for the larger deviations is mainly the
restricted choice of thin and thick disc parameters (for a constant SFR
and the adopted vertical density-profile shapes).

The \tri code is a powerful tool for testing Milky Way models against star
count data. The investigation of star number densities in
the CMD (Hess diagrams) are particularly helpful in distinguishing the impacts of the different stellar populations.

Some more progress may be possible by varying the thin disc parameters
further, but the main disagreements are due to fundamental limitations of the models. In the \tri model, the range of input parameters and functions is quite
restricted. More freedom in choosing the SFH, the vertical profiles of the
thin and thick discs and for halo profile would be desirable when
simulating high Galactic-latitude data. The comparison
with fiducial star cluster isochrones has additionally shown that an improvement in
the predicted colours of stars for late stages of stellar evolution in
the $ugriz$ filter system is desirable.

\subsection{The \besan model}

We investigated in a multi-colour analysis the properties of the
\besan model by comparing with SDSS data of the NGP field.  We generated
mock stellar catalogues with the \besan model. The output in $u^*$,
$g'$, $r'$, $i'$, and $z'$ photometries was converted into the SDSS
$u$, $g$, $r$, $i$, and $z$ system using the linear relations provided by
\citet{Tucker2006} and \citet{Ju08}.

The comparison of the colour-absolute magnitude CMDs of the model
output of thick disc and halo with the fiducial globular clusters
covering the respective metallicity ranges, uncovered significant
differences. The MS and turnoff colours are blue-shifted by from 0.05 mag to
0.3 mag, increasing from the red to the blue colour bands. The large
offset in $u-g$ is mainly caused by a zero-point problem in the $u'$
filter of the CFHT-MegaCam reported by \citet{An08}. The giant
branches are fitted well but the models do not extend far enough to the red end of the
CMDs.

A comparison of the luminosity functions in the different filters
shows that there is a good general agreement between the data and the
model. Only for magnitudes $<15$ mag, the star counts are
overestimated. In the $r$ band, there is a deficit of stars at the
faint end at $r>19$ mag. The  contributions of the different
components (thin disc, thick disc, and halo) as a function of apparent
magnitude were plotted to identify the dominating populations. The
fraction of giants are generally of the order of a few percent, which
may be neglected for the sake of our tests.

The \besan model output and the differences from the data were plotted in
Hess diagrams of four colours [$(u-g, g)$, $(g-r, g)$, $(r-i, r)$, and
$(i-z, i)$].  To analyse the impact of the different Galactic
components, the individual Hess diagrams of the three components were
also shown.  The relative difference between the data and the model
cover a range larger than $-100\%$ to $100\%$ in each Hess diagram.
Except for some well-defined features caused by the colour offset of
the isochrones, the median relative differences in the Hess diagrams
are between 0.2 and 0.5 (ignoring the strongly perturbed $(u-g,g)$ diagram).
The largest discrepancies in the Hess diagrams are the follows:
\begin{description} 
   \item[-] The MS turn-off regions of thick disc and halo are
underestimated in terms of the observed colours, meanwhile they are overestimated in terms of the modelled colours owing to the blue-shifted MS in the \besan model.
   \item[-] The halo is strongly underestimated, which is compensated for
in the luminosity functions by M dwarfs of the thin disc, which fall
in the model in the investigated colour range but do not do so in the
data.  
   \item[-] The excess of observed stars in the transition between the
thin disc and the thick disc compared to the model in terms of blue colours
may also be due to the shifts in the synthetic colours produced by.
   \item[-] The prominent gap in the MS of the thin disc at $r\sim
0.45$, which is not present in the SDSS data. This feature appears in
all of the thin disc realizations of the Hess diagrams, but the gap in
the $r-i$ diagram is the most significant one.  
\end{description}

\subsection{General conclusions}

We have investigated the ability of the \tri model to reproduce star
counts in the SDSS filters $(g-r,g)$ and the quality of the \besan
model in the full-filter system $ugriz$ for the NGP field. We have
analysed the number distributions of stars in the CMDs (via Hess
diagrams) instead of only using luminosity functions. We have used the
\jj model \citep{jj,jgv} to quantify the quality of the fits. In
comparison to the \jj model with a median deviation between data and
model of 5.6\%, the \tri model with deviations of the order of 26\%
and the \besan model with median deviations reaching from 20\% to 53\% show
significant deficits. In the \tri model, the main reason for the larger
deviations is the restricted choice of the density profiles for the thin
disc, thick disc, and halo. In the \besan model, the synthetic colours
in the $ugriz$ filters of the stellar populations need first of all to be improved, before the other properties of the Milky Way components
can be investigated further.

In general, Milky Way models based on synthetic stellar populations
such as the \tri and \besan models are powerful tools for producing mock
catalogues of stellar populations for comparison with data. Both models,
\tri and \besan, reproduce the observed SDSS luminosity functions very well.
However, the luminosity functions are insensitive to the assumed structural
properties of the Milky Way components. To improve the
modelled Hess diagrams, the synthetic colours of the stellar
populations in $ugriz$ filters need to be upgraded in both the \tri and the
\besan models. We are looking forward to applying the \tri, the \besan, or
the new Galaxia \citep{Galaxia} tool to the full SDSS data set.

\begin{acknowledgements}

We thank the anonymous referee for thoroughly reading the paper and
providing comments that helped us to improve the manuscript. We are
grateful to Dr. L. Girardi for important discussions about the \tri
model, and to Dr. S. Vidrih for his contributions to the codes used
for the comparisons presented in this paper. SG thanks Dr. A. Robin
for her discussions on the local calibrations of the \besan model.

SG was supported by a scholarship of the China Scholarship Council
(CSC). This research was also supported by the Collaborative Research
Centre (Sonderforschungsbereich SFB) 881 ``The Milky Way System''. 

Funding for the SDSS and SDSS-II has been provided by the Alfred P.
Sloan Foundation, the Participating Institutions, the National Science
Foundation, the U.S. Department of Energy, the National Aeronautics
and Space Administration, the Japanese Monbukagakusho, the Max Planck
Society, and the Higher Education Funding Council for England. The
SDSS Web Site is http://www.sdss.org/.

The SDSS is managed by the Astrophysical Research Consortium for the
Participating Institutions. The Participating Institutions are the
American Museum of Natural History, Astrophysical Institute Potsdam,
University of Basel, University of Cambridge, Case Western Reserve
University, University of Chicago, Drexel University, Fermilab, the
Institute for Advanced Study, the Japan Participation Group, Johns
Hopkins University, the Joint Institute for Nuclear Astrophysics, the
Kavli Institute for Particle Astrophysics and Cosmology, the Korean
Scientist Group, the Chinese Academy of Sciences (LAMOST), Los Alamos
National Laboratory, the Max-Planck-Institute for Astronomy (MPIA),
the Max-Planck-Institute for Astrophysics (MPA), New Mexico State
University, Ohio State University, University of Pittsburgh,
University of Portsmouth, Princeton University, the United States
Naval Observatory, and the University of Washington.
\end{acknowledgements}

\bibliographystyle{aa}

\end{document}